\documentclass[12pt,a4paper,footinclude=true,headinclude=true]{book} % KOMA-Script book
\usepackage[utf8]{inputenc}              
\usepackage{lipsum}
\PassOptionsToPackage{
	drafting=false,
	dottedtoc=true,
	eulerchapternumbers=true,
	linedheaders=true,
	floatperchapter=true,
	eulermath=true,
	style=classicthesis
}{classicthesis}
\usepackage[linedheaders,parts,pdfspacing]{classicthesis} % ,manychapters
\usepackage{comment} 
\usepackage{amsmath}
\usepackage{mathtools, nccmath}
\usepackage{amsfonts}
\usepackage{amssymb}
\usepackage{graphicx}
\usepackage{longtable}
\usepackage{enumitem}
\usepackage{todonotes}
\usepackage{physics}
\usepackage{amsthm}
\usepackage[margin=1 in]{geometry}
\usepackage{physics}
\usepackage{verbatim}
\usepackage{titlesec}
\usepackage{lipsum}
\titleformat*{\chapter}{\fontsize{15}{15}\selectfont}
\titleformat*{\section} {\Large}
\titleformat*{\subsection}{\large}

\newcommand{\unmed}{\frac{1}{2}}

\definecolor{CTsemi}{named}{Maroon}

\begin{document}
	
	\begin{titlepage}
		\setlength{\parindent}{0pt} \setlength{\parskip}{0pt}
	
		\begin{center}
			\vfill 
			
			\begin{minipage}{\textwidth}

				\begin{center}
					\includegraphics[width=4.5cm]{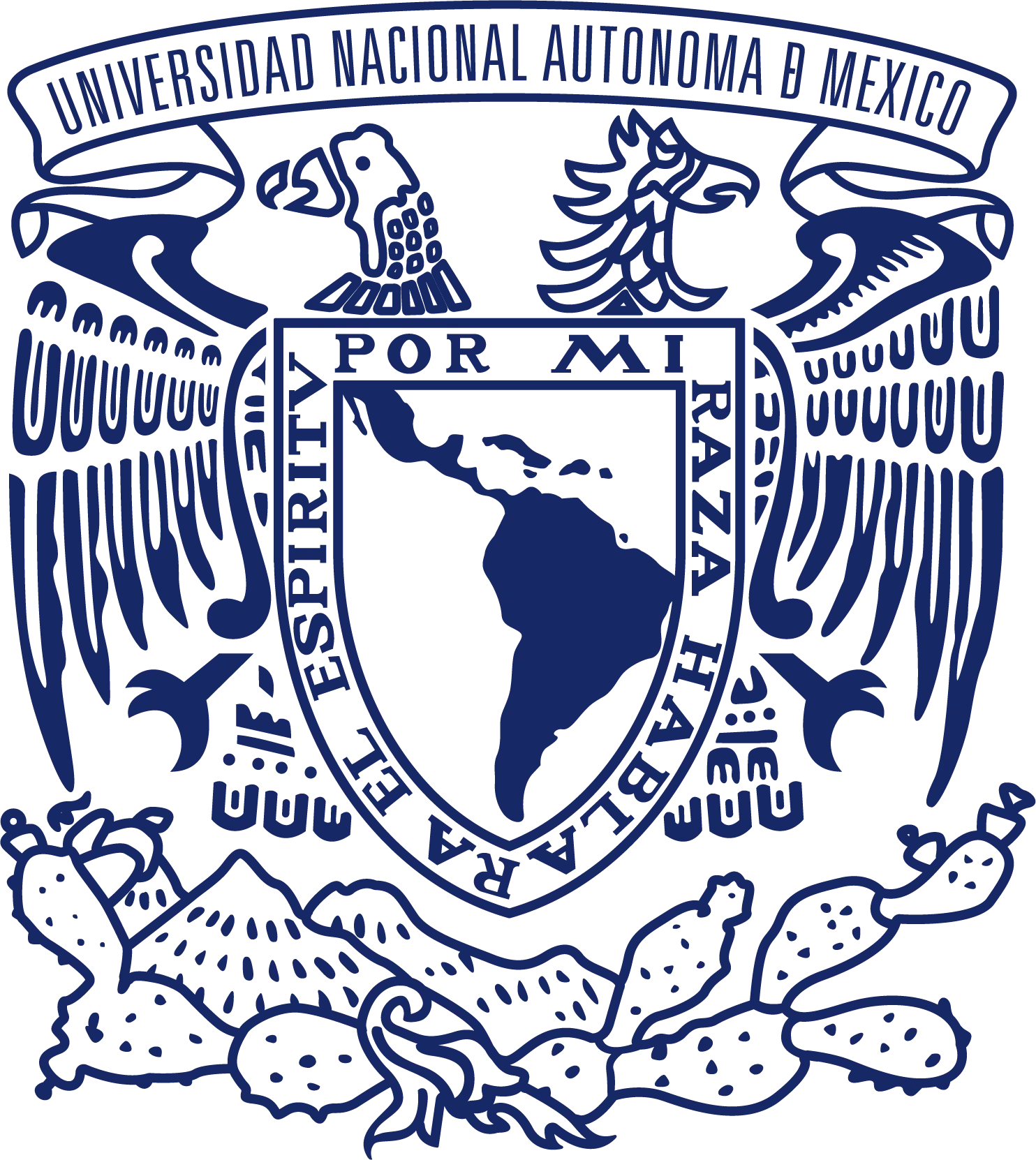}	
				\end{center} 	
				\begin{center}
					\Huge\textbf{UNIVERSIDAD NACIONAL AUT\'ONOMA DE M\'EXICO}
				\end{center} 
				
			\end{minipage}
				
			\vfill
			
			\begin{minipage}{0.7\textwidth}
				\begin{center}
					\large POSGRADO EN CIENCIAS FÍSICAS\\
					\large INSTITUTO DE CIENCIAS NUCLEARES
				\end{center}
			\end{minipage}
			\begin{center}
				
				\vfill
				%\medskip \rule{.9\textwidth}{2pt}
				\large\textbf{QUANTUM INFORMATION GEOMETRY AND ITS CLASSICAL ASPECT}				%{\Large \mathbfseries  \title   }
				
				%\medskip \rule{.9\textwidth}{2pt}
			\end{center}
			
			\begin{center}
				\vspace*{0.7cm}
				{\huge T E S I S}\\
				\vspace*{0.7cm}
				que para optar por el grado de \\\bigskip
				{\large {MAESTRO EN CIENCIAS (FÍSICA)}} \\\bigskip
				
				%PRESENTA:\\\medskip
					PRESENTA: \\
					\large SERGIO JAVIER BUSTOS JUÁREZ
				
				\bigskip
				
				TUTOR PRINCIPAL: \\
				DR. JOSÉ DAVID VERGARA OLIVER, INSTITUTO DE CIENCIAS NUCLEARES, UNAM\\
				%Coadvisors
				\vspace{0.5cm}
				COMITÉ TUTOR:\\
				DR. ÁNGEL SÁNCHEZ CECILIO, FACULTAD DE CIENCIAS, UNAM\\
				DR. YURI BONDER GRIMBERG, INSTITUTO DE CIENCIAS NUCLEARES, UNAM\\
				%\coadvisor1 \\\coadvisor2 \\\coadvisor3 
			\end{center}
			
			\vfill
			
			\begin{center}
				{Ciudad Universitaria, Cd. Mx., junio de 2022}\\
			\end{center}
			\cleardoublepage
		\end{center}
	\end{titlepage}
	
		\newpage
	\vspace*{0.3\textheight}
	\begin{flushright}%
		\Large\emph{Dedicado a mis papás Ana y Arturo.}\\
		\textit{Uno de mis primeros logros,}\\
		\textit{todo gracias a su esfuerzo.}
		\thispagestyle{empty}
	\end{flushright}
	%Contents

	\tableofcontents

	\chapter*{Abstract}  \chaptermark{Abstract} 
	\addcontentsline{toc}{chapter}{Abstract}

        Throughout this work, we will study some of the most important concepts in the area of quantum information geometry as well as the relations between them. We will emphasize the characteristics that arise because they were defined using a quantum mechanical framework and highlight which parts of them cannot be attained under a classical treatment. However, we will show that when the state is Gaussian, we can use classical analogs to obtain the same mathematical results, thus creating a tool that facilitates calculations for such cases since with them we only need to manipulate classical functions.\newline

	First, we introduce some ideas from quantum field theory that will serve as a base for the proofs behind the expressions given in the rest of the work. Then we examine the structure of parameter space utilizing the fidelity and the Quantum Geometric Tensor, composed of the Quantum Metric Tensor and the Berry curvature. The former gives us a way to measure distances between quantum states in parameter space, and the latter is related to Berry's phase, which governs quantum interference.\newline
	
    We then present the quantum covariance matrix, show how it can be linked to the QGT, and discuss how it can be used to study entanglement between quantum systems by obtaining the purity, linear entropy, and von Neumann entropy. As examples, we calculate all these quantities for several systems, including the Stern-Gerlach, a two qubits system, two symmetrically coupled harmonic oscillators, and N coupled harmonic oscillators. \newline

	To commence the final part of this thesis, which is focused on classical analogs, we discuss why certain quantum phenomena cannot be replicated when using a classical framework and the differences that arise when one concept is used in a classical or quantum context. With this in mind, we analyze how the aforementioned quantum concepts could be applied in a classical sense, in the same way as Hannay did in 
 \cite{classical Hannay} with the Berry phase.\newline
	
	Particularly we examine classical analogs of the Quantum Geometric Tensor, containing within it those for the Quantum Metric Tensor and Berry's curvature (which, in this case, its analog will be related to Hannay's angle), and also one for the quantum covariance matrix. At this point, we use the fact that when our state is Gaussian, all the
    information needed to produce the purity, linear entropy, and von Neumann entropy is contained within the quantum covariance matrix, so using its classical analog as a starting point, we generate classical analogs for each of these derived quantities, which in turn will yield information of the separability of our classical systems.\newline
	
	We conclude this work with calculations of these classical analogs for the same harmonic systems that we examined using the quantum formalism; we obtain the exact same results given that the studied states are Gaussian.

	\chapter*{Resumen en español}  \chaptermark{Resumen en español} 
		\addcontentsline{toc}{chapter}{Resumen en español}

    A lo largo de este trabajo estudiaremos a profundidad algunos de los conceptos más importantes del área de geometría de la información cuántica así como las relaciones que tienen entre ellos. Haciendo énfasis en discutir las características que poseen debido a ser cantidades definidas dentro de un marco teórico cuántico y resaltar las partes de ellos que no es posible obtener si se les estudia bajo un tratamiento clásico. Sin embargo, mostraremos que si el estado en cuestión es Gaussiano podremos usar análogos clásicos para obtener los mismos resultados matemáticos, creando así una herramienta matemática que nos facilita el cálculo para tales situaciones, en el sentido de que solamente será necesario manipular funciones clásicas. \newline

Primero introduciremos ideas provenientes de la Teoría Cuántica de Campos las cuales nos servirán como base para las demostraciones de las expresiones utilizadas en el resto del trabajo. Posteriormente examinaremos la estructura del espacio de parámetros utilizando la fidelidad y el Tensor Geométrico Cuántico, el cual se compone del Tensor Métrico Cuántico y la curvatura de Berry. La primera nos proporciona una manera de medir distancias entre estados en el espacio de parámetros mientras que la segunda está relacionada con la fase de Berry, la cual gobierna la interferencia cuántica.\newline

 Luego introducimos la matriz de covarianza cuántica, mostrando como se puede asociar al TGC, y discutimos cómo se puede utilizar para estudiar el entrelazamiento entre sistemas cuánticos obteniendo de ella la pureza, entropía lineal y la entropía de von Neumann. Como ejemplos calculamos todas estas cantidades para distintos sistemas, incluyendo el Stern-Gerlach, uno descrito utilizando dos qubits, dos osciladores armónicos simétricamente acoplados y N osciladores simétricamente acoplados.\newline

Para comenzar la última parte de la tesis, la cual se centra en los análogos clásicos, discutimos primeramente porque ciertos fenómenos cuánticos no pueden ser replicados al utilizar un marco teórico clásico, así como las diferencias que surgen en un concepto cuando se le utiliza bajo un contexto ya sea clásico o cuántico. Con esto en mente analizamos cómo utilizar  las cantidades cuánticas discutidas previamente dentro de un tratamiento clásico, del mismo modo que lo hizo Hannay  \cite{classical Hannay} con la fase de Berry. \newline
    
Examinaremos análogos cuánticos del Tensor Geométrico Cuántico, el cual ya contiene los del Tensor Métrico Cuántico y el de la curvatura de Berry (que en este caso se relaciona con el ángulo de Hannay), así como uno para la matriz de covarianza cuántica. En este punto utilizamos el hecho de que cuando nuestro estado  es Gaussiano, toda la información necesaria para generar la pureza, la entropía lineal y la entropía de von Neumann, está contenida dentro de la matriz de covarianza cuántica, por lo que partiendo de su análogo clásico podemos generar análogos clásicos para cada una de estas cantidades, y estas a su vez nos proporcionarán información acerca de la separabilidad de nuestros sistemas clásicos.\newline

Concluimos el trabajo con el cálculo de estos análogos para los mismos sistemas tratados bajo el formalismo cuántico, obteniendo exactamente los mismos resultados si nuestro estado es Gaussiano.

\chapter*{Agradecimientos} \chaptermark{Agradecimientos} 
		\addcontentsline{toc}{chapter}{Agradecimientos}

Agradezco a mis papás Ana y Arturo por su enorme apoyo y cariño. Soy infinitamente afortunado al tener unos padres que han logrado construir un hogar lleno de amor. Conforme más crezco, más reconozco y valoro los grandes sacrificios que han hecho por mi y para que tenga la mejor vida posible, todos mis logros siempre serán también suyos.\newline

Luis, gracias por siempre recordarme que no todo en la vida son los estudios ni la investigación. Cada vez que estoy contigo mis días se vuelven muy divertidos, no podría existir un mejor hermano para mi.\newline

Dr. David Vergara, sin su paciencia, experiencia y guía no sería la persona que soy ahora. Como mi mentor, espero poder retribuir todo el tiempo y esfuerzo que ha invertido en mi con trabajos de calidad y siendo el mejor físico que pueda. Siempre tendré en cuenta todas sus enseñanzas tanto profesionales como personales.\newline

Dra. Gabriela Murguía, gracias por permitirme crecer como docente a su lado y por enseñarme lo que es apoyar a los estudiantes incluso fuera del salón de clases.\newline

Cursar una maestría en física durante la pandemia no fue nada fácil, pero me considero dichoso de haber podido contar con mis amigos aunque sea para platicar un rato de la vida. Muchas gracias Mariana, Alejandro, Rodo, Pepe, Teo y Dulce, por hacer estos años difíciles más felices.\newline

Muchísimas gracias a los Dres. Ángel Sánchez Cecilio y Yuri Bonder Grimberg por estar al pendiente de mi, tanto personalmente como de mi avance académico a lo largo de toda la maestría.\newline

Agradezco de sobremanera a los Dres. Alberto Martín, Isaac Pérez, Saúl Ramos, y Andrea Valdés, por sus comentarios, sugerencias y pláticas que me permitieron mejorar ampliamente la calidad de este trabajo.\newline

Gracias a CONACyT por la beca número $758382$ concedida por dos años para realizar mi maestría.\newline

Gracias al Proyecto UNAM-PAPIIT IN$105422$ "Información cuántica en teoría de campos y sistemas afines" por la beca de maestría otorgada para la elaboración de esta tesis.\newline

	\chapter*{Introduction}  \chaptermark{Introduction} 
	\addcontentsline{toc}{chapter}{Introduction}

Entanglement is the quintessential quantum effect since there is no equivalence for it in classical mechanics, and it tells us that even if parts of our system are non-interacting and light-years apart, when they are entangled one can affect the measurement of the other. \newline

What began as a thought experiment in the famous Einstein-Podolsky-Rosen paper \cite{EPR}, has sparked several decades of research which continue up to this day. Although it should be noted that the implications of entanglement on the foundations of quantum mechanics remained mostly in the realm of philosophy for almost 30 years until John Bell's insightful paper \cite{Bell} (and its complete experimental verification by A. Aspect and his team \cite{Aspect1, Aspect2, Aspect3}) showed us in pure mathematical form that there cannot exist a theory of local hidden variables (such as the one desired by EPR) that successfully reproduces all the predictions of quantum mechanics, making it impossible to construct a classical theory that triumphantly describes our universe. This is one of the few ideas (with their corresponding experimental verification) that have imposed such revolutionary changes to our philosophical understanding of our natural world, since it tempers with concepts such as realism and locality, things that we take for granted in our classical intuition.\newline

Within this last century our perspective on these "quantum only" phenomena has changed from an undesired byproduct to a fully exploitable resource studied scrupulously in their own branch of physics, \textit{Quantum Information Theory}, while also being used to generate futuristic quantum technologies including\textit{ quantum teleportation }and \textit{quantum computing} \cite{Nielsen}. \newline

The ideas of quantum entanglement are regularly understood in simple systems involving just a few qubits (as we will see in Chapter 3), but they can also be present in continuum systems. In recent years there has been an increasing interest in studying entanglement between quantum fields in the context Gauge/Gravity duality, including the ideas
 of Reeh and Schlieder \cite{Reeh, and all that}, Srednicki \cite{Srednicki} and Bombelli \cite{Bombelli} that use the entanglement between quantum coupled oscillators as a stepping stone to characterize the entropy of a black hole.\newline

 The main tools to measure entanglement between subsystems are the purity and von Neumann entropy. There are several different ways to calculate them, the standard one is using the density matrix, but if our state is Gaussian we can use the quantum covariance matrix. We will see that the Quantum Geometric Tensor, which incorporates all the information about distances between states in parameter space and quantum interference, is closely related to the quantum covariance matrix and thus to the purity and entropy of Gaussian states. Taking as inspiration \cite{classical Hannay, classical QMT}  and \cite{classical QGT} we will construct classical analogs of the quantum covariance matrix, and from it classical analogs of the purity and von Neumann entropy. We generate these mathematical tools longing for them to be able to get accurate results even when the objects of study are quantum fields, with the only condition being that the state in question is Gaussian. In this work we do not get that far, but we settle all the basis needed in order to do so, following closely the steps taken by Srednicki in \cite{Srednicki}.\newline

 This thesis is divided in 5 distinct chapters:
 
 \begin{itemize}
     \item Starting with chapter 1 where we discuss the main mathematical tools that are needed in order to understand the concepts and proofs that will come in the following chapters.
     
     \item In chapter 2 we introduce the central concepts of quantum information geometry for this thesis, such as the fidelity, Quantum Metric Tensor, Berry's curvature and the Quantum Geometric Tensor. We also see how they can be used to predict quantum phase transitions within our physical systems.
     
     \item Chapter 3 is the core of the work, in it we present the quantum covariance matrix and how to relate it to the QGT. We present the purity, linear entropy and von Neumann entropy and how to obtain them using the density matrix. We also show that if the state in question is Gaussian (which we also define here) its possible to calculate them using only the quantum covariance matrix. To close this chapter we meticulously calculate all these important quantities using both the aforementioned methods for 4 distinct examples, consisting of the Stern-Gerlach, a two qubit system, two coupled harmonic oscillators and the N coupled harmonic oscillators that Srednicki uses to calculate the entropy of black holes, showing the advantages and disadvantages of each of the techniques.
     
     \item With chapter 4 we initiate the second part of the thesis, in which we construct our classical analogs. We commence it with a discussion on why several properties of our universe only emerge through a quantum framework and not in a classical one, even if the concepts used in both are the same. Then we make a brief review of the action-angle variables since they will be our main tool to generate the classical analogs for all the previously mentioned quantities. With them we study two of the most important previously stablished classical analogs, the one of Berry's Phase, Hannay's angle  \cite{classical Hannay}, and the one for the QGT which encompasses it \cite{classical QMT, classical QGT}.
     
     \item We close this work with chapter 5, in which we construct our classical analogs for the quantum covariance matrix and its derived quantities in the case that our state of study is a Gaussian states, the purity, linear entropy and von Neumann entropy, showing that we get the exact same mathematical results that were obtained with the quantum calculation.

 \end{itemize}

\part{Quantum Information Geometry}

\chapter{Path Integrals, Green functions and generating functionals}

In order to fully understand the QGT and its link with the quantum covariance matrix we need to be familiar with some of the most important ideas used in Quantum Field Theory. The first concept that we will study in this thesis is Feynman's path integral since it will be fundamental to construct the Hamiltonian formulation of the QGT. Then we will focus on Green's functions, generating functionals and finally the perturbative approach to calculate Green's functions. This last method will be useful to show the power of the previously mentioned formulation of the QGT.

\section{Path Integrals and Green Functions}

One of the main problems in quantum mechanics is finding out the transition probability amplitude of a particle that has an initial position $q_0$ at time $t_0$ and will later be found at $q$ with time $t$. Perhaps the most ingenious way to solve it is by the path integral approach, which we can obtain by taking the braket of its initial state $\ket{q_0,t_0}$ and the final state $\ket{q,t}$, i.e. $\braket{q ,t}{q_0 ,t_0}$ \footnote{This is also known as the Kernel of Schrödinger's equation since $\left(i \hslash \frac{\partial}{\partial t}-\hat{H}\right)\braket{q,t}{q_0, t_0} = 0 $.} and divide the time interval between the initial and final time by introducing complete sets of coordinate basis states for every intermediate time point \footnote{A beautiful explanation of the idea behind the path integral can be found in Zee's book of QFT \cite{Zee}.}. What this really does is take into account every possible path between the initial and final states, but each one is weighted by the particular action that governs the system \cite{A Das}.\newline

By using the identity
\begin{equation}
\int_{-\infty}^{\infty} d q_{i}\left|q_{i}, t_{i}\right\rangle\left\langle q_{i}, t_{i}\right|=\mathbb{I}
\end{equation}
we can write the braket of our initial and final states as
\begin{equation}
 \braket{q ,t}{q_0 ,t_0}=\int_{-\infty}^{\infty} d q_{1}\left\langle q, t \mid q_{1}, t_{1}\right\rangle\left\langle q_{1}, t_{1} \mid q_{0}, t_{0}\right\rangle,
\end{equation}
with the condition that $t > t_1 > t_0$. Repeating this process $N$ times, meaning that we partition the time interval $[t_0,t]$ in $N+1$ equal parts such that 
\begin{equation}t > t_N > t_{N-1} > \cdots > t_2 > t_1 > t_0\end{equation}
where $ \Delta t = t_{j+1}- t_j $ for every $j$, allows us to formulate
\begin{equation} \label{2 intermedio}
\begin{aligned}
 & \braket{q ,t}{q_0 ,t_0}\\&=\int_{-\infty}^{\infty} d q_{N} \cdots d q_{1}\left\langle q, t \mid q_{N}, t_{N}\right\rangle\left\langle q_{N}, t_{N} \mid q_{N-1}, t_{N-1}\right\rangle \cdot\left\langle q_{2}, t_{2} \mid q_{1}, t_{1}\right\rangle\left\langle q_{1}, t_{1} \mid q_{0}, t_{0}\right\rangle.
\end{aligned}
\end{equation}

Now we need to simplify each of these terms. For example if we focus our attention to
\begin{equation}
\left\langle q_{j+1}, t_{j+1} \mid q_{j}, t_{j}\right\rangle=\left\langle q_{j+1}\left|e^{\frac{-i}{\hslash} \Delta t \hat{H}}\right| q_{j}\right\rangle
\end{equation}
we must introduce another identity operator, but this time in terms of the conjugate momenta $p$, and then use the Taylor series of the exponential to apply the Hamiltonian operator, this is
\begin{align}
\left\langle q_{j+1}\left|e^{\frac{-i}{\hslash} \Delta t \hat{H}}\right| q_{j}\right\rangle &=\int_{-\infty}^{\infty} d p_{j}\left\langle q_{j+1} \mid p_{j}\right\rangle\left\langle p_{j}\left|e^{\frac{-i}{\hslash} \Delta t \hat{H}}\right| q_{j}\right\rangle \\
&=\int_{-\infty}^{\infty} d p_{j}\left\langle q_{j+1} \mid p_{j}\right\rangle\left\langle p_{j}\left|1-\frac{i}{\hslash} \Delta t \hat{H}\right| q_{j}\right\rangle \\
&=\int_{-\infty}^{\infty} \frac{d p}{2 \pi \hslash} e^{\frac{i}{\hslash} p\left(q_{j+1}-q_{j}\right)}\left(1-\frac{i}{\hslash} \Delta t H\left(q_{j}, p_{j}\right)\right) \\
&=\int_{-\infty}^{\infty} \frac{d p}{2 \pi \hslash} e^{\frac{-i}{\hslash} \Delta t H\left(q_{j}, p_{j}\right)+\frac{i}{\hslash} p_{j}\left(q_{j+1}-q_{j}\right)}.
\end{align}
What we have accomplished here is that since $H\left(q_j,p_j\right)$ is the eigenvalue of the Hamiltonian operator, we got rid of every operator in the integral.\newline

By repeating this process $N$ more times, one for each braket in \eqref{2 intermedio}, we arrive at
\begin{equation}\label{3 intermedio}
\begin{aligned}
\left\langle q, t \mid q_{0}, t_{0}\right\rangle &= \int_{-\infty}^{\infty} d q_{1} \cdots d q_{N} \frac{d p_{0}}{2 \pi \hslash} \cdots \frac{d p_{N}}{2 \pi \hslash} e^{\frac{i}{\hslash} p_{N}\left(q-q_{N}\right)} e^{\frac{i}{\hslash} p_{N-1}\left(q_{N}-q_{N-1}\right)} \ldots \\
& \cdots e^{\frac{i}{\hslash} p_{1}\left(q_{2}-q_{1}\right)} e^{\frac{i}{\hslash} p_{0}\left(q_{1}-q_{0}\right)} e^{\frac{-i}{\hslash} \Delta t H\left(q_{N}, p_{N}\right)} \cdots e^{\frac{-i}{\hslash} \Delta t H\left(q_{0}, p_{0}\right)},
\end{aligned}
\end{equation}
and since our time interval is $\Delta t = t_{j+1}- t_j$ we can express $q_{j+1}$ in the form
\begin{equation}
q_{j+1}=q\left(t_{j+1}\right)=q\left(t_{j}+\Delta t\right),
\end{equation}
which we can expand as
\begin{align}
q_{j+1} &=q\left(t_{j}\right)+\Delta t \dot{q}\left(t_{j}\right) +  \frac{1}{2}\Ddot{q}(t_j){\Delta t}^2+\cdots\\
&\approx q_{j}+\Delta t \dot{q}_{j}.
\end{align}
Now, to get the continuous limit of the partition we let $N\to \infty$ in \eqref{3 intermedio} as
\begin{equation}
\left\langle q, t \mid q_{0}, t_{0}\right\rangle=\lim _{N \rightarrow \infty} \int_{-\infty}^{\infty} d q_{1} \cdots d q_{N} \frac{d p_{0} \cdots d p_{N}}{(2 \pi \hslash)^{N+1}} e^{\frac{i}{\hslash} \sum_{j=0}^{N} p_{j}\left(q_{j+1}-q_{j}\right)} e^{\frac{-i}{\hslash} \Delta t \sum_{j=0}^{N} H\left(q_{j}, p_{j}\right)}.
\end{equation}
and by defining
\begin{equation}
\lim _{N \rightarrow \infty} d q_{1} \cdots d q_{N}=\mathcal{D} q
\end{equation}
\begin{equation}
\lim _{N \rightarrow \infty} \frac{d p_{0} \cdots d p_{N}}{(2 \pi \hslash)^{N+1}}=\mathcal{D} p
\end{equation}
we write it in the compact fashion
\begin{align}
\left\langle q, t \mid q_{0}, t_{0}\right\rangle &=\int_{-\infty}^{\infty} \mathcal{D} q \mathcal{D} p \lim _{N \rightarrow \infty} e^{\frac{i}{\hslash} \sum_{j=0}^{N}\left(p_{j} \dot{q}_{j}-H\left(q_{j}, p_{j}\right)\right) \Delta t} \\
&=\int_{-\infty}^{\infty} \mathcal{D} q \mathcal{D} p e^{\frac{i}{\hslash} \int_{t_{0}}^{t} d t(p \dot{q}-H(q, p))}
\end{align}
Since the Hamiltonian is the Legendre transformation of the Lagrangian
\begin{equation}
H=p \dot{q}-L
\end{equation}
and the action is defined as 
\begin{equation}S =\int L dt,\end{equation}
equivalently we formulate
 
\begin{equation}
\left\langle q, t \mid q_{0}, t_{0}\right\rangle=\int_{-\infty}^{\infty} \mathcal{D} q \mathcal{D} p e^{\frac{i}{\hslash} \mathcal{S}}.
\end{equation}
which is the \textit{path integral formulation} in terms of $\mathcal{D}q$ and $\mathcal{D}p$. However, it is possible to leave it only in terms of $\mathcal{D}q$ by considering in \eqref{3 intermedio} that our Hamiltonian is $H(q, p)=\frac{p^{2}}{2 m}+V(q)$, then:
\begin{align}
\left\langle q, t \mid q_{0}, t_{0}\right\rangle &=\lim _{N \rightarrow \infty} \int_{-\infty}^{\infty} d q_{1} \cdots d q_{N} \frac{d p_{0}}{2 \pi \hslash} \cdots \frac{d p_{N}}{2 \pi \hslash} e^{\frac{i}{\hslash} \sum_{j=0}^{N}\left(p_{j}\left(q_{j+1}-q_{j}\right)-\Delta t H(p, q)\right)} \\
&=\lim _{N \rightarrow \infty} \int_{-\infty}^{\infty} d q_{1} \cdots d q_{N} \frac{d p_{0}}{2 \pi \hslash} \cdots \frac{d p_{N}}{2 \pi \hslash} e^{\frac{i}{\hslash} \Delta t \sum_{j=0}^{N}\left\{p_{j} \dot{q}_{j}-\left(\frac{p_{j}^{2}}{2 m}+V(q)\right)\right\}},
\end{align}
and using the following result
\begin{equation}
\int_{-\infty}^{\infty} \frac{d p_{j}}{2 \pi \hslash} e^{\frac{i}{\hslash} \Delta t\left(p_{j} \dot{q}_{j}-\frac{p_{j}^{2}}{2 m}\right)}=\sqrt{\frac{2 m \pi \hslash}{i \Delta t}} e^{\frac{i \Delta t}{\hslash} \frac{m \dot{q}_{j}^{2}}{2}},
\end{equation}
we get
\begin{align}
\left\langle q, t \mid q_{0}, t_{0}\right\rangle &=\lim _{N \rightarrow \infty} \int_{-\infty}^{\infty} \frac{d q_{1} \cdots d q_{N}}{(2 \pi \hslash)^{N+1}}\left(\frac{2 m \pi \hslash}{i \Delta t}\right)^{\frac{N+1}{2}} e^{\frac{i \Delta t}{\hslash}\left\{\sum_{j=0}^{N}\left(\frac{m}{2}{q_{j}}^{2}-V(q)\right)\right\}} \\
&=\int_{-\infty}^{\infty} d q_{1} \cdots d q_{n}\left(\frac{m}{2 i \pi \hslash \Delta t}\right)^{\frac{N+1}{2}} e^{\frac{i}{\hslash} \int_{t_{0}}^{t} d \tau L(q(\tau), \dot{q}(\tau))}.
\end{align}
The difference in this process is that we redefine the measure of the path integral as 
\begin{equation}
\mathcal{D} q=d q_{1} \cdots d q_{n}\left(\frac{m}{2 j \pi \hslash \Delta t}\right)^{\frac{N+1}{2}},
\end{equation}
which leaves us at the most common expression for Feynman's path integral:
\begin{equation}\label{path integral}
\left\langle q, t \mid q_{0}, t_{0}\right\rangle=\int_{q\left(t_{0}\right)=q_{0}}^{q(t)=q} \mathcal{D} q e^{\frac{i}{h} \int_{t_{0}}^{t} d \tau L(q(\tau), \dot{q}(\tau))}
\end{equation}

It is also important to note that in the case that we have a position operator acting on our ket $\left\langle q, t\left|\hat{q}\left(t_{j}\right)\right| q_{0}, t_{0}\right\rangle$ we can follow the same procedure, i.e. inserting identity operators as
\begin{align}
\left\langle q, t\left|\hat{q}\left(t_{j}\right)\right| q_{0}, t_{0}\right\rangle &\label{sandwich campos}=\int_{-\infty}^{\infty} d q_{j}\left\langle q, t \mid q_{j}, t_{j}\right\rangle\left\langle q_{j}, t_{j}\left|\hat{q}\left(t_{j}\right)\right| q_{0}, t_{0}\right\rangle \\
&=\int_{-\infty}^{\infty} d q_{j}\left\langle q, t \mid q_{j}, t_{j}\right\rangle q\left(t_{j}\right)\left\langle q_{j}, t_{j} \mid q_{0}, t_{0}\right\rangle
\end{align}
to get
\begin{equation}
\left\langle q, t\left|\hat{q}\left(t_{j}\right)\right| q_{0}, t_{0}\right\rangle=\int \mathcal{D} q(\tau) q\left(t_{j}\right) e^{\frac{i}{\hslash} \int_{t_{0}}^{t} d \tau L},
\end{equation}
where $t_{0} \leq \tau_{1} \leq t_{j}$ and $t_{j} \leq \tau_{2} \leq t$, also $\mathcal{D} q\left(\tau_{1}\right) d q_{j} \mathcal{D} q\left(\tau_{2}\right)=\mathcal{D} q(\tau)$. We can generalize this result in the way that if we have $n$ operators inside \eqref{sandwich campos} we obtain
\begin{equation}\label{n campos ordenados}
\left\langle q, t\left|T\left(\hat{q}\left(t_{1}\right) \ldots \hat{q}\left(t_{n}\right)\right)\right| q_{0}, t_{0}\right\rangle=\int \mathcal{D} q(\tau) q\left(t_{1}\right) \ldots q\left(t_{n}\right) e^{\frac{i}{\hslash} \int_{t_{0}}^{t} d \tau L},
\end{equation}
where $T \left(\hat{q}\left(t_1\right) \ldots \hat{q} \left(t_n\right)\right)$ stands for the temporal ordered (or normal ordered) operators.

\subsection{Green's Functions}

The Green's function of a system is denoted by
\begin{equation}\label{Green particulas}
G_{n}\left(t_{1}, \cdots, t_{n}\right)=\left\langle 0\left|T \hat{q}\left(t_{1}\right) \ldots \hat{q}\left(t_{n}\right)\right| 0\right\rangle,
\end{equation}
where $\ket{0}$ is the ground state said system, with the lower energy possible as its eigenvalue
\begin{equation}
\hat{H}|0\rangle=E_{0}|0\rangle.
\end{equation}
The rest of the eigenstates of $H$ are denoted by $|n\rangle$ such that $H|n\rangle=E_{n}|n\rangle$ with $E_{n}>E_{n-1}$, and with all these states we construct the identity operator $\mathbb{I}=\sum_{n}|n\rangle\langle n|$.\newline

There is however an alternative expression for the Green's function in terms of the path integral. To get it we once again consider the transition amplitude $\braket{q,t}{q_0,t_0}$ and extract from it the time dependence in terms of the Hamiltonian,
\begin{equation}
    \left\langle q, t \mid q_{0}, t_{0}\right\rangle =\left\langle q\left|e^{-\frac{i}{\hslash}\left(t-t_{0}\right) \hat{H}}\right| q_{0}\right\rangle 
\end{equation}
then we expand it in terms of its eigenvectors and the energy eigenvalues as follows:
\begin{align}
\left\langle q, t \mid q_{0}, t_{0}\right\rangle 
&=\sum_{n}\left\langle q\left|e^{-\frac{i}{\hslash}\left(t-t_{0}\right) \hat{H}}\right| n\right\rangle\left\langle n \mid q_{0}\right\rangle \\
&=\sum_{n} e^{-\frac{i}{\hslash}\left(t-t_{0}\right) E_{n}}\langle q \mid n\rangle\left\langle n \mid q_{0}\right\rangle \\
&=e^{-\frac{i}{\hslash}\left(t-t_{0}\right) E_{0}}\left[\langle q \mid 0\rangle\left\langle 0 \mid q_{0}\right\rangle+\sum_{n \geq 1}\langle q \mid n\rangle\left\langle n \mid q_{0}\right\rangle e^{-\frac{i}{\hslash}\left(t-t_{0}\right)\left(E_{n}-E_{0}\right)}\right].
\end{align}
By doing the change of variable $\mathcal{T}= t- t_{0} \rightarrow  \mathcal{T}_{\eta}=\mathcal{T}(1-i \eta)$, where $\eta$ is real and $\eta>0$, and taking the limit $\mathcal{T}\to\infty$, the terms in the exponential tend to zero and we are left with only the first term. Simplifying this last equation into
\begin{equation}\label{i time}
   \left\langle q, t \mid q_{0}, t_{0}\right\rangle_{\eta} \approx\langle q \mid 0\rangle\left\langle 0 \mid q_{0}\right\rangle e^{-\frac{i}{\hslash} E_{0} T_{\eta}}.
\end{equation}
Now let us consider two particular times $t_a$ and $t_b$ such that $t_0 < t_a < t_1 <\cdots < t_n < t_b <t$, which allows us to use
\begin{equation}
\begin{aligned}
&\mathbb{I}=\int_{-\infty}^{\infty} d q_{a}\left|q_{a}, t_{a}\right\rangle\left\langle q_{a}, t_{a}\right| \\
&\mathbb{I}=\int_{-\infty}^{\infty} d q_{b}\left|q_{b}, t_{b}\right\rangle\left\langle q_{b}, t_{b}\right|
\end{aligned}
\end{equation}
and by following what we did in the last section we obtain
\begin{equation}
\begin{aligned}
\left\langle q, t\left|T \hat{q}\left(t_{1}\right) \ldots \hat{q}\left(t_{n}\right)\right| q_{0}, t_{0}\right\rangle=\int_{-\infty}^{\infty} d q_{a}\left\langle q, t\left|T \hat{q}\left(t_{1}\right) \ldots \hat{q}\left(t_{n}\right)\right| q_{a}, t_{a}\right\rangle\left\langle q_{a}, t_{a} \mid q_{0}, t_{0}\right\rangle \\
=\int_{-\infty}^{\infty} d q_{a} \int_{-\infty}^{\infty} d q_{b}\left\langle q, t \mid q_{b}, t_{b}\right\rangle\left\langle q_{b}, t_{b}\left|T \hat{q}\left(t_{1}\right) \ldots \hat{q}\left(t_{n}\right)\right| q_{a}, t_{a}\right\rangle\left\langle q_{a}, t_{a} \mid q_{0}, t_{0}\right\rangle
\end{aligned}
\end{equation}
which can be simplified using \eqref{i time} in these final terms dependent on $t_a$ and $t_b$,
\begin{equation}
\left\langle q, t \mid q_{b}, t_{b}\right\rangle \approx\langle q \mid 0\rangle\left\langle 0 \mid q_{b}\right\rangle e^{-\frac{i}{\hslash} E_{0}\left(t-t_{b}\right)(1-i \eta)},
\end{equation}

\begin{equation}
\left\langle q_{a}, t_{a} \mid q_{0}, t_{0}\right\rangle \approx\left\langle q_{a} \mid 0\right\rangle\left\langle 0 \mid q_{0}\right\rangle e^{-\frac{i}{\hslash} E_{0}\left(t_{a}-t_{0}\right)(1-i \eta)}.
\end{equation}
Therefore we can write \eqref{n campos ordenados} as

\begin{align}
&\nonumber\left\langle q, t\left|T \hat{q}\left(t_{1}\right) \ldots \hat{q}\left(t_{n}\right)\right| q_{0}, t_{0}\right\rangle\\ \approx &\nonumber \int_{-\infty}^{\infty} d q_{a} \int_{-\infty}^{\infty} d q_{b}\langle q \mid 0\rangle\left\langle 0 \mid q_{b}\right\rangle e^{-\frac{i}{\hslash} E_{0}\left(t-t_{b}\right)(1-i \eta)}\\
&\left\langle q_{b}, t_{b}\left|T \hat{q}\left(t_{1}\right) \ldots \hat{q}\left(t_{n}\right)\right| q_{a}, t_{a}\right\rangle\left\langle q_{a} \mid 0\right\rangle\left\langle 0 \mid q_{0}\right\rangle e^{-\frac{i}{\hslash} E_{0}\left(t_{a}-t_{0}\right)(1-i \eta)}\\
\approx& \nonumber \int_{-\infty}^{\infty} d q_{a} \int_{-\infty}^{\infty} d q_{b}\langle q \mid 0\rangle\left\langle 0\left|e^{\frac{i}{\hslash} \hat{H} t_{b}}\right| q_{b}\right\rangle e^{-\frac{i}{\hslash} E_{0} t(1-i \eta)}\\
&\left\langle q_{b}, t_{b}\left|T \hat{q}\left(t_{1}\right) \ldots \hat{q}\left(t_{n}\right)\right| q_{a}, t_{a}\right\rangle\left\langle q_{a}\left|e^{\frac{i}{\hslash} \hat{H} t_{a}}\right| 0\right\rangle\left\langle 0 \mid q_{0}\right\rangle e^{\frac{i}{\hslash} E_{0} t_{0}(1-i \eta)}\\
=&\nonumber\int_{-\infty}^{\infty} d q_{a} \int_{-\infty}^{\infty} d q_{b}\langle q \mid 0\rangle e^{-\frac{i}{\hslash} E_{0} t(1-i \eta)}\\
&\left\langle 0 \mid q_{b}, t_{b}\right\rangle\left\langle q_{b}, t_{b}\left|T \hat{q}\left(t_{1}\right) \ldots \hat{q}\left(t_{n}\right)\right| q_{a}, t_{a}\right\rangle\left\langle q_{a}, t_{a} \mid 0\right\rangle\left\langle 0 \mid q_{0}\right\rangle e^{\frac{i}{\hslash} E_{0} t_{0}(1-i \eta)}\\
=&\label{ordenados con exp temporal}\langle q \mid 0\rangle\left\langle 0 \mid q_{0}\right\rangle e^{-\frac{i}{\hslash} E_{0}\left(t-t_{0}\right)(1-i \eta)}\left\langle 0\left|T \hat{q}\left(t_{1}\right) \ldots \hat{q}\left(t_{n}\right)\right| 0\right\rangle_{\eta}
\end{align}
where we recognize the final term of \eqref{ordenados con exp temporal} as the Green's function.\newline

We can use the exponential on the RHS of \eqref{ordenados con exp temporal} to get the times inside the braket, and summarising we got
\begin{equation}
\left\langle q, t\left|T \hat{q}\left(t_{1}\right) \ldots \hat{q}\left(t_{n}\right)\right| q_{0}, t_{0}\right\rangle_{\eta} \approx\left\langle q, t \mid q_{0}, t_{0}\right\rangle_{\eta} \cdot\left\langle 0\left|T \hat{q}\left(t_{1}\right) \ldots \hat{q}\left(t_{n}\right)\right| 0\right\rangle_{\eta}
\end{equation}
then we just divide everything by $\left\langle q, t \mid q_{0}, t_{0}\right\rangle_{\eta}$ to get the final expression of our Green's function
\begin{align}
G_{n}\left(t_{1}, \ldots, t_{n}\right) &=\left\langle 0\left|T \hat{q}\left(t_{1}\right) \ldots \hat{q}\left(t_{n}\right)\right| 0\right\rangle_{\eta} \\
&=\lim _{t_{0} \rightarrow-\infty ; t \rightarrow+\infty} \frac{\left\langle q, t\left|T \hat{q}\left(t_{1}\right) \ldots \hat{q}\left(t_{n}\right)\right| q_{0}, t_{0}\right\rangle_{\eta}}{\left\langle q, t \mid q_{0}, t_{0}\right\rangle_{\eta}} \\
&\label{green path}=\lim _{t_{0} \rightarrow-\infty ; t \rightarrow+\infty} \frac{\int \mathcal{D} q(\tau) q\left(t_{1}\right) \ldots q\left(t_{n}\right) e^{-\frac{i}{\hslash} S}}{\int \mathcal{D} q(\tau) e^{-\frac{i}{\hslash} S[q]}}
\end{align}
where $S$ stands for the action of the system
\begin{equation}
S=\int_{t_{0}(1-i \eta)}^{t(1-i \eta)} d \tau L(q, \dot{q}),
\end{equation}
and it is satisfied that
\begin{equation}
\lim _{t \rightarrow+\infty(1-i \eta)} q(t)=0,
\end{equation}
\begin{equation}
\lim _{t_{0} \rightarrow-\infty(1-i \eta)} q\left(t_{0}\right)=0.
\end{equation}

\subsection{Path integrals with quantum fields}

Up to this point we have only considered systems with one degree of freedom, but everything can be generalized to many degrees of freedom, or even infinite as in the case of a field theory, without too much difficulty.\newline

Let us begin with the simpler case of finite degrees of freedom, suppose that we are dealing with a system that has $n$-degrees of freedom, which we can characterize with the coordinates $q^{\alpha}(t)$ where $\alpha=1,2, \cdots, n$, then the transition amplitude \eqref{path integral} becomes
\begin{equation}
\left\langle q^{\alpha}, t \mid q_{i}^{\alpha}, t_{i}\right\rangle= \int \mathcal{D} q^{\alpha} e^{\frac{i}{h} S},
\end{equation}
and the action is
\begin{equation}
S=\int_{t_{i}}^{t_{f}} \mathrm{~d} t L\left(q^{\alpha}, \dot{q}^{\alpha}\right).
\end{equation}
It should be noticed that the integration in this path integral considers  once again all the paths starting at $q_{i}^{\alpha}$ at $t=t_{i}$ and ending at $q^{\alpha}$ at $t$.\newline

Now for the perhaps more interesting case of field theories with infinite degrees of freedom we must recall that here space gets demoted from an operator to a label, so that in conjunction with time we can form space-time. Also, that we are not so much interested in describing the motion of a single particle anymore, but rather the dynamics of the field itself. These changes can be accommodated in the formulation of the path integral so that it works for fields as well.\newline

For the $0+1$ dimensional case that we did before, we constructed the path integral by dividing the time interval into infinitesimal parts. We can do the same for a $1+1$ space-time by additionally partitioning the space interval 
\begin{equation}
-\frac{L}{2} \leq q \leq \frac{L}{2}
\end{equation}
into $N$ equal pieces of length $\epsilon$ such that
\begin{equation}
N \epsilon=L,
\end{equation}
keeping in mind that we will let $L \rightarrow \infty$ and $N \rightarrow \infty$ at the end. \newline

This effectively divides space-time in infinitesimal boxes which we can label with an index "i". If $\phi(x, t)$ is a field permeating this $1+1$ space-time, then its average value within each $i$-th box of infinitesimal area $\delta A_{i}$ is
\begin{equation}
\phi_{i}=\frac{1}{\delta A_{i}} \int_{\delta A_{i}} \mathrm{~d} t \mathrm{~d} q \phi(t, q)
\end{equation}
and with it we can define the path integral measure:
\begin{equation}
\int \mathcal{D} \phi=\int \prod_{i} \mathrm{~d} \phi_{i}.
\end{equation}

It should be noted that when dealing with fields, we cannot explicitly carry out these path integrals because they diverge. Green's functions however, can be calculated without this problem since they are defined as ratios between path integrals and thus the divergences cancel each other out \cite{A Das}.\newline

For the rest of this chapter we will continue to work with quantum fields, as most of their applications are in the areas of quantum filed theory and particle physics.

\section{Generating Functionals}

There is a great way to calculate Green's functions using currents. Consider the action in the presence of an external classical source $J(x)$. The vacuum amplitude in the presence of this source is then a functional called the generating functional and is denoted by $Z[J]$:

\begin{equation}
Z[J] =\int \mathcal{D}\phi(x) e^{i\int\limits_{-\infty(1-i\eta)}^{\infty(1-i\eta)} d^dx [\mathcal{L}(\phi,\partial_\mu {\phi})+J(x)\phi(x)] } .
\end{equation} 
If we expand the $e^{i\int\limits_{-\infty(1-i\eta)}^{\infty(1-i\eta)} d^dx J(x)\phi(x)}$ term we get,
\begin{align}\label{zjdes}
 Z[J] =	&\nonumber	\int \mathcal{D}\phi(x) e^{i\int\limits_{-\infty(1-i\eta)}^{+\infty(1-i\eta)} d^dx \mathcal{L}(\phi,\partial_\mu \phi) }
\\&\nonumber[ 1 + i\int d^dx J(x)\phi(x) +
\frac{i^2}{2!} 
\int\limits_{-\infty}^{\infty} d^dx_1 d^dx_2 J(x_1)\phi(x_1) J(x_2)\phi(x_2) \\&+\cdots+
\frac{i^n}{n!} 
\int\limits_{-\infty}^{\infty} d^dx_1 \cdots d^dx_n J(x_1)\cdots J(x_n) \phi(x_1) \cdots \phi(x_n) ].
\end{align}
Defining $Z[0]=\int \mathcal{D}\phi(x) e^{i\int\limits_{-\infty(1-i\eta)}^{\infty(1-i\eta)} d^dx \mathcal{L}(\phi,\partial_\mu \phi)} $ and taking into account our result from the previous section \eqref{green path}, we can write it the first term in the expansion (\ref{zjdes}) as
\begin{equation}
i\int d^d x_1 J(x_1) \int \mathcal{D}\phi(x)  e^{i\int\limits_{-\infty(1-i\eta)}^{+\infty(1-i\eta)} d^dx \mathcal{L}(q,\partial_\mu \phi) }
\phi(x_1)= Z[0]i \int d^d x_1 J(x_1) G_1(x_1),
\end{equation}
where 
\begin{equation}
G_1(x_1)= Z[0]^{-1} \int \mathcal{D}\phi(x)  e^{i\int\limits_{-\infty(1-i\eta)}^{+\infty(1-i\eta)} d^dx \mathcal{L}(\phi,\partial_\mu \phi) }.
\phi(x_1),
\end{equation}
Then the nth term is
\begin{multline*}
i^n\int \mathcal{D}\phi(x) d^dx_1\dots d^dx_n e^{i\int\limits_{-\infty(1-i\eta)}^{+\infty(1-i\eta)} d^dx \mathcal{L}(\phi,\partial_\mu \phi) }
J(x_1) \cdots J(x_n)\phi(x_1) \cdots \phi(x_n)
  \\
= Z[0] i^n \int d^dx_1 \dots d^dx_n J(x_1) \cdots J(x_n) G_n(x_1, \ldots, x_n),
\end{multline*}
which gives us the general expression in terms of the following sum:
\begin{equation}
\begin{split}
Z[J] = &  Z[0]\left[1 + \sum\limits_{n=1}^{\infty}\frac{(i)^n}{n!} \int d^dx_1\dots d^dx_n J(x_1) \cdots J(x_n) G_n(x_1, \dots,x_n)\right].
\end{split}
\end{equation}

To discern the utility of $Z[J]$ in this form, first we must obtain its first functional derivative
\begin{equation}\label{primera derivada funcional}
\begin{split}
\frac{\delta Z[J]}{\delta J(x_1)} & =\frac{\delta}{\delta J(x_1)} \int \mathcal{D}\phi(x)
e^{i\int d^dx [\mathcal{L}(\phi,\partial_\mu \phi)+J(x)\phi(x)] } \\
& = \int \mathcal{D}\phi(x) \frac{\delta }{\delta J(x_1)}\left[e^{i\int d^dx J(x)\phi(x)}\right] e^{i\int d^dx \mathcal{L}(\phi,\partial_\mu \phi) },
\end{split}
\end{equation}
 from where we see that the only factor that is modified is the one with the source $J(x)$, and evaluating it we get
\begin{equation}
\begin{split}
\frac{\delta }{\delta J(x_1)}\left[e^{i\int d^dx J(x)\phi(x)}\right]
& = e^{i\int d^dx J(x)\phi(x)}
\frac{\delta }{\delta J(x_1)} \left[i\int d^dx J(x)\phi(x)\right] \\
& = e^{i\int d^dx J(x)\phi(x)}  i\int d^dx \frac{\delta J(x)}{\delta J(x_1)} \phi(x) \\
& = e^{i\int d^dx J(x)\phi(x)}  i\int d^dx \delta (x-x_1) \phi(x) \\
& = e^{i\int d^dx J(x)\phi(x)}  i \phi(x_1) ,
\end{split}
\end{equation}
plugging this result in \eqref{primera derivada funcional} gives us
\begin{equation}
\frac{\delta Z[J]}{\delta J(x_1)} 
= \int \mathcal{D}\phi(x) i \phi(x_1) e^{i\int d^dx J(x)\phi(x)}
e^{i\int d^dx \mathcal{L}(\phi,\partial_\mu \phi) },
\end{equation}
and evaluating $\frac{\delta Z[J]}{\delta J(x_1)}$ in $J=0$ we reach

\begin{equation}
\frac{\delta Z[J]}{\delta J(x_1)} \Big|_{J=0}
= \int \mathcal{D}\phi(x) i \phi(x_1) 
e^{i\int d^dx \mathcal{L}(\phi,\partial_\mu \phi)} = i G_1(x_1)\cdot Z[0].
\end{equation}

From this procedure we learn that all the possible Green's Functions can be obtained by a succession of functional derivatives applied to the generating functional, i.e.
\begin{equation}\label{gnconZ}
G_n(x_1,\ldots,x_n)=\left. \frac{1}{i^nZ[0]}\cdot\frac{\delta}{\delta J(x_1)  \cdots\delta J(x_n)} Z[J]\right|_{J=0},
\end{equation}	
but it is important to remark that sometimes, depending on the Lagrangian and what we want to extract from it, we might take $J$ as an arbitrary constant instead of $0$ in the evaluation.

\section{Perturbative Approach to Green's functions}

In  it is far more complicated to work in a theory with an arbitrary potential $V(\phi)$ and most of the time we cannot obtain an exact solution, thus we need to use perturbation theory to do calculations. Fortunately, as we will see in the this section, the path integral approach gives a robust process for computing the much needed expectation values.\newline

Let us assume that we have a Lagrangian density with the form

\begin{equation}
\mathcal{L}(\phi, \partial_\mu \phi)=\frac{1}{2}\left(\partial_\mu \phi\partial^\mu\phi -\alpha \phi^{2}\right)-\lambda V(\phi),
\end{equation}
then we make a Wick rotation, taking $t \to -i\tau $, then the action becomes
\begin{align}
    S&=\int d^{d-1} x \int dt \Bigg[\frac{1}{2}\left(\partial_\mu \phi\partial^\mu\phi -\alpha \phi^{2}\right)-\lambda V(\phi) \Bigg]
    \\&=\int d^{d-1} x \int dt \Bigg[\frac{1}{2}\left(\left(\frac{\partial\phi}{\partial t}\right)^2 - (\nabla \phi)^2 -\alpha \phi^{2}\right)-\lambda V(\phi) \Bigg]
    \\&=-i\int d^{d-1} x \int d\tau \Bigg[ \frac{1}{2}\left(-\left(\frac{\partial\phi}{\partial\tau}\right)^2 - (\nabla \phi)^2 -\alpha \phi^{2}\right)-\lambda V(\phi) \Bigg],
\end{align}
and the generating functional $Z[J]$ now has a real exponent in the form

\begin{equation}
Z[J]=\int \mathcal{D} \phi(x) e^{-\int d x\left[\mathcal{L}_{E}+J(x) \phi(x)+\lambda V(\phi)\right]},
\end{equation}
where $\mathcal{L}_E$ stands for the Euclidean Lagrangian of the free scalar field,  
i.e.
\begin{equation}
\mathcal{L}_{E}=\frac{1}{2}\left(\left(\frac{\partial \phi}{\partial \tau}\right)^{2}+\left(\nabla \phi\right)^2 +\alpha \phi^{2}\right).
\end{equation}

If we take the n-point functional derivative of the generating functional
\begin{align}
&\nonumber\frac{\delta^{n} Z[J]}{\delta J\left(x_{1}\right) \ldots \delta J\left(x_{n}\right)}\\&\label{deriv func generadora wick}=(-1)^{n} \int \mathcal{D} \phi(x) \phi\left(x_{1}\right) \ldots \phi\left(x_{n}\right)  \exp \left[-\int d x\left(\mathcal{L}_{E}+J(x) \phi(x)+\lambda V(\phi)\right)\right],
\end{align}
then we can write the Green's function as

\begin{equation}
\left\langle 0 |T \hat{\phi}\left(x_{1}\right) \ldots \hat{\phi}\left(x_{n}\right)|0\right\rangle=\left.(-1)^{n}\left(\frac{1}{Z[J]} \frac{\delta^{n} Z[J]}{\delta J\left(x_{1}\right) \ldots \delta J\left(x_{n}\right)}\right)\right|_{J=0} \equiv G_{n}^{i n t}\left(x_{1}, \ldots, x_{n}\right) .
\end{equation}
where the "int" label means interaction, since it considers our potential $V(\phi)$.\newline

In order to simplify our equations, from now on we will use the shorthand notation of the expectation values:

\begin{equation}
  \left\langle 0 |T \hat{\phi}\left(x_{1}\right) \ldots \hat{\phi}\left(x_{n}\right)|0\right\rangle = \left\langle  \phi\left(x_{1}\right) \ldots \phi\left(x_{n}\right)\right\rangle .
\end{equation}

If we use \eqref{deriv func generadora wick}, we can rewrite the Green's functions $G_n^{int}(x_n,\dots,x_1)$ in terms of the action of the system $S = \int dx ( (\mathcal{L}_E +V(\phi))$ as

\begin{equation}
G_{n}^{i n t}\left(x_{1}, \ldots, x_{n}\right)=\frac{\int \mathcal{D} \phi(x) \phi\left(x_{1}\right) \ldots \phi\left(x_{n}\right) e^{-S[\phi(x)]}}{\int \mathcal{D} \phi(x) e^{-S[\phi(x)]}},
\end{equation}

where our action can be separated $S_0=\int dx \mathcal{L}_E$ and $S_1=\int dx \lambda V(\phi)$, and expanding the exponential corresponding to $S_1$ we attain

\begin{equation}
e^{-S}=e^{-S_{0}-S_{1}}=e^{-S_{0}}\left[1+\sum_{m=1}^{\infty} \frac{(-1)^{m}}{m !} S_{1}^{m}\right],
\end{equation}

plugging this in equation \eqref{deriv func generadora wick} we arrive at

\begin{equation}\label{G int expandido}
G_{n}^{i n t}\left(x_{1}, \ldots, x_{n}\right)=\frac{\int \mathcal{D} \phi(x) \phi\left(x_{1}\right) \ldots \phi\left(x_{n}\right) e^{-S_{0}[\phi(x)]}\left[1+\sum_{m=1}^{\infty} \frac{(-1)^{m}}{m !} S_{1}^{m}\right]}{\int \mathcal{D} \phi(x) e^{-S_{0}[\phi(x)]}\left[1+\sum_{m=1}^{\infty} \frac{(-1)^{m}}{m !} S_{1}^{m}\right]}.
\end{equation}

To simplify this expression and get a more applicable result we restrict ourselves to the case that the potential has the form

\begin{equation}
V(\phi)=\frac{\phi^{k}}{k !},
\end{equation}
then we can rewrite equation \eqref{G int expandido} in terms of the Green's functions of the free scalar field

\begin{align}
&\nonumber G_{n}^{i n t}\left(x_{1}, \ldots, x_{n}\right)= \\
&\label{green interaccion euc}\frac{G_{n}\left(x_{1}, \ldots, x_{n}\right)+\sum_{m=1}^{\infty} \frac{(-\lambda / k !)^{m}}{m !} \int d s_{1} \ldots d s_{m} G_{n+m k}\left(x_{1}, \ldots, x_{n}, s_{1}^{k}, \ldots, s_{m}^{k}\right)}
{1+\sum_{m=1}^{\infty} \frac{(-\lambda / k !)^{m}}{m !} \int d y_{1} \ldots d y_{m} G_{m k}\left(y_{1}^{k}, \ldots, y_{m}^{k}\right)},
\end{align}
where all the green functions on the RHS are of the euclidean Lagrangian. Making use of the binomial theorem as

\begin{align}
(1+x)^{-1}&\nonumber=1+(-1)(x)+\frac{(-1)(-2)}{2 !}(x)^{2}+\frac{(-1)(-2)(-3)}{3 !}(x)^{3} 
+\cdots \\
&=1-x+x^{2}-x^{3}+x^{4}+\cdots,
\end{align}
we can expand the denominator of \eqref{green interaccion euc},

\begin{align}
&\nonumber G_{n}^{i n t}\left(x_{1}, \ldots, x_{n}\right)\approx \\
&\nonumber \left( G_{n}\left(x_{1}, \ldots, x_{n}\right)+\sum_{m=1}^{\infty} \frac{(-\lambda / k !)^{m}}{m !} \int d s_{1} \ldots d s_{m} G_{n+m k}\left(x_{1}, \ldots, x_{n}, s_{1}^{k}, \ldots, s_{m}^{k}\right)\right)\\
&\nonumber\times\Bigg[ 1-\sum_{m=1}^{\infty} \frac{(-\lambda / k !)^{m}}{m !} \int d y_{1} \ldots d y_{m} G_{m k}\left(y_{1}^{k}, \ldots, y_{m}^{k}\right) 
\\&\nonumber + \left(\sum_{m=1}^{\infty} \frac{(-\lambda / k !)^{m}}{m !} \int d y_{1} \ldots d y_{m} G_{m k}\left(y_{1}^{k}, \ldots, y_{m}^{k}\right)\right)
\\&\times\left(\sum_{w=1}^{\infty} \frac{(-\lambda / k !)^{w}}{w !} \int d z_{1} \ldots d z_{w} G_{w k}\left(z_{1}^{k}, \ldots, z_{w}^{k}\right)\right)+ \cdots\Bigg]
\end{align}
expanding the product, and keeping the terms up to second order in lambda we approximate it to
\begin{align}
&\nonumber G_{n}^{i n t}\left(x_{1}, \ldots, x_{n}\right)\approx 
\\&\nonumber
G_{n}\left(x_{1}, \ldots, x_{n}\right)+ (-\lambda / k !) \int d s_{1} G_{n+ k}\left(x_{1}, \ldots, x_{n}, s_{1}^{k}\right)
\\&\nonumber
+ \frac{(-\lambda / k !)^{2}}{2 !} \int d s_{1} d s_{2} G_{n+2 k}\left(x_{1}, \ldots, x_{n}, s_{1}^{k}, s_{2}^{k}\right)
\\&\nonumber -G_{n}\left(x_{1}, \ldots, x_{n}\right) (-\lambda / k !)\int d y_{1} G_{ k}\left(y_{1}^{k}\right)
\\&\nonumber -G_{n}\left(x_{1}, \ldots, x_{n}\right) \frac{(-\lambda / k !)^{2}}{2 !} \int d y_{1} d y_{2} G_{2 k}\left(y_{1}^{k},  y_{2}^{k}\right)
\\&\nonumber- (-\lambda / k !) \int d s_{1} G_{n+1 k}\left(x_{1}, \ldots, x_{n}, s_{1}^{k}\right) (-\lambda / k !) \int d y_{1} G_{k}\left(y_{1}^{k}\right) 
\\&+G_{n}\left(x_{1}, \ldots, x_{n}\right)
\left((-\lambda / k !) \int d y_{1} G_{ k}\left(y_{1}^{k}\right)\right)
\left((-\lambda / k !) \int d z_{1} G_{ k}\left(z_{1}^{k}\right)\right)
\end{align}
distributing the product, and keeping the terms up to second order in lambda we approximate it to

\begin{align}
&\nonumber G_{n}^{i n t}\left(x_{1}, \ldots, x_{n}\right)\approx G_{n}\left(x_{1}, \ldots, x_{n}\right)
\\&\nonumber 
+ (-\lambda / k !) \left[\int d s_{1} G_{n+ k}\left(x_{1}, \ldots, x_{n}, s_{1}^{k}\right) - \int d y_{1} G_{n}\left(x_{1}, \ldots, x_{n}\right) G_{ k}\left(y_{1}^{k}\right)  \right] 
\\&\nonumber
+ \frac{(-\lambda / k !)^{2}}{2 !} \Bigg[ \int d s_{1} d s_{2} G_{n+2 k}\left(x_{1}, \ldots, x_{n}, s_{1}^{k}, s_{2}^{k}\right)
\\&\nonumber
 - \int d y_{1} d y_{2} G_{n}\left(x_{1}, \ldots, x_{n}\right) G_{2 k}\left(y_{1}^{k},  y_{2}^{k}\right) - 2 \int d s_{1}  d y_{1} G_{n+1 k}\left(x_{1}, \ldots, x_{n}, s_{1}^{k}\right)   G_{k}\left(y_{1}^{k}\right) 
\\&+ 2
 \int d y_{1} d z_{1}  G_{n}\left(x_{1}, \ldots, x_{n}\right)  G_{ k}\left(y_{1}^{k}\right)
  G_{ k}\left(z_{1}^{k}\right)\Bigg] + \cdots,
\end{align}

where we have defined

\begin{equation}
G_{n+m}^{i n t}\left(x_{1}^{n}, x_{2}^{m}\right) \equiv\left\langle \phi^{n}\left(x_{1}\right) \phi^{m}\left(x_{2}\right)\right\rangle,
\end{equation}

and similarly for functions including powers of $x$ \cite{QGT from generating}. \newline

Having studied these techniques we are ready to apply them within the context of quantum information theory.

\chapter{The Quantum Geometric Tensor in Quantum Mechanics}

 In this chapter our main focus of study will be the \textit{Quantum Geometric Tensor} (QGT) with its real part being the \textit{Quantum Metric Tensor} and imaginary component which is related to \textit{Berry's phase}. First we introduce the concept of \textit{fidelity between quantum states}, from which the QGT emerges, and gives the concept its experimental context.\\

\section{Quantum Fidelity}

Normally in quantum mechanics (with Dirac notation) we use the braket $\braket{\Psi^\prime}{\Psi}$, or overlap between the two quantum states $\ket{\Psi^\prime}$ and $\ket{\Psi}$, to denote the transition amplitude from one to the other, which is a complex number that when we square its absolute value (or modulus) we obtain the probability of the system going from the initial state $\ket{\Psi}$ to the final state $\ket{\Psi^\prime}$ (as we saw in the the previous chapter). \newline

On the other hand, there is a complementary second point of view and it tells us that the overlap measures the similarity between the two states, meaning that the operation returns a $1$ if two states are exactly the same, $0$ if they are orthogonal, or any complex value with norm in between these values if we are dealing with two states that are not completely indistinguishable (such as the case when comparing a pure state with a mixed one as we shall see) \cite{Nielsen}.\newline

This interpretations is crucial in quantum information theory since experimentally we would like to transfer quantum states  over long distances without any loss of information. Meaning that if we encode our information within a quantum state and transfer it through any mechanism, it would be ideal for our initial input state to be indistinguishable from the output one. In this sense we can use the overlap between the input and output states to measure how much information was lost in the process. However, a global phase difference between the states can alter the overlap, so we need to find another approach to measure distances (or similarity) in such a way that this does not happen. With this purpose in mind we will use the overlap to define the fidelity.\newline

Mathematically we define the overlap between two states as
\begin{equation}
f\left(\Psi^{\prime}, \Psi\right)=\left\langle\Psi^{\prime} \mid \Psi\right\rangle,
\end{equation}
and the fidelity will be the modulus of the overlap, i.e.,

\begin{equation}
F\left(\Psi^{\prime}, \Psi\right)=\left|\left\langle\Psi^{\prime} \mid \Psi\right\rangle\right|,
\end{equation}
where $\ket{\Psi}$, $\ket{\Psi'}$ are the input and output states respectively, and both of them are normalized.\newline

The fidelity possesses the following properties:
\begin{equation}
0  \leq F\left(\Psi^{\prime}, \Psi\right) \leq 1 
\end{equation}
\begin{equation}
	F\left(\Psi^{\prime}, \Psi\right) =F\left(\Psi, \Psi^{\prime}\right)
\end{equation}
\begin{equation}
F\left(U \Psi^{\prime}, U \Psi\right) =F\left(\Psi^{\prime}, \Psi\right)
\end{equation}
\begin{equation}
F\left(\Psi_{1} \otimes \Psi_{2}, \Psi_{1}^{\prime} \otimes \Psi_{2}^{\prime}\right) =F\left(\Psi_{1}^{\prime}, \Psi_{1}\right) F\left(\Psi_{2}^{\prime}, \Psi_{2}\right),
\end{equation}
where $U$ stands for an unitary transformation.\newline

The fidelity between two mixed states $\left(\rho, \rho^{\prime}\right)$, were both are normalized and semi-positive defined with $\operatorname{tr} \rho=\operatorname{tr} \rho^{\prime}=1$, is
\begin{equation}
F\left(\rho, \rho^{\prime}\right)=\operatorname{tr} \sqrt{\rho^{1 / 2} \rho^{\prime} \rho^{1 / 2}},
\end{equation}
and regularly finding it is not a trivial thing to do. However we can use simplified expressions when we are dealing with these special cases \cite{Gu}: 
\begin{itemize}
    \item When the two states are pure
    \begin{equation}
    F\left(\rho, \rho^{\prime}\right)=\left|\left\langle\Psi^{\prime} \mid \Psi\right\rangle\right|,
    \end{equation}
\item If at least one state is pure, meaning that $\rho=|\Psi\rangle\langle\Psi|$, then 
\begin{equation}
F\left(\rho, \rho^{\prime}\right)=\sqrt{\left\langle\Psi\left|\rho^{\prime}\right| \Psi\right\rangle}.
\end{equation}

\item If both of states are diagonal in the same basis
\end{itemize}
\begin{equation}
F\left(\rho, \rho^{\prime}\right)=\sum_{j} \sqrt{\rho_{j j} \rho_{j j}^{\prime}}.
\end{equation}

One of the main uses of the fidelity is that it can predict quantum phase transitions when we look at it in the context of parameter space. Suppose we have two systems described by $H$ and $H^\prime$ respectively, where the only difference between them is that the parameters of $H^\prime$ are slightly different from those of $H$. Once we have the fidelity between the ground states of these systems, wherever there are abrupt changes it will indicate a quantum phase transition; but since is not always easy to obtain we will have to use a perturbative approach, or specifically the Quantum Geometric Tensor.\newline

\section{Quantum Geometric Tensor}

It should be noted that the fidelity itself is not a metric but from it emerges the Quantum Metric Tensor \cite{Berry1989b} and as an extension, the Quantum Geometric Tensor. With it, it is possible to study several properties of physical systems including the fidelity but also entanglement, entanglement entropy and quantum phase transitions as we will see in the following sections.\newline 

To see the origin of this improved concept we will follow a perturbative approach in parameter space first proposed by Provost and Vallee in \cite{Provost} (although we are going to use a more familiar notation).\newline

Let us consider two infinitesimally separated states $|\Psi(\lambda)\rangle$ and $|\Psi(\lambda + d\lambda)\rangle$, that depend on the $n$-dimensional parameter $\lambda=(\lambda_{1}, \ldots, \lambda_{n})\in \mathbb{R}^n$, then calculate the norm of the difference, i.e. the overlap of the difference between them which up to second order is: 

\begin{align}
    \|\psi(\lambda+\mathrm{d} \lambda)-\psi(\lambda)\|^{2}
   & =(\langle\Psi(\lambda+d\lambda)|-\langle\Psi(\lambda)|) \cdot(|\Psi(\lambda+d\lambda)\rangle-|\Psi(\lambda)\rangle)
    \\&=\langle\partial_{\mu} \Psi \mid \partial_{\nu} \Psi\rangle d \lambda^{\mu} d \lambda^{\nu},
\end{align}
where the partials $\partial_i$ are taken with respect to the parameters. Since the overlap is a complex number we can separate its real and imaginary parts,
\begin{equation}
\left\langle\partial_{\mu} \Psi \mid \partial_{\nu} \Psi\right\rangle=\zeta_{\mu\nu}+i \sigma_{\mu\nu},
\end{equation}
with the real part being symmetric 
\begin{equation}
\zeta_{\mu \nu}(\lambda)=\frac{1}{2}\left(\left\langle\partial_{\mu} \Psi \mid \partial_{\nu} \Psi\right\rangle+\left\langle\partial_{\nu} \Psi \mid \partial_{\mu} \Psi\right\rangle\right)=\zeta_{\nu\mu}(\lambda) ,
\end{equation}
and the imaginary one antisymmetric
\begin{equation}
    \sigma_{\mu\nu}(\lambda)=\frac{1}{2i}\left(\left\langle\partial_{\mu} \Psi \mid \partial_{\nu} \Psi\right\rangle-\left\langle\partial_{\nu} \Psi \mid \partial_{\mu} \Psi\right\rangle\right)=-\sigma_{\nu \mu}(\lambda),
\end{equation}
then
\begin{equation}
\left\langle\partial_{\mu} \psi \mid \partial_{\nu} \psi\right\rangle \mathrm{d} \lambda^{\mu} \mathrm{d} \lambda^{\nu}=\left(\zeta_{\mu \nu}+i \sigma_{\mu \nu}\right) \mathrm{d} \lambda^{\mu} \mathrm{d} \lambda^{\nu}.
\end{equation}

At this point we notice that $\sigma_{\mu \nu} \mathrm{d} \lambda^{\mu} \mathrm{d} \lambda^{\nu}$ vanishes since $\sigma_{\mu \nu}$ is antisymmetric and $\mathrm{d} \lambda^{\mu} \mathrm{d} \lambda^{\nu}$ is symmetric, leaving the quantum distance as 
\begin{equation}
\|\psi(\lambda+\mathrm{d} \lambda)-\psi(\lambda)\|^{2}=\zeta_{\mu \nu} \mathrm{d} \lambda^{\mu} \mathrm{d} \lambda^{\nu}.
\end{equation}

Nevertheless, it would be wrongful to define $\zeta_{\mu \nu}$ as a metric tensor since it is not gauge invariant, which is one of the requirements to be so. In order to fix this problem we apply the gauge transformation 
\begin{equation}
\left|\psi^{\prime}(\lambda)\right\rangle=\exp ^{i \alpha(\lambda)}|\psi(\lambda)\rangle,
\end{equation} 
and follow the same procedure as before, defining $\left\langle\partial_{\mu} \psi^{\prime} \mid \partial_{\nu} \psi^{\prime}\right\rangle=\zeta_{\mu \nu}^{\prime}+i \sigma_{\mu \nu}^{\prime}$, which yields
\begin{equation}
\zeta_{\mu \nu}^{\prime} =\zeta_{\mu \nu}-\beta_{\mu} \partial_{\nu} \alpha-\beta_{\nu} \partial_{\mu} \alpha+\partial_{\mu} \alpha \partial_{\nu} \alpha,
\end{equation}
\begin{equation}
\sigma_{\mu \nu}^{\prime} =\sigma_{\mu \nu},
\end{equation}
where 
\begin{equation}
\beta_{\mu}(\lambda)=-i\left\langle\psi(\lambda) \mid \partial_{\mu} \psi(\lambda)\right\rangle,
\end{equation}
which is real because of the normalization of our quantum state $\langle\psi(\lambda) \mid \psi(\lambda)\rangle=1$ and it is called the \textit{Berry connection} \cite{Berry}.\newline

If we apply the gauge transformation to the Berry connection it changes as
\begin{equation}
\begin{aligned}
\beta_{\mu}^{\prime}(\lambda) &=-i\left\langle\Psi^{\prime}(\lambda) \mid \partial_{\mu} \Psi^{\prime}(\lambda)\right\rangle \\
&=-i\left\langle\Psi\left|e^{-i \alpha} e^{i \alpha}\left(\left|\partial_{\mu} \Psi\right\rangle+i|\Psi\rangle \partial_{\mu} \alpha\right)\right)\right.\\
&=\beta_{\mu}+\partial_{\mu} \alpha .
\end{aligned}
\end{equation}
Thus we define our proper and well defined invariant metric, the \textit{Quantum Metric Tensor}, as:
\begin{equation}
g_{\mu \nu}(\lambda)=\zeta_{\mu \nu}(\lambda)-\beta_{\mu}(\lambda) \beta_{\nu}(\lambda)
\end{equation}
where if a gauge transformation is applied, the changes from the $\beta$'s counteract the ones originating from the $\zeta$:
\begin{equation}
\begin{aligned}
\zeta_{\mu \nu}^{\prime}-\beta_{\mu}^{\prime} \beta_{\nu}^{\prime}=& \zeta_{\mu \nu}-\beta_{\mu} \beta_{\nu}+\partial_{\mu} \alpha \beta_{\nu}+\beta_{\mu} \partial_{\nu} \alpha+\partial_{\mu} \alpha \partial_{\nu} \alpha \\
&-\beta_{\mu} \partial_{\nu} \alpha-\beta_{\nu} \partial_{\mu} \alpha -\partial_{\mu} \alpha \partial_{\nu} \alpha \\
=& \zeta_{\mu \nu}-\beta_{\mu} \beta_{\nu}
\end{aligned}
\end{equation}
meaning that $g_{\mu \nu}^{\prime}(\lambda)=g_{\mu \nu}(\lambda)$.\newline

We can extend this concept to the \textit{Quantum Geometric Tensor}:
\begin{equation}\label{QGT}
G_{\mu \nu}(\lambda)=\left\langle\partial_{\mu} \psi(\lambda) \mid \partial_{\nu} \psi(\lambda)\right\rangle-\left\langle\partial_{\mu} \psi(\lambda) \mid \psi(\lambda)\right\rangle\left\langle\psi(\lambda) \mid \partial_{\nu} \psi(\lambda)\right\rangle
\end{equation}
where its real part is our Quantum Metric Tensor, 
\begin{equation}\label{QMT}
\operatorname{Re} G_{\mu \nu}=g_{\mu \nu} = \frac{1}{2}\left(\left\langle\partial_{\mu} \Psi \mid \partial_{\nu} \Psi\right\rangle+\left\langle\partial_{\nu} \Psi \mid \partial_{\mu} \Psi\right\rangle\right)-\left\langle\partial_{\mu} \Psi \mid \Psi\right\rangle\left\langle\Psi \mid \partial_{\nu} \Psi\right\rangle, 
\end{equation}
and the imaginary part is related to the \textit{Berry Curvature} $F_{\mu\nu}$ as
\begin{equation}\label{BC}
\frac{1}{2} F_{\mu \nu}=\operatorname{Im} G_{\mu \nu}= \sigma_{\mu \nu} =\frac{1}{2 i}\left(\left\langle\partial_{\mu} \Psi \mid \partial_{\nu} \Psi\right\rangle-\left\langle\partial_{\nu} \Psi \mid \partial_{\mu} \Psi\right\rangle\right).
\end{equation}
 This quantity contains additional information not present in the QMT, related to the interference between states since with it we can obtain Berry's phase. This is an extra phase that emerges in the wave function when we vary its parameters adiabatically forming a cyclic circuit in parameter space. Its an example of an anholonomy, the inability of certain variables describing the system to return to their original values when traversing any closed path, another example of anholonomy would be the parallel transport of General Relativity \cite{Berry1989b, Berry}. \newline
 
 It can also be used to explain specific quantum phenomena present in systems whose environment
undergoes a periodic change, for example neutrons
 passing through a helical magnetic
field, or polarized light in a coiled
optic fiber or charged particles circling
an isolated magnetic field \cite{Berry Scientific American}.
\newline
 
 Specifically,  the Berry curvature $F_{\mu \nu}$ is related to 
 \textit{Berry's phase} $\gamma_B$ by an integral in parameter space
\begin{equation}
\gamma_{\mathrm{B}}(C)=\int_{\Sigma} \frac{1}{2} \mathrm{~F}_{\mu \nu} \mathrm{d} \lambda^{\mu} \wedge \mathrm{d} \lambda^{v}
\end{equation}
where $C$ is the closed path that we traveled over  adiabatically in  parameter space and $\Sigma $ is the area enclosed by it, both associated as $\partial \Sigma = C$ \cite{GPhases}.

\subsection{Fidelity, QGT and the line element}

Over this section we have only worked with the overlap, not the fidelity. To see more explicitly how these concepts are related and how the QMT $g_{\mu \nu}$ plays the role of a metric we begin by expanding $|\psi(\lambda+\mathrm{d} \lambda)\rangle$ with respect to $\lambda$ as

 \begin{equation}
 |\psi(\lambda+\mathrm{d} \lambda)\rangle= \ket{\psi(\lambda)} + \ket{\partial_\mu \psi (\lambda)}d \lambda^\mu + \frac{1}{2} \ket{\partial_\mu \partial_\nu \psi (\lambda)}d \lambda^\mu d\lambda^\nu + \cdots
 \end{equation} 

and keeping the terms up to second order in $\mathrm{d} \lambda$, we take its inner product with the state $|\psi(\lambda)\rangle$ getting
\begin{align}
  \langle\psi(\lambda) \mid \psi(\lambda+\mathrm{d} \lambda)\rangle&
  =1+ \left\langle\psi(\lambda) \mid \partial_{\mu} \psi(\lambda)\right\rangle \mathrm{d} \lambda^{\mu}+\frac{1}{2}\left\langle\psi(\lambda) \mid \partial_{\mu} \partial_{\nu} \psi(\lambda)\right\rangle \mathrm{d} \lambda^{\mu}\mathrm{d} \lambda^{\nu}\\
  &=1+i \beta_{\mu}(\lambda) \mathrm{d} \lambda^{\mu}+\frac{1}{2}\left\langle\psi(\lambda) \mid \partial_{\mu} \partial_{\nu} \psi(\lambda)\right\rangle \mathrm{d} \lambda^{\mu} \mathrm{d} \lambda^{\nu}  
\end{align}
then the fidelity between the states(or modulus of the overlap) is
\begin{align}
    \mid \langle\psi(\lambda) \mid \psi(\lambda+\mathrm{d} \lambda)\rangle \mid
  &=\sqrt{1+ \operatorname{Re}\left\langle\psi(\lambda) \mid \partial_{\mu} \partial_{\nu}  \psi(\lambda)\right\rangle \mathrm{d} \lambda^{\mu} \mathrm{d} \lambda^{\nu} + \beta_{\mu}\beta_{\nu}(\lambda) \mathrm{d} \lambda^{\mu}(\lambda) \mathrm{d} \lambda^{\nu} }
  \\&=1+\frac{1}{2} ( \operatorname{Re}\left\langle\psi(\lambda) \mid \partial_{\mu} \partial_{\nu}  \psi(\lambda)\right\rangle \mathrm{d} \lambda^{\mu} \mathrm{d} \lambda^{\nu} + \beta_{\mu}\beta_{\nu}(\lambda) \mathrm{d} \lambda^{\mu}(\lambda) \mathrm{d} \lambda^{\nu})  
\end{align}
where we used the binomial series in the last line to get rid of the square root. Since $\left\langle\psi \mid \partial_{\mu} \psi\right\rangle$ is an imaginary number, then we know that $\left\langle\partial_{\mu} \psi \mid \partial_{\nu} \psi\right\rangle+\left\langle\psi \mid \partial_{\mu} \partial_{\nu} \psi\right\rangle$ is also imaginary, therefore
\begin{equation}
\operatorname{Re}\left\langle\psi \mid \partial_{\mu} \partial_{\nu} \psi\right\rangle=-\operatorname{Re}\left\langle\partial_{\mu} \psi \mid \partial_{\nu} \psi\right\rangle=-\gamma_{\mu \nu},
\end{equation}
leaving our expression of the fidelity in terms of the QMT
\begin{equation}
\begin{aligned}
|\langle\psi(\lambda) \mid \psi(\lambda+\mathrm{d} \lambda)\rangle| &=1-\frac{1}{2}\left(\gamma_{\mu \nu}(\lambda)-\beta_{\mu}(\lambda) \beta_{\nu}(\lambda)\right) \mathrm{d} \lambda^{\mu} \mathrm{d} \lambda^{\nu} \\
&=1-\frac{1}{2} g_{\mu \nu}(\lambda) \mathrm{d} \lambda^{\mu} \mathrm{d} \lambda^{\nu}
\end{aligned}.
\end{equation}

Finally, the line element is defined by two infinitesimally separated states on the Hilbert space as
\begin{equation}
d s^{2}= g_{\mu\nu} d \lambda_{\mu} d \lambda_{\nu}=\langle\delta \psi\mid \delta \psi\rangle-|\langle\psi\mid \delta \psi\rangle|^{2} \\
\end{equation}
where $\delta \psi(\lambda)=\psi(\lambda+d\lambda)-\psi(\lambda)$. However, if we want it to work in the Hilbert space of rays instead of the Hilbert space of States, so it is gauge invariant we would rather define it as 
\begin{equation}
d s^{2}=2-2\|\langle\psi(\lambda) \mid \psi(\lambda+\mathrm{d} \lambda)\rangle\|
\end{equation}
leaving us once again with
\begin{equation}
    d s^{2}= g_{\mu\nu} d \lambda_{\mu} d \lambda_{\nu}
\end{equation}
but this time it is gauge invariant \cite{Provost}.

\subsection{Quantum Phase Transitions}

The quantum states obtained from a Hamiltonian depend on a set of parameters (such as coupling constants, angular frequencies, etc) which present the structure of a manifold that can be partitioned in regions,  in each we will be able to move adiabatically from one point to another without encountering divergences in the expectation values of any observable. The boundaries between regions are called critical points since when we cross them our observables experience a \textit{quantum phase transition} which is a non-analytical behavior \cite{Zanardi}, and as we will see it causes the Quantum Metric Tensor to stop being analytic. To understand this process we will need to arrive at the metric from a different path than the one we have used so far. \newline 

Let us consider a system described by the Hamiltonian $H(\lambda)$, where $\lambda$ once again denotes the parameters that govern our system. If we take the variation  $\lambda \rightarrow \lambda+\delta \lambda$, then we will have
\begin{equation}
H(\lambda+\delta \lambda)=H(\lambda)+\partial_{\lambda} H(\lambda) \delta \lambda .
\end{equation}
If we consider our variation $\delta \lambda$ small we can apply perturbation theory, and defining $H_{I}=\partial_{\lambda} H(\lambda)$ the ground state of our system in the point $\lambda+\delta \lambda$ is
\begin{equation}
\left|\Psi_{0}(\lambda+\delta \lambda)\right\rangle=\left|\Psi_{0}(\lambda)\right\rangle+\delta \lambda \sum_{n \neq 0} \frac{\left\langle\Psi_{n}(\lambda)\left|H_{I}\right| \Psi_{0}(\lambda)\right\rangle\left|\Psi_{n}(\lambda)\right\rangle}{E_{0}(\lambda)-E_{n}(\lambda)}+O\left(\delta \lambda^{2}\right)
\end{equation}
If we normalize $\left|\Psi_{0}(\lambda+\delta \lambda)\right\rangle$, then the fidelity squared is
\begin{equation}
F^{2}=1-\delta \lambda^{2} \sum_{n \neq 0} \frac{\left|\left\langle\Psi_{n}(\lambda)\left|H_{I}\right| \Psi_{0}(\lambda)\right\rangle\right|^{2}}{\left(E_{0}(\lambda)-E_{n}(\lambda)\right)^{2}}
\end{equation}
and applying the same expansion for the square root as before we get
\begin{equation}
F=1-\frac{\delta \lambda^{2}}{2} \sum_{n \neq 0} \frac{\left|\left\langle\Psi_{n}(\lambda)\left|H_{I}\right| \Psi_{0}(\lambda)\right\rangle\right|^{2}}{\left(E_{0}(\lambda)-E_{n}(\lambda)\right)^{2}},
\end{equation}
where the second order term is the perturbative form of the  \textit{Fidelity Susceptibility}
\begin{equation}
\chi_{F}(\lambda)=\sum_{n \neq 0} \frac{\left|\left\langle\Psi_{n}(\lambda)\left|H_{I}\right| \Psi_{0}(\lambda)\right\rangle\right|^{2}}{\left(E_{0}(\lambda)-E_{n}(\lambda)\right)^{2}} .
\end{equation}
Then we can write the QMT as
\begin{equation}
g_{a b}=\sum_{n \neq 0} \frac{\left\langle\Psi_{0}(\lambda)\left|\partial_{a} H\right| \Psi_{n}(\lambda)\right\rangle\left\langle\Psi_{n}(\lambda)\left|\partial_{b} H\right| \Psi_{0}(\lambda)\right\rangle}{\left(E_{0}(\lambda)-E_{n}(\lambda)\right)^{2}}
\end{equation}
and from which we can see that as the energy difference between the ground and excited states decreases, our metric diverges and we will find a quantum phase transition in our system \cite{Tesis Fancy, QGT from generating}.

\section{QGT in Quantum Mechanics for the n-th excited state}

There exist a different formulation of the QGT, which we derived in a previous work \cite{Tesis licenciatura}, here we will only remark the key points of the procedure. The main advantage of this new method is that it only requires the Hamiltonian of the system, getting rid of the necessity of having to work with its wave function; also it can expand the concept to include variation of the phase space if we choose an ordering rule (e.g. normal ordering) for the $\hat{p}$, $\hat{q}$ operators.\newline 

Suppose that our system is given initially, from $t=-\infty$ to $t=0$, by the Hamiltonian $H_i=H$ and after $t=0$ it has been perturbed and its now described by $H_f=H + \delta H$ where

\begin{equation}
\delta H = \frac{\partial H}{\partial z^A}\delta z^A
\end{equation}
with
\begin{equation}
z^A=(q^i,p_i, \lambda_a).
\end{equation}

 Since the QGT is related to the overlap between states let us begin with the braket $\braket{q_f,t_f}{q_i, t_i}$, where the subscripts $i$ and $f$ denote the initial and final Hamiltonians respectively, then we introduce inside of it identity operators in the same fashion as the path integral:

\begin{equation}
\braket{q_f,t_f}{q_i, t_i} = 	 \sum_{m_f,m_i } \bra{q_f} e^{\frac{-it_f E_m^f}{\hslash}}\ket{m_f}\braket{m_f}{m_i }\bra{m_i }e^{\frac{it_i E_{m }^i}{\hslash}}\ket{q_i},
\end{equation}
assuming orthogonality between the states ($\braket{a}{b}= \delta_{a,b}$) and multiplying for the energy exponentials for the nth state we get

\begin{align}
&\nonumber e^{\frac{it_f (E_n^f)}{\hslash}}e^{\frac{-it_i (E_n^i )}{\hslash}}\braket{q_f,t_f}{q_i, t_i} 
\\&\nonumber=\bra{q_f} e^{\frac{-it_f (E_0^f - E_n^f )}{\hslash}}\ket{0_f}\braket{0_f}{0_i}\bra{0_i}e^{\frac{it_i (E_{0'}^i - E_n^i )}{\hslash}}\ket{q_i} + \dots
\\&\nonumber
+ \bra{q_f} e^{\frac{-it_f (E_{n-1}^f- E_n^f )}{\hslash}}\ket{(n-1)_f}\braket{(n-1)_f}{(n-1)_i}\bra{(n-1)_i}e^{\frac{it_i (E_{(n-1)'}^i - E_n^i )}{\hslash}}\ket{q_i}
\\&\nonumber +
\braket{q_f}{n_f}\braket{n_f}{n_i }\braket{n_i }{q_i} 
\\& +
\bra{q_f} e^{\frac{-it_f (E_{n+1}^f - E_n^f )}{\hslash}}\ket{(n+1)_f}\braket{(n+1)_f}{(n+1)_i}\bra{(n+1)_i}e^{\frac{it_i (E_{(n+1)'}^i-E_n^i )}{\hslash}}\ket{q_i} 
+ \cdots .
\end{align}	

To regularize these exponential terms we introduce the prescription:  $E_n^f\to E_n^f+i\epsilon$, $E_n^i\to E_n^i+i\epsilon$ and we let $t_i \to -\infty$ and $t_f \to \infty$, assuming that $q(\infty)=q(-\infty)=0$. This yields

\begin{align}
&\nonumber \lim\limits_{t_f\to\infty, t_i\to -\infty}e^{\frac{it_f (E_n^f + i\epsilon)}{\hslash}}e^{\frac{-it_i (E_n^i )}{\hslash}}\braket{q_f,t_f}{q_i, t_i}  \Big\rvert_{E_n^f \to E_n^f + i\epsilon,E_n^i \to E_n^i + i\epsilon} 
\\&\nonumber=\bra{q_f} e^{\frac{-it_f (E_0^f - E_n^f)}{\hslash}} e^{-\infty}\ket{0_f}\braket{0_f}{0_i}\bra{0_i}e^{\frac{it_i (E_{0'}^i - E_n^i)}{\hslash}}e^{-\infty}\ket{q_i} + \dots
\\&\nonumber
+ \bra{q_f} e^{\frac{-it_f (E_{n-1}^f- E_n^f)}{\hslash}}e^{-\infty}\ket{(n-1)_f}\braket{(n-1)_f}{(n-1)_i}\bra{(n-1)_i}e^{\frac{it_i (E_{(n-1)'}^i - E_n^i)}{\hslash}}e^{-\infty}\ket{q_i}
\\&\nonumber +
\big[\braket{q_f}{n_f}\braket{n_f}{n_i }\braket{n_i }{q_i} \big] \Big\rvert_{E_n^f \to E_n^f + i\epsilon,E_n^i \to E_n^i + i\epsilon}
\\&\nonumber +
\bra{q_f} e^{\frac{-it_f (E_{n+1}^f - E_n^f)}{\hslash}}e^{-\infty}\ket{(n+1)_f}\braket{(n+1)_f}{(n+1)_i}\bra{(n+1)_i}e^{\frac{it_i (E_{(n+1)'}^i-E_n^i)}{\hslash}}e^{-\infty}\ket{q_i} 
\\&+ \dots  ,
\end{align}
from where we see that the exponential factors dependent on $\epsilon$ go to zero leaving only

\begin{align}
&\nonumber \lim\limits_{t_f\to\infty, t_i\to -\infty}e^{\frac{it_f (E_n^f + i\epsilon)}{\hslash}}e^{\frac{-it_i (E_n^i )}{\hslash}}\braket{q_f,t_f}{q_i, t_i} \Big\rvert_{E_n^f \to E_n^f + i\epsilon,E_n^i \to E_n^i + i\epsilon} 
\\&=\big[ \braket{q_f}{n_f}\braket{n_f}{n_i }\braket{n_i }{q_i}\big]\Big\rvert_{E_n^f \to E_n^f + i\epsilon,E_n^i \to E_n^i + i\epsilon},
\end{align}
and if we divide by the brakets $\braket{q_f}{n_f}\braket{n_i}{q_i}$ and we introduce in them the exponentials we can write

\begin{equation}
	\braket{n_f}{n_i} \Big\rvert_{E_n^f \to E_n^f + i\epsilon,E_n^i \to E_n^i + i\epsilon}=\frac{ \braket{q_f,\infty}{q_i, -\infty}}{\braket{q_f,\infty}{n_f}\braket{n_i}{q_i,-\infty}}\Big\rvert_{E_n^f \to E_n^f + i\epsilon,E_n^i \to E_n^i + i\epsilon}.
\end{equation}

Now we will deal with each term separately to leave them in terms of the Hamiltonian by inserting infinite identity operators to get path integrals

\begin{align}
&\nonumber \braket{q_f,\infty}{q_i,-\infty}\Big\rvert_{E_n^f \to E_n^f + i\epsilon,E_n^i \to E_n^i + i\epsilon}
\\&\label{4 paso medio overlap}
=\int \mathcal{D}q \mathcal{D}p e^{\frac{i}{\hslash}\int_{-\infty}^\infty dt (p\dot{q}-H_i)  -\frac{i}{\hslash}\int_{0}^\infty dt \delta H }\Big\rvert_{E_n^f \to E_n^f + i\epsilon,E_n^i \to E_n^i + i\epsilon}.
\end{align}

At this point we will restrict ourselves to perturbations in the form of translations to be able to pull them out out the path integral. Introducing the notation
\begin{equation}
\mathcal{O}_A= \frac{\partial H}{\partial z^A},
\end{equation}
equation \eqref{4 paso medio overlap} is rewritten as
\begin{align}\label{4 paso medio overlap 2}
&\nonumber \braket{q_f,\infty}{q_i,-\infty}\Big\rvert_{E_n^f \to E_n^f + i\epsilon,E_n^i \to E_n^i + i\epsilon}
\\&=\int \mathcal{D}q \mathcal{D}p e^{\frac{i}{\hslash}\int_{-\infty}^\infty dt (p\dot{q}-H_i) -\frac{i}{\hslash}\int_{0}^\infty dt \mathcal{O}_A(t)\delta z^A }\Big\rvert_{E_n^f \to E_n^f + i\epsilon,E_n^i \to E_n^i + i\epsilon}.
\end{align}

Remembering that the generating functions in this case are

\begin{equation}
Z_j=\int \mathcal{D}q\mathcal{D}p e^{\frac{i}{\hslash}\int_{-\infty}^\infty dt (p\dot{q} - H_j)},
\end{equation}	
and since the theory is time reversible the denominator factors become
\begin{equation}
\braket{q_f , \infty}{n_f}\Big\rvert_{E_n^f \to E_n^f + i\epsilon}=\sqrt{Z_f}\Big\rvert_{E_n^f \to E_n^f + i\epsilon},
\end{equation}

\begin{equation}
\braket{n_i}{q_i , -\infty}\Big\rvert_{E_n^i \to E_n^i + i\epsilon}=\sqrt{Z_i}\Big\rvert_{E_n^i \to E_n^i + i\epsilon},
\end{equation}
leaving \eqref{4 paso medio overlap 2} in the new form

\begin{equation}\label{4 antes del ve}
\braket{n_f}{n_i}\Big\rvert_{E_n^f \to E_n^f + i\epsilon,E_n^i \to E_n^i + i\epsilon}= \frac{\int \mathcal{D}q e^{\frac{i}{\hslash}\int_{-\infty}^\infty dt (p\dot{q} -H_i)  -\frac{i}{\hslash}\int_{0}^\infty dt \delta z^A\mathcal{O}_A }}{\sqrt{Z_i Z_f}}\Big\rvert_{E_n^f \to E_n^f + i\epsilon,E_n^i \to E_n^i + i\epsilon}.
\end{equation}

To simplify this result we write it in terms of mean values with the help of the expression

\begin{equation}\label{meanval sin ops}
\expval{A}_n=\bra{n_i}A\ket{n_i}=\frac{1}{Z_i}\int\mathcal{D}q\mathcal{D}p e^{\frac{i}{\hslash}\int_{-\infty}^\infty d t (p\dot{q} -H_i) }A(q)\Big\rvert_{E_n^i \to E_n^i + i\epsilon}
\end{equation}
where we are taking the mean value with respect to the initial Hamiltonian $H_i$. In consequence equation \eqref{4 antes del ve} develops into

\begin{equation}
\braket{n_f}{n_i}\Big\rvert_{E_n^f \to E_n^f + i\epsilon,E_n^i \to E_n^i + i\epsilon}
=\frac{\expval{e^{-\frac{i}{\hslash}\int_{0}^\infty dt \delta z^A\mathcal{O}_A  }}_n}{\sqrt{\expval{e^{-\frac{i}{\hslash}\int_{-\infty}^\infty dt \delta z^P\mathcal{O}_P }}_n}}\Big\rvert_{E_n^f \to E_n^f + i\epsilon,E_n^i \to E_n^i + i\epsilon},
\end{equation}
and the square of its modulus is then
\begin{equation}
\abs{\braket{n_f}{n_i}}^2=\frac{\expval{e^{-\frac{i}{\hslash}\int_{0}^\infty dt \delta z^A\mathcal{O}_A  }}_n\expval{e^{-\frac{i}{\hslash}\int_{-\infty}^0 dt \delta z^B\mathcal{O}_B  }}_n}{\expval{e^{-\frac{i}{\hslash}\int_{-\infty}^\infty dt \delta z^P\mathcal{O}_P  }}_n}.
\end{equation}

To facilitate this expression we use the Maclaurin (Taylor in $0$) series of the exponential up to second order in the perturbation, getting 
\begin{align}
&\expval{e^{-\frac{i}{\hslash}\int_{0}^\infty dt \delta z^A\mathcal{O}_A (t) }}_n 
\\&=
1 - \frac{i}{\hslash}\expval{\int_{0}^\infty dt \delta z^A\mathcal{O}_A (t)}_n - \frac{1}{2\hslash^2}\expval{\int_{0}^\infty dt_1 \delta z^A\mathcal{O}_A (t_1)  \int_{0}^\infty dt_2 \delta z^B\mathcal{O}_B (t_2)  }_n,
\end{align}
which can be simplified further utilizing the binomial theorem, leaving us the result

\begin{equation}
\abs{\braket{n_f}{n_i}}^2=1 - G_{A B}^{(n)}\delta z^A \delta z^B,
\end{equation}
or in terms of the quantum fidelity
\begin{equation}
F(z, z + \delta z) = \abs{\braket{n_f}{n_i}} = 1 - \unmed G_{A B}^{(n)}\delta z^A \delta z^B,
\end{equation}
where
\begin{equation}\label{TGC general}
G_{A B}^{(n)}=\frac{-1}{\hslash^2}\int_{-\infty}^{t_0}dt_1\int_{t_0}^{\infty}dt_2[\expval{\mathcal{O}_A(t_1)\mathcal{O}_B(t_2)}_n-\expval{\mathcal{O}_A(t_1)}_n\expval{\mathcal{O}_B(t_2)}_n],
\end{equation}
it should be noted that in this result the mean values are for any n-th excited state, not just the ground state, and also that they are taken using \eqref{meanval sin ops}, however they can be taken in the regular way using operators.

\subsection{Equivalence between the methods}

To gain confidence in the previous result we will show that it can be translated into a more familiar expression regularly used to obtain the QGT. We begin with our equation \eqref{TGC general} 

\begin{equation}\nonumber
G_{A B}^{(n)}=\frac{-1}{\hslash^2}\int_{-\infty}^{t_0}dt_1\int_{t_0}^{\infty}dt_2[\expval{\mathcal{O}_A(t_1)\mathcal{O}_B(t_2)}_n-\expval{\mathcal{O}_A(t_1)}_n\expval{\mathcal{O}_B(t_2)}_n],
\end{equation}
and we will take it to the Schrödinger picture where the operators are time independent, meaning that if we have an arbitrary operator $\hat{A}(t)$, it should be written as  $\hat{A}(t)=e^{\frac{i}{\hslash}Ht}\hat{A}e^{-\frac{i}{\hslash}Ht}$ where $\hat{A}$ is in the Schrödinger picture and it can be thought as the one in the Heisenberg picture at time $t=0$. First we will focus our attention to the second term in  \eqref{TGC general}

\begin{equation}
\expval{\mathcal{O}_A(t_1)}_n = \bra{n}e^{\frac{i}{\hslash}Ht_1}\mathcal{O}_Ae^{-\frac{i}{\hslash}Ht_1}\ket{n}=\bra{n}\mathcal{O}_A\ket{n},
\end{equation}
where we applied the first exponential to the bra, and the second to the ket, then the results cancel each other. Now if we do the same manipulation to the first term we find that

\begin{equation}
\expval{\mathcal{O}_A(t_1)\mathcal{O}_B(t_2)}_n=e^{\frac{i}{\hslash}E_n (t_2-t_1)}\bra{n}\mathcal{O}_Ae^{-\frac{i}{\hslash}Ht_1}e^{\frac{i}{\hslash}Ht_2}\mathcal{O}_B\ket{n},
\end{equation}
and inserting an identity operator in the energy basis between the exponentials we get 

\begin{equation}
\expval{\mathcal{O}_A(t_1)\mathcal{O}_B(t_2)}_n
=\sum_{m}e^{\frac{i}{\hslash}(E_m-E_n) (t_2-t_1)}\bra{n}\mathcal{O}_A\ket{m}\bra{m}\mathcal{O}_B\ket{n}.
\end{equation}

If we extract the nth term from the sum as

\begin{align}
&\nonumber\sum_{m}e^{\frac{i}{\hslash}(E_m-E_n) (t_2-t_1)}\bra{n}\mathcal{O}_A\ket{m}\bra{m}\mathcal{O}_B\ket{n}
\\&=\sum_{m\neq n}e^{\frac{i}{\hslash}(E_m-E_n) (t_2-t_1)}\bra{n}\mathcal{O}_A\ket{m}\bra{m}\mathcal{O}_B\ket{n} + \bra{n}\mathcal{O}_A\ket{n}\bra{n}\mathcal{O}_B\ket{n},
\end{align}
the integrand of  \eqref{TGC general} becomes

\begin{equation}
\expval{\mathcal{O}_A(t_1)\mathcal{O}_B(t_2)}_n-\expval{\mathcal{O}_A(t_1)}_n\expval{\mathcal{O}_B(t_2)}_n = \sum_{m\neq n}e^{\frac{i}{\hslash}(E_m-E_n) (t_2-t_1)}\bra{n}\mathcal{O}_A\ket{m}\bra{m}\mathcal{O}_B\ket{n},
\end{equation}
and then

\begin{equation}
G_{A B}^{(n)}=\frac{-1}{\hslash^2}\sum_{m\neq n}\Big[\int_{-\infty}^{t_0}dt_1\int_{t_0}^{\infty}dt_2e^{\frac{i}{\hslash}(E_m-E_n) (t_2-t_1)}\Big]\bra{n}\mathcal{O}_A\ket{m}\bra{m}\mathcal{O}_B\ket{n},
\end{equation}
where we have isolated the time dependence. To integrate it, keeping in mind the ranges of $t_1$ and $t_2$, we establish the prescription
\begin{align}
	\int_{-\infty}^{t_0}dt_1\int_{t_0}^{\infty}dt_2e^{\frac{i}{\hslash}(E_m-E_n) (t_2-t_1)} &= \lim\limits_{\epsilon\to 0^+}\int_{-\infty}^{t_0}dt_1\int_{t_0}^{\infty}dt_2e^{\frac{i}{\hslash}(E_m-E_n+i\epsilon) (t_2-t_1)} 
	\\&= \frac{-\hslash^2}{(E_m-E_n)^2}
\end{align}
and therefore

\begin{equation}
G_{A B}^{(n)}=\sum_{m\neq n}\frac{\bra{n}\mathcal{O}_A\ket{m}\bra{m}\mathcal{O}_B\ket{n}}{(E_m-E_n)^2}.
\end{equation}

In the case that we are only interested in perturbing the parameters $\lambda^a$ of the Hamiltonian, the above expression becomes

\begin{equation}\label{f pert TGC}
G_{ij}^{(n)}=\sum_{m\neq n}\frac{\bra{n}\partial_i H\ket{m}\bra{m}\partial_j H\ket{n}}{(E_m-E_n)^2}.
\end{equation}
which is the perturbative formula for the Quantum Geometric Tensor \cite{Gu, Zanardi}.\\

From \eqref{f pert TGC}, we can easily arrive to Provost's and Vallee's expression \cite{Provost}. First we differentiate the Schrödinger equation as
\begin{equation}
	\braket{\partial_i m }{n}=\frac{\bra{n}\partial_i H\ket{m}}{(E_n-E_m)},
\end{equation}
where  $n\neq m$, then
\begin{equation}
G_{ij}^{(n)}=\sum_{m\neq n}\braket{\partial_i n}{m}\braket{m}{\partial_j n} = \sum_{m}\braket{\partial_i n}{m}\braket{m}{\partial_j n} - \braket{\partial_i n}{n}\braket{n}{\partial_j n},
\end{equation}
where we added a zero in the form of the missing term needed to complete the sum. From here we observe that there is an identity operator in the energy basis ($\mathbb{I}=\sum_{m}\ket{m}\bra{m}$), and therefore
\begin{equation}
G_{ij}^{(n)}=\bra{\partial_i n}\ket{\partial_j n} - \braket{\partial_i n}{n}\braket{n}{\partial_j n}.
\end{equation}

From these result we can see that our equation \eqref{TGC general} has the same validity (when dealing with perturbations of the parameters) as the one by Provost and Vallee, but it does not carry the burden of having to deal with, or even know, the wave function in the sense that we can obtain our expectation values using perturbation theory in the same way as with our Green functions from chapter 1.\newline

It can also be expanded to consider variations of the phase space with translations as the perturbation and an ordering rule. With this procedure we can calculate the purity of systems as we will see in the next chapter.

\chapter{Quantum covariance matrix, purity and entropy}

When dealing with Gaussian states in quantum mechanics, the quantum covariance matrix completely determines several properties of said states, like the purity, linear quantum entropy, and von Neumann entropy, which are intimately related to the entanglement between them. For this reason, in this chapter we are going to link the QGT with the Quantum Covariance Matrix to generate a new way to calculate all these previously mentioned properties of the states.

\section{Quantum Covariance Matrix}

The \textit{quantum covariance matrix} is the generalization of the probabilistic covariance matrix (for a review of several probability and statistics concepts see Appendix A), which is the square matrix that contains the covariance between the elements conforming a vector. Any covariance matrix is symmetric, semipositive defined, and on the diagonal it contains the variances.\newline

Let $Z=\left(Z_{1}, \ldots, Z_{m}\right)$ be a random variable of $\mathbb{R}^{m}$, then its covariance matrix of $m \times m$ entries will be $\sigma=\left(\operatorname{Cov}\left(Z_{j}, Z_{k}\right)\right)_{1 \leq j, k \leq m}$. The principal diagonal of the covariance matrix will consist of the variances denoted by $\sigma_{Z_{j}}^{2}$.\newline

For example, if $m=2$, taking our vector as $Z=(Q, P)$ the covariance matrix is:
\begin{equation}
\sigma=\left[\begin{array}{cc}
\operatorname{Cov}(X, X)  & \operatorname{Cov}(X, P) \\
\operatorname{Cov}(X, P) & \operatorname{Cov}(P, P)
\end{array}\right], 
\end{equation}
with
\begin{equation}
     \operatorname{Cov}(X, P)=\langle X P \rangle-\langle X \rangle \langle P \rangle ,
\end{equation}
which can be generalized if we now we expand our vector $Z$ as
\begin{equation}
Z=\left(Q_{1}, \ldots, Q_{n} ; P_{1}, \ldots, P_{n}\right)
\end{equation}
where $Q_{j}: \mathbb{R}_{x}^{n} \longrightarrow \mathbb{R}$ and $P_{j}: \mathbb{R}_{p}^{n} \longrightarrow \mathbb{R} $. In this case the quantum covariance matrix $\sigma$ of $Z$ is the symmetric matrix
\begin{equation}
\sigma=\left[\begin{array}{cc}
\sigma_{Q Q} & \sigma_{Q P} \\
\sigma_{P Q} & \sigma_{P P}
\end{array}\right] \quad \text { with } \quad \sigma_{P Q}=\sigma_{Q P}^{T}
\end{equation}
where $\sigma_{Q Q}$ and $\sigma_{P P}$ stand for the covariance matrices of $Q=\left(Q_{1}, \ldots, Q_{n}\right)$ and $P=\left(P_{1}, \ldots, P_{n}\right)$ respectively, and 
\begin{equation}
\sigma_{Q P}=\left(\operatorname{Cov}\left(Q_{j}, P_{k}\right)\right)_{1 \leq j, k \leq n}.
\end{equation}

Usually, as the notation implies, the $Q_{j}$ and $P_{k}$ functions represent the position $q_{j}$, and momentum $p_{k}$ coordinates \cite{de Gosson}. Taking into account that our coordinates would not always commute (in general everywhere within our matrix except for the diagonal) in quantum mechanics we define the \textit{quantum covariance matrix} as

\begin{equation}\label{qcov matrix}
\sigma_{\alpha \beta} = \frac{1}{2} \langle \hat{\mathbf{z}}_{\alpha} \hat{\mathbf{z}}_{\beta} + \hat{\mathbf{z}}_{\beta} \hat{\mathbf{z}}_{\alpha}  \rangle_{m}-\langle \hat{\mathbf{z}}_{\alpha}\rangle_{m} \langle \hat{\mathbf{z}}_{\beta} \rangle_{m}\,, 
\end{equation}
with
\begin{equation}
    \hat{\mathbf{z}} = (\hat{q}_1, \hat{q}_2, \cdots, \hat{q}_n, \hat{p}_1, \hat{p}_2, \cdots, \hat{p}_n).
\end{equation}

\section{Relationship between the covariance matrix and the QGT}

To relate the phase space part of the QGT  $G_{AB}^{(m)}$  with the quantum covariance matrix we will start with the previously obtained perturbative form of the QGT \eqref{f pert TGC} 
\begin{equation}\nonumber
G_{AB}^{(m)}=\sum_{n\neq m}\frac{\langle m|\boldsymbol{\hat{\mathcal{O}}}_{A}|n\rangle\langle n|\boldsymbol{\hat{\mathcal{O}}}_{B}|m\rangle}{(E_{n}-E_{m})^{2}}\,,
\end{equation}
and to  focusing on its phase space part 
\begin{equation}
G_{ab}^{(m)}=\sum_{n\neq m}\frac{\langle m|\boldsymbol{\hat{\mathcal{O}}}_{a}|n\rangle\langle n|\boldsymbol{\hat{\mathcal{O}}}_{b}|m\rangle}{(E_{n}-E_{m})^{2}}\,.
\end{equation}

Now we will first consider the operator $\hat{\mathbf{q}}_a$, then the Schrödinger equation in configuration space is
\begin{equation}
\hat{\mathbf{H}} \left( q,-i\hslash\frac{\partial}{\partial q} \right) \psi_{n}(q) = E_{n}\psi_{n}(q)\,,
\end{equation}
with $\psi_{n}(q)=\langle q|n \rangle $. If we apply on it the operator $-i\hslash\frac{\partial}{\partial q_a}$, multiply it by the wavefunction $\psi_{m}^{*}(q)$, and integrate it with respect to $q$, we find
\begin{equation}
-i\hslash \int \mathrm{d}^N q\, \psi_{m}^{*}(q)\frac{\partial \hat{\mathbf{H}}}{\partial q_a}\psi_{n}(q) = (E_{n} - E_{m}) \int \mathrm{d}^N q\, \psi_{m}^{*}(q)\left(-i\hslash\frac{\partial \psi_{n}(q)}{\partial q_a}\right) \, ,
\end{equation}
or equivalently,
\begin{equation}
-i\hslash \langle m| \boldsymbol{\hat{\mathcal{O}}}_{q_a} |n \rangle = (E_{n} - E_{m}) \langle m| \hat{\mathbf{p}}_a |n \rangle \,. 
\end{equation}
Utilizing these results we can see that
\begin{equation}
G_{q_{a}q_{b}}^{(m)}=\sum_{n\neq m}\frac{\langle m|\boldsymbol{\hat{\mathcal{O}}}_{q_a}|n\rangle\langle n|\boldsymbol{\hat{\mathcal{O}}}_{q_b}|m\rangle}{(E_{n}-E_{m})^{2}}=\sum_{n\neq m}\frac{1}{\hslash^2}\langle m| \hat{\mathbf{p}}_a |n \rangle \langle n| \hat{\mathbf{p}}_b |m \rangle\,.
\end{equation}
Now we sum a zero by adding and substracting the missing term with $n=m$ and identifyng the emerging completeness relation we arrive at the expression:
\begin{equation}
G_{q_{a}q_{b}}^{(m)}=\frac{1}{\hslash^2}\big{(} \langle \hat{\mathbf{p}}_{a} \hat{\mathbf{p}}_{b} \rangle_{m} - \langle \hat{\mathbf{p}}_{a} \rangle_{m} \langle \hat{\mathbf{p}}_{b} \rangle_{m} \big{)}\,. \label{gtqq}
\end{equation}
At this point we take the real part of \eqref{gtqq} to find the Quantum Metric Tensor of configuration space, yielding the result
\begin{equation}\label{qcov qgt 1}
g_{q_{a}q_{b}}^{(m)}=\frac{1}{\hslash^2}\left( \frac{1}{2}\langle \hat{\mathbf{p}}_{a} \hat{\mathbf{p}}_{b}+\hat{\mathbf{p}}_{b} \hat{\mathbf{p}}_{a} \rangle_{m} - \langle \hat{\mathbf{p}}_{a} \rangle_{m} \langle \hat{\mathbf{p}}_{b} \rangle_{m} \right)\,.
\end{equation}

On the other hand, if instead we utilized the Schrödinger equation in momentum space, we would have found the following relation for $p_a$:
\begin{equation}
i\hslash \langle m| \boldsymbol{\hat{\mathcal{O}}}_{p_a} |n \rangle = (E_{n} - E_{m}) \langle m| \hat{\mathbf{q}}_a |n \rangle \,. 
\end{equation}
Which allows us to proceed in a similar manner as before, finding that for the terms $g_{q_{a}p_{b}}^{(m)}$ and $g_{p_{a}p_{b}}^{(m)}$ we have
\begin{equation}\label{qcov qgt 2}
g_{q_{a}p_{b}}^{(m)}=-\frac{1}{\hslash^2}\left( \frac{1}{2}\langle \hat{\mathbf{p}}_{a} \hat{\mathbf{q}}_{b}+\hat{\mathbf{q}}_{b} \hat{\mathbf{p}}_{a} \rangle_{m} - \langle \hat{\mathbf{p}}_{a} \rangle_{m} \langle \hat{\mathbf{q}}_{b} \rangle_{m} \right)\,,
\end{equation}
\begin{equation}\label{qcov qgt 3}
g_{p_{a}p_{b}}^{(m)}=\frac{1}{\hslash^2}\left( \frac{1}{2}\langle \hat{\mathbf{q}}_{a} \hat{\mathbf{q}}_{b}+\hat{\mathbf{q}}_{b} \hat{\mathbf{q}}_{a} \rangle_{m} - \langle \hat{\mathbf{q}}_{a} \rangle_{m} \langle \hat{\mathbf{q}}_{b} \rangle_{m} \right)\,.
\end{equation}

With this we have discovered that the Quantum Metric Tensor for the phase space is intimately related to the quantum covariance metric \eqref{qcov matrix} (it is important to remark that it contains the $1/\hslash^2$ factor and the sign of the components $g_{q_{a}p_{b}}^{(m)}$).

\section{Density operator, purity and entropy}

When we work with ensembles where all the individual systems that constitute it are in the same state $|\psi\rangle$, we refer to them as an ensemble of \textit{pure states} and  can be described as,
\begin{equation}\label{pure state}
|\psi\rangle=\sum_{i}^N c_{i}|i\rangle
\end{equation}
where $|i\rangle$ are the basis kets of our $N$-dimensional Hilbert space (we are restricting ourselves to the case where the elements of the basis  $|i\rangle$ are orthonormal); however they are hard to come by in nature since most of our systems will be interacting with their surroundings and will become altered. It is more common to encounter \textit{mixed states}, which as the name implies could be in any of a varied collection of possible states that characterize the complete system but just with them alone, we can not describe our complete pure state, i.e. \eqref{pure state}.\newline

A convenient way to encapsulate all the information of the ensemble, pure or not, is in the form of the \textit{density operator} (sometimes called density matrix since it can take that form in some basis):
\begin{equation}
\rho=\sum_{i}^N p_{i}|i\rangle\langle i|
\end{equation}
where $p_{i}$ is the probability that a state picked randomly out of the ensemble is in the state $|i\rangle$. There is a simple way to verify if our state is pure or mixed, by taking the trace of density operator squared. If our state is pure then 
\begin{equation}
    \operatorname{Tr}(\rho^2) = 1
\end{equation}
otherwise, if it is mixed
\begin{equation}
    \frac{1}{d} \leq\operatorname{Tr}(\rho^2)  < 1,
\end{equation}
where $d$ is the dimension of the Hilbert space (if it is infinite the lower bound is $0$).\newline

Our system can be composed by two subsystems $A$ and $B$,  we will call a \textit{bipartite system} one with a Hilbert space that can be written as,
\begin{equation}
\mathcal{H}_{A B}=\mathcal{H}_{A} \otimes \mathcal{H}_{B}
\end{equation}
but most of the time we wont be able to separate our subsystems so easily. In general, we study the state of a subsystem (pure or mixed) by obtaining its \textit{reduced density matrix}, which is defined by the partial trace (a generalization of the regular trace) of the state density matrix of whole system as
\begin{equation}
\rho_{A}=\operatorname{tr}_{B} \rho,
\end{equation}
meaning that if we want the reduced density matrix of the state in which subsystem $A$ is, we must take the partial trace of our whole density matrix with respect to the rest, in this case, the subsystem $B$.\newline

Specifically, any density operator of a bipartite system in the Hilbert space $\mathcal{H}_{A} \otimes \mathcal{H}_{B}$ can then be decomposed as \begin{equation}
\rho_{A B}=\sum_{i j k l} c_{i j k l}\left|a_{i}\right\rangle\left\langle a_{j}|\otimes| b_{k}\right\rangle\left\langle b_{l}\right|
\end{equation}
where  $\left\{\left|a_{i}\right\rangle\right\}$ and  $\left\{\left|b_{i}\right\rangle\right\}$ are the basis of $\mathcal{H}_{A}$ and $\mathcal{H}_{B}$ respectively. Then, we will define its \textit{partial trace} as:
\begin{equation}
\rho_A =\operatorname{tr}_{B} \rho_{A B}=\sum_{i j k l} c_{i j k l}\left|a_{i}\right\rangle\left\langle a_{j}\right|\left\langle b_{l} \mid b_{k}\right\rangle
\end{equation}
which lives on $\mathcal{H}_{A}$. It should be noted that  $\operatorname{Tr}\left|b_{k}\right\rangle b_{l} \mid=\sum_{n}\left\langle n \mid b_{k}\right\rangle\left\langle b_{e} \mid n\right\rangle=\left\langle b_{l} \mid b_{k}\right\rangle$ is a complex number.\newline

Now since our density matrix is itself a sum of operators, we can define the average expected value of an arbitrary operator $\alpha$ as:
\begin{equation}\label{expect val densop}
\langle\bar{\alpha}\rangle=\sum_{i} p_{i}\langle i|\alpha| i\rangle
\end{equation}
where the bar on $\langle\bar{\alpha}\rangle$ is there to reminds us that two kinds of averaging have been carried out. First we obtained the expectation value $\langle i|\alpha| i\rangle$ for each possible state $|i\rangle$ and then we took the average over these results with the sum and our $p_i$ factor. This operation can be also expressed as

\begin{align}
\operatorname{Tr}(\alpha \rho) &=\sum_{j}\langle j|\alpha \rho| j\rangle \\
&=\sum_{j} \sum_{i}\langle j|\alpha| i\rangle\langle i \mid j\rangle p_{i}=\sum_{i} \sum_{j}\langle i \mid j\rangle\langle j|\alpha| i\rangle p_{i} \\
&=\sum_{i}\langle i|\alpha| i\rangle p_{i} \\
&\label{(4.2.22)}=\langle\bar{\alpha}\rangle.
\end{align}
With this result we can obtain the following important properties of the density operator \cite{Shankar}: 
\begin{itemize}
    \item $\rho^{\dagger}=\rho$
\item $\operatorname{Tr} \rho=1$
\item $\rho^{2}=\rho$ for a pure ensemble
\item $\rho=(1 / k) I$
for an ensemble uniformly distributed over $k$ states
\item $\frac{1}{d} \leq \operatorname{Tr} \rho^{2} \leq 1$ (upper equality holds for a pure ensemble)
\end{itemize}

Given the usefulness of the trace of $\rho^{2}$ and its particular properties when dealing with pure states, it seems convenient to give it its own name. We define the \textit{purity} $\mu$ of a state $\rho$ as,
\begin{equation}
\mu(\rho)=\operatorname{Tr} \rho^{2},
\end{equation}
and it is one of the possible ways to measure the amount of information in a system.\newline

When our Hilbert space $\mathcal{H}$ is of $N$ dimensions it has the range 
\begin{equation}
\frac{1}{N} \leq \mu \leq 1 \text ,
\end{equation}
attaining the value of $1/N$ when dealing with a totally random mixture of states, and $1$ in the case of pure states. It should be noted that when the dimension of our space goes to infinity $N \rightarrow \infty$ (or in the limit of continuous systems), the minimum value of the purity tends to zero.\newline

Now that we can measure how "pure" is a state, we can construct a way to quantify how "impure" or "how mixed" a quantum state is. The simplest way to do so is with the \textit{linear entropy} defined as:
\begin{equation}
S_{L}(\rho)=\frac{N}{N-1}(1-\mu)=\frac{N}{N-1}\left(1-\operatorname{Tr} \rho^{2}\right) ,
\end{equation}
and has the possible values
\begin{equation}
0 \leq S_L (\rho) \leq 1 \text.
\end{equation}
 
 The linear entropy, as the name suggest, is only a first-order approximation of a more powerful type of entropy, the von Neumann entropy. 

\subsection{Shannon and von Neumann entropies}

To fully understand the von Neumann entropy we must first study its classical counterpart, the Shannon entropy. To do so, we need to learn how to mathematically represent how much information of our space of probabilities we gain after one measurement.\newline

The basic unit of classical information is the \textit{bit} (sometimes called the \textit{shannon}), and $1$ bit can be understood as the information gained from a measurement that cuts the space of possibilities in half, meaning that if prior to a measurement we have $6$ possibilities and after it we have $3$ remaining, we gained $1$ bit of information of our system. If instead a measurement reduces the space of possibilities to $1/4$ of the  original, we say that the it gave us $2$ bits of information, and so on or and so forth. In this way of thinking we can characterize the \textit{information} $I$ in terms of the probability as follows:
\begin{equation}
  p = \left(\frac{1}{2}\right)^I 
\end{equation}
rearranging we get
\begin{equation}
    I = \log_2\left(\frac{1}{p}\right) = -\log_2(p)
\end{equation}
which can be thought as how many times a measurement cuts our possibilities in half\footnote{For a detailed yet easy to understand explanation about this topic watch \textit{Solving Wordle using information theory} by 3Blue1Brown \cite{3b1b info}.}\footnote{It should be noted that this last expression can be represented in different bases of the logarithm, if we have base 2 we are dealing with \textit{bits}, base $e$ gives us \textit{natural units} or \textit{nat}, and base 10 is represented with \textit{dits}.}.\newline

Most of the time, a measurement has an arrange of possible outcomes each with different probabilities of occurring, so in reality what we need to define is the expected value of the information that we might get from that measurement, and we do so in the usual way:
\begin{equation}\label{Shannon}
S_S=-\sum_{k=1}^{N} p_{k} \log p_{k}
\end{equation}
which we will call \textit{Shannon entropy} \cite{Shannon}. It can also be interpreted as our ignorance of the system prior to the measurement.\newline

 The Shannon entropy was built for classical information theory, in quantum information theory the probabilities $\left\{p_{k}\right\}$ that we utilize in \eqref{Shannon} need to be substituted by the eigenvalues of the density matrix $\rho$, giving us the \textit{von Neumann entropy} \cite{von Neumann}, 
\begin{equation}
S_{V}=-\operatorname{Tr}[\rho \log \rho]=-\sum p_{k} \log p_{k}.
\end{equation}
which deals with \textit{qubits} as the unit of \textit{quantum information}. The main difference between qubits and regular bits is that instead of dealing with absolute answers in the form of ones or zeroes, our state that carries information can remain in a superposition of states while we make operations on it.\newline

The von Neumann entropy has the following properties:

\begin{itemize}
    \item Concavity    
    \begin{equation}
S_{V}\left(\lambda_{1} \rho_{1}+\ldots+\lambda_{n} \rho_{n}\right) \geq \lambda_{1} S_{V}\left(\rho_{1}\right)+\ldots+\lambda_{n} S_{V}\left(\rho_{n}\right),
\end{equation}
with $\lambda_{i} \geq 0, \sum_{i} \lambda_{i}=1$. Meaning that the Von Neumann entropy increases with mixed states.

\item Subadditivity. 
Consider a system $\mathcal{S}$ with two subsystems $\mathcal{S}_{1,2}$, then
\begin{equation}\label{subadditivity}
S_{V}(\rho) \leq S_{V}\left(\rho_{1}\right)+S_{V}\left(\rho_{2}\right),
\end{equation}
where $\rho_{1}=\operatorname{Tr}_2 \rho $ is the reduced density matrix  of the subsystem $1$, and analogously for the subsystem $2$. We get the equality when the density operator is directly the tensor product of our two subsystems
\begin{equation}
S_{V}\left(\rho_{1} \otimes \rho_{2}\right)=S_{V}\left(\rho_{1}\right)+S_{V}\left(\rho_{2}\right) .
\end{equation}
Which is not the same as with the purity, that instead of having the sum of the individual entropies, we had the multiplication on the purities of the states
\begin{equation}
\mu\left(\rho_{1} \otimes \rho_{2}\right)=\mu\left(\rho_{1}\right) \cdot \mu\left(\rho_{2}\right),
\end{equation}
since  the trace of a product equates the product of the traces.

\item Araki Lieb inequality
\begin{equation}
S_{V}(\rho) \geq\left|S_{V}\left(\rho_{1}\right)-S_{V}\left(\rho_{2}\right)\right| .
\end{equation}
\item Triangle inequality
\begin{equation}
\left|S_{V}\left(\rho_{1}\right)-S_{V}\left(\rho_{2}\right)\right| \leq S_{V}(\rho) \leq S_{V}\left(\rho_{1}\right)+S_{V}\left(\rho_{2}\right) .
\end{equation}
\end{itemize}

Within \eqref{subadditivity} there is an important difference between the classical and quantum information theories, since when we deal with the analogous property of classical Shannon entropy we find that the global entropy is bigger than that of the parts, i.e.
\begin{equation}
S(X, Y) \geq S(X),
\end{equation}
\begin{equation}
S(X, Y) \geq S(Y).
\end{equation}
This implies that there is more information in a composite classical system than in any of its parts; however, when we consider the von Neumann entropy this does not occur. For example, suppose that we have a  bipartite quantum system in a pure state $\rho=|\psi\rangle\langle\psi|$, then its von Neumann entropy is $S_{V}(\rho)=0$, while for the individual subsystems would be $S_{V}\left(\rho_{1}\right) = S_{V}\left(\rho_{2}\right) \geq 0$. Meaning that even if our original global system $\rho$ was prepared in a well defined and completely known way, when we measure local observables  on each of the subsystems, we can not avoid the randomness in our results since some unpredictability is intrinsic to our quantum systems.\newline

It is imperative to recognize that we cannot always reconstruct our whole system described by $\rho$ (apart from the trivial instance of $\rho =\rho_{1} \otimes \rho_{2}$ ), with only the information provided separately by the two subsystems. There is some non-local and non-factorizable information encoded in quantum correlations between the two subsystems, in this way we say that they are \textit{entangled}, which is something completely new and different with respect to the classical counterpart.\newline

With the von Neumann entropy we can study the entanglement between subsystems utilizing the reduced density matrix:
\begin{equation}
S_{V(A)} \equiv-\operatorname{tr} \rho_{A} \log \rho_{A}
\end{equation}

Finally, we should remark that all the quantum quantities defined in this section, specifically the purity $\mu$, linear entropy $S_{L}$ and Von Neumann entropy $S_{V}$, are invariant under unitary transformations since they only depend on the eigenvalues of $\rho$ \cite{Tesis Italiana}.

\section{Wigner functions and Gaussian States}

When we deal with a quantum mechanical system, we normally describe it either in the configuration space or momentum space, and therefore it would seem desirable to define a quantum version of the phase space, but we cannot do so in the same way as in classical mechanics since the Heisenberg uncertainty relation is ingrained within our framework. To generalize the idea of phase  space, allowing it to handle the probabilistic interpretation of quantum mechanics we need to define what is known as the Wigner function of our system, and if the form of this function turns out to be Gaussian, we will label our state as a Gaussian state.\newline

Gaussian states have been used as a tool to research the nature of entanglement in continuous systems and have gained popularity inside the field of Quantum Optics because they can be represented by a relatively simple algebraic formalism and even in the case of systems with infinite dimensions can be entirely described with a finite number of parameters. They are also easy to work with experimentally since their corresponding physical states are manageable to prepare and control in the laboratory by means of standard quantum optics techniques \cite{Gaussian Quantum Information}. \newline

\subsection{Wigner function}

In quantum mechanics, the position and momentum do not have common eigenstates, and given the commutator of these operators $[q_i,p_j]=i \delta_{i,j}\hslash$ (which implies the uncertainty relation), we can conclude that these observables cannot take definite values, and therefore, well defined trajectories in phase space do not exist for quantum mechanical systems.\newline

As we have mentioned before, here the Heisenberg uncertainty principle gets in our way to define a quantum phase space, but if we built one that could operate the uncertainties we may naively expect it to simply blur the classical trajectories in terms of probability amplitudes. However, this will not be the case since we need negative probability distributions to account for all quantum phenomena, implying that our quantum uncertainties are far more subtle  and rich than regular classical noise.\newline

If we have a well defined system described by the density operator $\hat{\rho}$, the best thing we can use to describe its position or momentum are its probability density functions that detail the statistics of a measurement, mathematically we represent them by $\langle q|\hat{\rho}| q\rangle$ and $\langle p|\hat{\rho}| p\rangle$. \newline

Therefore, to successfully generalize the idea of phase space, what we really need is a joint probability distribution of position and momentum, that we will call the \textit{Wigner function} denoted by $W_{\rho}(\mathbf{r})$ where $\mathbf{r}=\left(\mathbf{q}_{1}, \ldots, \mathbf{q}_{N}, \mathbf{p}_{1}, \ldots, \mathbf{p}_{N}\right)^{T}$ are our phase space variables, and the subindex indicates that it belongs to our quantum state $\hat{\rho}$. What characterizes this particular function is that its marginals correspond to the position and momentum probability density functions, that is
\begin{equation}\label{marginals Wigner}
\langle q|\hat{\rho}| q\rangle=\int_{\mathbb{R}} d p W_{\rho}(\mathbf{r}) \quad \text { and } \quad\langle p|\hat{\rho}| p\rangle=\int_{\mathbb{R}} d q W_{\rho}(\mathbf{r}),
\end{equation}
and one can prove that this function is uniquely defined in this way, meaning that the whole information of our system will be contained within its Wigner function and all the results that can be obtained with it are equivalent to those obtained with the density operator.\newline

Before giving an explicit form of it, we need to encode all our position and momentum operators (in this context they are regularly called \textit{cuadrature operators}) in the single vector $\hat{\mathbf{R}}=\left(\hat{q}_{1}, \ldots, \hat{q}_{N}, \hat{p}_{1}, \ldots, \hat{p}_{N}\right)$, then we can write the canonical commutation relations in the particular form
\begin{equation}
\left[\hat{R}_{m}, \hat{R}_{n}\right]= \mathrm{i}\hslash \Omega_{m n}, 
\end{equation}
where
\begin{equation}
\Omega=\bigoplus_{j=1}^{N} \Omega_{1}, \quad
\end{equation}
with
\begin{equation}
    \Omega_{1}=\left(\begin{array}{cc}
0 & 1 \\
-1 & 0
\end{array}\right)
\end{equation}
is \textit{symplectic form}, which satisfies $\Omega^{T}=-\Omega=\Omega^{-1} .$ Also, with the quantum covariance matrix it fulfills the following relation:
\begin{equation}\label{symplectic uncertainty}
\sigma+i \Omega \geq 0
\end{equation}

We will define the \textit{displacement operator} as
\begin{equation}
\hat{D}(\mathbf{r})=\exp \left[\frac{\mathrm{i}}{2\hslash} \hat{\mathbf{R}}^{T} \Omega \mathbf{r}\right]=\exp \left[\frac{\mathrm{i}}{2\hslash}(p \hat{q}-q \hat{p})\right]
\end{equation}
and its expectation value, called the \textit{characteristic function}
\begin{equation}
\chi_{\rho}(\mathrm{s})=\operatorname{tr}\{\hat{\rho} \hat{D}(\mathrm{~s})\}=\langle\hat{D}(\mathrm{~s})\rangle
\end{equation}
and with them we can now give an explicit, or more useful form of the Wigner function as the Fourier transform of the characteristic function:
\begin{equation}
W_{\rho}(\mathbf{r})=\int_{\mathbb{R}^{2}} \frac{d^{2} \mathbf{s}}{(4 \pi \hslash)^{2}} e^{-\frac{i}{2\hslash} \mathbf{r}^{T} \Omega \mathbf{s}} \chi_{\rho}(\mathbf{s}).
\end{equation}
It should be noted that the Wigner function has an alternative formulation (the original formulation proposed by Wigner), that is useful for certain calculations:

\begin{equation}\label{Wigner Original}
W_{\rho}(\mathbf{r})=\int_{\mathbb{R}} \frac{d y}{4 \pi \hslash} e^{-\frac{i}{2\hslash} p y}\left\langle q+\frac{y}{2}|\hat{\rho}| q-\frac{y}{2}\right\rangle
\end{equation}

For completeness, we will now enlist some of the most useful properties of the Wigner function:

\begin{itemize}

\item The marginals can be expressed as \eqref{marginals Wigner}. To see this we begin by integrating the Wigner function with respect to the momentum
\begin{align}
    \int_{\mathbb{R}} d p W_{\rho}(\mathbf{r})&=\int_{\mathbb{R}^{2}} \frac{d^{2} \mathbf{r}^{\prime}}{(4 \pi \hslash)^{2}} \left[\int_{\mathbb{R}} d p e^{\frac{i}{2 \hslash} q^{\prime} p}\right] e^{-\frac{i}{2 \hslash} p^{\prime} q} \chi\left(\mathbf{r}^{\prime}\right)
    \\&=\int_{\mathbb{R}^{2}} \frac{d^{2} \mathbf{r}^{\prime}}{(4 \pi \hslash)^{2}} [2\hslash ( 2\pi) \delta\left(q^{\prime}\right)] e^{-\frac{i}{2} p^{\prime} q} \chi\left(\mathbf{r}^{\prime}\right)
    \\&
    =\int_{\mathbb{R}} \frac{d p^{\prime}}{4 \pi \hslash} e^{-\frac{i}{2} p^{\prime} q} \chi\left(0, p^{\prime}\right)
    \\&=\int_{\mathbb{R}} \frac{d p^{\prime}}{4 \pi \hslash} e^{-\frac{i}{2\hslash} p^{\prime} q} \operatorname{tr}\left\{\hat{\rho} \hat{D}\left(0, p^{\prime}\right)\right\}
\end{align}
where we used that the Dirac's delta function can be written as
\begin{equation}
\delta(q)=\frac{1}{2 \pi} \int_{-\infty}^{\infty} e^{i pq} d p
\end{equation}
and that it has the property
\begin{equation}
\delta(\alpha q)=\frac{\delta(q)}{|\alpha|}.
\end{equation}
Now we put the trace in terms of the position eigenstates as a basis to get
\begin{align}
    \int_{\mathbb{R}} d p W_{\rho}(\mathbf{r})&=\int_{\mathbb{R}} d y \int_{\mathbb{R}} \frac{d p^{\prime}}{4 \pi \hslash} e^{-\frac{i}{2 \hslash} p^{\prime} q}\left\langle y\left|\hat{\rho} e^{\frac{i}{2} p^{\prime} \hat{Q}}\right| y\right\rangle
    \\&=\int_{\mathbb{R}} d y \left[\int_{\mathbb{R}} \frac{d p^{\prime}}{4 \pi \hslash} e^{-\frac{i}{2 \hslash} p^{\prime}(q-y)}\right]\langle y|\hat{\rho}| y\rangle
    \\&=\int_{\mathbb{R}} d y  \delta(q-y)\langle y|\hat{\rho}| y\rangle
    \\&=\langle q|\hat{\rho}| q\rangle, 
\end{align}

which is the probability density function of the position measurements. To obtain the second marginal we need to follow an analogous procedure but instead of integrating with respect to the momenta we integrate with respect to the position.

\item It is real for every point in phase space: $W_{\rho}(\mathbf{r}) \in \mathbb{R}$.

\item It is normalized:
\begin{equation}
\int_{\mathbb{R}^{d}} d^{d} \mathbf{r} W_{\rho}(\mathbf{r})=1
\end{equation}
\item Quantum expectation values can be related to averages in phase space with symmetrically-ordered operators, meaning that
\begin{equation}
\left\langle\left(\hat{q}^{m} \hat{p}^{n}\right)^{(s)}\right\rangle=\int_{\mathbb{R}^{d}} d^{d} \mathbf{r} W_{\rho}(\mathbf{r}) q^{m} p^{n}
\end{equation}
where $\left(\hat{Q}^{m} \hat{P}^{n}\right)^{(s)}$ refers to the symmetrized version of the product inside the expected value, for example
\begin{equation}\left(\hat{q}^{2} \hat{p}\right)^{(s)}=\frac{\left(\hat{q}^{2} \hat{p}+\hat{p} \hat{q}^{2}+\hat{q} \hat{p} \hat{q}\right)}{3} .\end{equation} This means that we can associate a classical observable with its quantum counterpart by first symmetrizing and then substituting the variables with the operators.

\item The trace of two states is the trace of their corresponding Wigner functions:
\begin{equation}\label{trace Wigner}
\operatorname{tr}\left\{\hat{\rho}_{1} \hat{\rho}_{2}\right\}=4 \pi \int_{\mathbb{R}^{d}} d^{d} \mathbf{r} W_{\rho_{1}}(\mathbf{r}) W_{\rho_{2}}(\mathbf{r})
\end{equation}
In the case where $\hat{\rho}_{1}=\hat{\rho}_{2} \equiv \hat{\rho}$, this can be reduced to
\begin{equation}
\int_{\mathbb{R}^{2}} d^{2} \mathbf{r} W_{\rho}^{2}(\mathbf{r})=\frac{1}{4 \pi} \operatorname{tr}\left\{\hat{\rho}^{2}\right\} \leq \frac{1}{4 \pi}
\end{equation}
since $\operatorname{tr}\left\{\hat{\rho}^{2}\right\} \leq 1$, and this implies that the Wigner function is bounded and does not diverge to $+\infty$ nor $-\infty$.

There is another special case of \eqref{trace Wigner} that is really important to look at, when we work with two pure states  $\hat{\rho}_{j}=\left|\psi_{j}\right\rangle\left\langle\psi_{j}\right|$ such that they are orthogonal with each other $\left\langle\psi_{1} \mid \psi_{2}\right\rangle=0$, it takes the form
\begin{equation}
\int_{\mathbb{R}^{2}} d^{2} \mathbf{r} W_{\left|\psi_{1}\right\rangle}(\mathbf{r}) W_{\left|\psi_{2}\right\rangle}(\mathbf{r})=\frac{1}{4 \pi}\left|\left\langle\psi_{1} \mid \psi_{2}\right\rangle\right|^{2}=0 ,
\end{equation}
and what it is really interesting here is that this result makes evident the need for the Wigner function to be negative at some points of phase space, as otherwise the product of the two would always add positively to the integral. This is why we call the Wigner function a \textit{quasiprobability density function}, since these negative values of probability would make no sense in the classical statistical approach.

\item We can uniquely describe a quantum state $\hat{\rho}$ as
\begin{equation}
\hat{\rho}=\int_{\mathbb{R}^{2}} \frac{d^{2} \mathbf{s}}{4 \pi} \hat{D}^{\dagger}(\mathbf{s}) \chi_{\rho}(\mathbf{s}).
\end{equation}

\end{itemize}

The only difference between the Wigner function and a regular probability density function is the fact that the former can obtain negative values. Physically this means that quantum mechanics cannot be simulated as classical noise and our quantum results would never be predicted utilizing only classical mechanics \cite{Notas Carlos}.

\subsection{Gaussian states}

 Gaussian states are common in Nature and easily attainable experimentally \cite{Tesis Italiana}, while also being easy to work with in the theoretical sense. This is because all their statistics are completely defined by first and second order products of our phase space operators such as $\left\langle\hat{R}_{j}\right\rangle$ and $\left\langle\hat{R}_{j} \hat{R}_{l}\right\rangle$. Formally we will define a \textit{Gaussian state} as one that its Wigner function takes the form of a Gaussian distribution, implying that we can express it as 
\begin{equation}
W_{\rho}(\mathbf{r})=\frac{1}{2 \pi \hslash \sqrt{\operatorname{det} \sigma}} \exp \left[-\frac{1}{2\hslash}(\mathbf{r}-\mathbf{d})^{T} \sigma^{-1}(\mathbf{r}-\mathbf{d})\right]
\end{equation}
where $\sigma$ is our quantum covariance matrix given by \eqref{qcov matrix} and $\mathrm{d}$ is called the \textit{mean vector}, which relates to our operators encoded in $\hat{R}_{j}$ as
\begin{equation}
d_{j}=\left\langle\hat{R}_{j}\right\rangle =\int_{\mathbb{R}^{2}} d^{2} \mathbf{r} W_{\rho}(\mathbf{r}) r_{j}.
\end{equation}

Since Gaussian states are always going to be represented by Gaussian functions, keeping the following integral at hand turns out to be quite useful:
\begin{equation}
\int_{\mathbb{R}^{N}} d^{N} \mathbf{r} \; e^{ \left(-\frac{1}{2} \mathbf{r}^{T} A \mathbf{r}+\mathbf{x}^{T} \mathbf{r}\right)}=\sqrt{\frac{(2 \pi)^{N}}{\operatorname{det} A}} \; e^{ \left(\frac{1}{2} \mathbf{x}^{T} A^{-1} \mathbf{x}\right)}
\end{equation}
where $\mathrm{x} \in \mathbb{R}^{N}$ and $\mathrm{A}$ is a non-singular $N \times N$ matrix. \newline

It turns out that Gaussian Wigner functions are positive everywhere, therefore they can be interpreted directly as probability density functions, meaning that the mean vector $\mathrm{d}$ will encode the information related to the expectation values of our observables, while the quantum covariance matrix describes the distribution of the measurements around the mean. In this interpretation, for a Gaussian State to be physical we will require that:
\begin{itemize}
    \item The mean vector and covariance matrix are real.
    \item The covariance matrix is symmetric and positive semidefinite (the variances along the diagonal cannot be negative)
    \item $\operatorname{det}\{ \sigma \} \geq 1$
\end{itemize}
it should be noted that this third requirement is imposed by quantum mechanics, coming directly from the uncertainty principle between our operators.

\subsection{Purity and entropy of Gaussian states}

Since Gaussian states are described with the first or second order products of our operators, an immediate and interesting application of them is the case of a system composed by several subsystems that interact with each other by the product of their separate operators. Here, all our physics will be contained within the quantum covariance matrix $\sigma_{(n)}$, and the Gaussian purity takes the particular form \cite{de Gosson, Purity Gaussian, Russian purity}:
\begin{equation}\label{purity gaussian}
\mu\left(a_{1}, a_{2}, \ldots, a_{n}\right)=\left(\frac{\hslash}{2}\right)^{n} \frac{1}{\sqrt{\operatorname{det} \sigma_{(n)}}} .
\end{equation}
where $n$ represents our individual subsystems $a_{1}, a_{2}, \ldots, a_{n}$, that compose our general system in a compact fashion.\newline

Before studying the form of the von Neumann entropy for Gaussian states we must first understand what a symplectic matrix is and what are its eigenvalues since we will write the former in terms of the latter.\newline

We will denote by \textit{symplectic matrix} any real matrix $S$ that its action preserves the symplectic form $\Omega$, i.e.
\begin{equation}\label{symplectic matrix}
S \Omega S^{T}=\Omega,
\end{equation}
given this particular form, all the symplectic matrices will have $det(S)=\pm 1$ and their inverses can be obtained by the relation
\begin{equation}
    S^{-1}=-\Omega S^{T} \Omega.
\end{equation}

With these symplectic matrices we can formally define our \textit{normal modes}, or how to decouple our subsystems (for example, utilizing a canonical transformation) with the \textit{Williamson theorem} which states that given a $2 N \times 2 N$ positive definite real matrix $M$ (such as our quantum covariance matrix) , there exists a symplectic transformation $S$ such that
\begin{equation}\label{Williamson}
S M S^{T}=D
\end{equation}
with
\begin{equation}
D=\bigoplus_{j=1}^{N} \nu_{j}\left(\begin{array}{ll}
1 & 0 \\
0 & 1
\end{array}\right)
\end{equation}
and $\nu_{j}>0$ for all $j \in\{1, \ldots, n\}$. The components of the set $\left\{\nu_{j}\right\}_{j=1}^{n}$ is what we will call the \textit{symplectic eigenvalues} of our matrix $M$.\newline

This theorem tells us exactly what the symplectic eigenvalues are, but enunciating it like this is not exactly illuminating on how to get them explicitly. To understand this procedure is better to look at the proof of the theorem rather than the theorem itself.\newline

\subsubsection{Proof of the Williamson theorem:}

Since $S$ can be any of the set of real  symplectic matrices that follows \eqref{Williamson} where $M$ invertible and with strictly positive eigenvalues, we can construct them  as 
\begin{equation}
S=D^{1 / 2} O M^{-1 / 2}
\end{equation}
so \eqref{Williamson} can be written as
\begin{align}
S M S^{T}&=
D^{1 / 2} O M^{-1 / 2} M^{1 / 2} M ^{1 / 2} (D^{1 / 2} O M^{-1 / 2})^{T}
\\&
=D^{1 / 2} O M^{-1 / 2} M^{1 / 2} M ^{1 / 2} M^{-1 / 2} O^T D^{1 / 2} 
\\&
=D^{1 / 2} O M^{-1 / 2} M^{1 / 2} M ^{1 / 2} M^{-1 / 2} O^{-1} D^{1 / 2} 
\\&
=D
\end{align}
for all $O \in O(2 N)$ (the set of $2N \times 2N$ orthogonal matrices), it should be noted that in the penultimate equation we utilized the fact that $M$ and $D$ are symmetric.\newline

To validate this construction, we need to guarantee that we can always find an orthogonal transformation $O$ such that the constructed form of the matrix $S$ is symplectic, utilizing \eqref{symplectic matrix} this means that
\begin{equation}
D^{1 / 2} O M^{-1 / 2} \Omega M^{-1 / 2} O^{\top} D^{1 / 2}=\Omega,
\end{equation}
but we will always be able to do so since the matrix
\begin{equation}\label{Omega prima}
    \Omega^{\prime}=M^{-1 / 2} \Omega M^{-1 / 2}
\end{equation}
is anti-symmetric and has full rank (its rank equals the largest possible for a matrix of the same dimensions) given that $\Omega$ and $M^{-1 / 2}$ have both full rank. With these conditions, for any $2 N \times$ $2 N$ real anti-symmetric matrix there exists an orthogonal transformation $O \in O(2 N)$ which puts it in a decoupled canonical form:
\begin{align}\label{evalues omega prime}
O \Omega^{\prime} O^{\top}&=\bigoplus_{j=1}^{N} \nu_{j}^{-1} \Omega_{1} \\&=\bigoplus_{j=1}^{N} \nu_{j}^{-1} \left(\begin{array}{cc}
0 & 1 \\
-1 & 0
\end{array}\right)
\end{align}
with $\nu_{j}$ real numbers, different from zero because $\Omega^{\prime}$ is full rank and strictly positive as result of $M$ being strictly positive. Therefore, if we now set $D=\operatorname{diag}\left(d_{1}, d_{1}, \ldots, d_{n}, d_{N}\right)$ we finish the proof since
\begin{align}
D^{1 / 2} O M^{-1 / 2} \Omega M^{-1 / 2} O^{\top} D^{1 / 2}&=D^{1 / 2} O \Omega^{\prime} O^{\top} D^{1 / 2}
\\&
=\bigoplus_{j=1}^{N} \nu_{j} \nu_{j}^{-1} \Omega_{1}
\\&
=\Omega.
\end{align}

\hfill$\square$\newline

What this proof makes evident is that to obtain our symplectic eigenvalues we must first construct the $\Omega'$ matrix, as stated in \eqref{evalues omega prime}, and the off diagonal entries will turn out to be the inverses of our symplectic eigenvalues \cite{Serafini}.\newline

To settle the operational ideas, let us look at the case of one degree of freedom, where our covariance matrix \eqref{qcov matrix} will be of the form

\begin{equation}
\sigma=\left[\begin{array}{cc}
\sigma_{qq} & \sigma_{qp} \\
\sigma_{qp} & \sigma_{pp}
\end{array}\right] 
\end{equation}
and its inverse is
\begin{equation}
\sigma ^{-1} =\left(\begin{array}{cc}
\frac{\sigma_{p p}}{-\sigma_{q p}^{2}+\sigma_{p p} \sigma_{q q}} & -\frac{\sigma_{q p}}{-\sigma_{q p}^{2}+\sigma_{p p} \sigma_{q q}} \\
-\frac{\sigma_{q p}}{-\sigma_{q p}^{2}+\sigma_{p p} \sigma_{q q}} & \frac{\sigma_{q q}}{-\sigma_{q p}^{2}+\sigma_{p p} \sigma_{q q}}
\end{array}\right)
\end{equation}
where we identify
\begin{equation}
    \operatorname{det}(\sigma) = -\sigma_{q p}^{2}+\sigma_{p p} \sigma_{q q}.
\end{equation}
Then, the square root of this inverse, i.e. $\sigma^{-1/2}$ has  entries:
\begin{equation}
{\sigma ^{-1/2}}_{xx} =\frac{\frac{\sigma_{p p}+\sqrt{4 \sigma_{qp}^{2}+\left(\sigma_{pp}-\sigma_{qq}\right)^{2}}-\sigma_{qq}}{\sqrt{\sigma_{p p}-\sqrt{4 \sigma_{qp}^{2}+\left(\sigma_{p p}-\sigma_{qq}\right)^{2}}+\sigma_{qq}}}+\frac{-\sigma_{pp}+\sqrt{4 \sigma_{qp}^{2}+\left(\sigma_{pp}-\sigma_{qq}\right)^{2}}+\sigma_{qq}}{\sqrt{\sigma_{pp}+\sqrt{4 \sigma_{qp}^{2}+\left(\sigma_{pp}-\sigma_{qq}\right)^{2}}+\sigma_{qq}}}}{\sqrt{2} \sqrt{4 \sigma_{q p}^{2}+\left(\sigma_{p p}-\sigma_{q q}\right)^{2}}}
\end{equation}

\begin{equation}
 \scalebox{1}{$
 {\sigma ^{-1/2}}_{xp}   =\frac{\sigma_{qp}\left(\sqrt{\sigma_{pp}-\sqrt{4 \sigma_{qp}^{2}\left(\sigma_{pp}-\sigma_{qq}\right)^{2}}+\sigma_{qq}}-\sqrt{\sigma_{\mathrm{pp}+} \sqrt{4 \sigma_{qp}^{2}+\left(\sigma_{pp}-\sigma_{qq}\right)^{2}}+\sigma_{qq}}\right)}{\sqrt{2} \sqrt{-\left(\left(4 \sigma_{qp}^{2}+\left(\sigma_{pp}-\sigma_{qq}\right)^{2}\right)\left(\sigma_{qp}^{2}-\sigma_{pp} \sigma_{qq}\right)\right)}}$
 }
\end{equation}

\begin{equation}
 {\sigma ^{-1/2}}_{pp}=   \frac{\frac{-\sigma_{p p}+\sqrt{4 \sigma_{q p}^{2}+\left(\sigma_{p p}-\sigma_{q q}\right)^{2}}+\sigma_{q q}}{\sqrt{\sigma_{p p}-\sqrt{4 \sigma_{q p}^{2}+\left(\sigma_{p p}-\sigma_{q q}\right)^{2}}+\sigma_{q q}}}+\frac{\sigma_{p p}+\sqrt{4 \sigma_{qp}^{2}+\left(\sigma_{p p}-\sigma_{q q}\right)^{2}}-\sigma_{q q}}{\sqrt{\sigma_{p p}+\sqrt{4 \sigma_{q p}^{2}+\left(\sigma_{p p}-\sigma_{q q}\right)^{2}}+\sigma_{q q}}}}{\sqrt{2} \sqrt{4 \sigma_{qp}^{2}+\left(\sigma_{p p}-\sigma_{qq}\right)^{2}}}.
\end{equation}
Now, with the matrix $\sigma^{-1/2}$ and \eqref{Omega prima} we construct

\begin{equation}
\Omega'=\left(\begin{array}{cc}
0 & \frac{1}{\sqrt{-\sigma_{qp}^{2}+\sigma_{pp} \sigma_{qq}}} \\
-\frac{1}{\sqrt{-\sigma_{qp}^{2}+\sigma_{pp} \sigma_{qq}}} & 0
\end{array}\right)
\end{equation}
which yields the symplectic eigenvalue of this particular case with one degree of freedom:
\begin{equation}\label{vn 1dof nu}
\nu = \sqrt{-\sigma_{qp}^{2}+\sigma_{pp}  \sigma_{qq}} = \sqrt{\operatorname{det}(\sigma)}
\end{equation}

In general, we can write the von Neumann entropy of a Gaussian state employing the symplectic eigenvalues $\nu_{k}$ of its correspondant quantum covariance matrix $\sigma_{(n)} / \hslash$ as: 

\begin{equation}\label{von Neumann gaussian}
S_V\left(a_{1}, a_{2}, \ldots, a_{n}\right)=\sum_{k=1}^{n} \mathcal{S}\left(\nu_{k}\right),
\end{equation}
with
\begin{equation}\label{vN symplectic}
\mathcal{S}\left(\nu_{k}\right)=\left(\nu_{k}+\frac{1}{2}\right) \ln \left(\nu_{k}+\frac{1}{2}\right)-\left(\nu_{k}-\frac{1}{2}\right) \ln \left(\nu_{k}-\frac{1}{2}\right),
\end{equation}
notice that $S\left(\nu_{k}\right)=0$ only if $\nu_{k}=\frac{1}{2}$ \cite{de Gosson}.\newline

With these tools at hand we can study several physical systems and the entanglement between their components.

\section{Examples: Stern-Gerlach, 2 qubits and the$\;$ \newline almighty oscillators}

\subsection{The Stern-Gerlach experiment}

This example will help us understand what are the properties of the density matrix and the meaning behind the concepts of mixed states and pure states, while also learning how to differentiate between them.\newline

If we consider a Stern-Gerlach experiment\footnote{For a beautiful explanation of this experiment consult \textit{Modern Quantum Mechanics} by J.J. Sakurai \cite{Sakurai}.}, prior to any measurement the silver atoms coming from the oven  do not have a definite spin orientation, implying that the two possible outcomes, positive or negative projections, are possible in any direction.\newline

We might be tempted to define the state of these atoms as
\begin{equation}
|\psi\rangle= \frac{1}{\sqrt{2}}|S_z, +\rangle + \frac{1}{\sqrt{2}}|S_z, -\rangle,
\end{equation}
or equivalently in terms of the density operator
\begin{equation}
\hat{\rho}=\left|\psi \rangle \langle \psi\right|=\frac{1}{2}\left|S_z, - \rangle \langle S_z, - \right|+\frac{1}{2}\left|S_z, - \rangle \langle S_z, +\right|+\frac{1}{2}\left|S_z, + \rangle \langle S_z, - \right|+\frac{1}{2} \left|S_z, +\rangle \langle S_z, +\right|
\end{equation}
where $\ket{S_z,+}$ represents the positive projection of the spin in the $Z$ axis and  $\ket{S_z,-}$ the negative one.  This can also be represented as a matrix if we take 
\begin{equation}
    \ket{S_z, +}= \begin{pmatrix}
1 \\
0 
\end{pmatrix}
\end{equation}
and
\begin{equation}
    \ket{S_z, -}= \begin{pmatrix}
0 \\
1 
\end{pmatrix}
\end{equation}
then we get
\begin{align}
    \hat{\rho}&= \unmed \left[ \begin{pmatrix}
0 \\
1 
\end{pmatrix}  \begin{pmatrix} 0 & 1 \end{pmatrix} +  \begin{pmatrix}
0 \\
1 
\end{pmatrix}  \begin{pmatrix} 1 & 0 \end{pmatrix} + \begin{pmatrix}
1 \\
0 
\end{pmatrix}  \begin{pmatrix} 0 & 1 \end{pmatrix} + \begin{pmatrix}
1 \\
0 
\end{pmatrix}  \begin{pmatrix} 1 & 0 \end{pmatrix}  \right]
\\& =\unmed \left[ \begin{pmatrix} 0 & 0 \\ 0 & 1\end{pmatrix}       + \begin{pmatrix} 0 & 0 \\ 1 & 0\end{pmatrix}  + \begin{pmatrix} 0 & 1 \\ 0 & 0\end{pmatrix}  + \begin{pmatrix} 0 & 0 \\ 0 & 1\end{pmatrix}        \right]
\\ & = \unmed \begin{pmatrix} 1 & 1 \\ 1 & 1\end{pmatrix}    .
\end{align}
However, this state does not represent the atoms coming out of the oven! These atoms truly have a $50\%$ chance of having either the spin up or down but only in the $Z$ direction. Notice that the spin state with the positive projection on the $X$  direction is defined exactly as we defined $\ket{\psi}$, i.e.
\begin{equation}
    \ket{S_x, +} =   \frac{1}{\sqrt{2}}|S_z, +\rangle + \frac{1}{\sqrt{2}}|S_z, -\rangle,
\end{equation}
while the one with the negative projection in $X$ is
\begin{equation}
    \ket{S_x, -} = -  \frac{1}{\sqrt{2}}|S_z, +\rangle + \frac{1}{\sqrt{2}}|S_z, -\rangle.
\end{equation}
 So if we were to measure the $\ket{\psi}$ state in the $X$ direction we would find that every atom comes out with the positive projection of spin in this direction. This can be shown using \eqref{expect val densop} and the \textit{Pauli matrices} \cite{Sakurai} which are
 \begin{equation}
\sigma_{1}=\sigma_{\mathrm{x}}=\left(\begin{array}{ll}
0 & 1 \\
1 & 0
\end{array}\right)
\end{equation}
\begin{equation}
\sigma_{2}=\sigma_{\mathrm{y}}=\left(\begin{array}{cc}
0 & -i \\
i & 0
\end{array}\right)
\end{equation}
 \begin{equation}
\sigma_{3}=\sigma_{\mathrm{z}}=\left(\begin{array}{cc}
1 & 0 \\
0 & -1
\end{array}\right).
\end{equation}

Starting with the $Z$ axis we find that
 \begin{equation}
\left\langle s_{z}\right\rangle=\operatorname{Tr}\left(\hat{\rho} s_{z}\right)=\frac{\hslash}{2} \operatorname{Tr}\left(\begin{array}{cc}
1 / 2 & 1 / 2 \\
1 / 2 & 1 / 2
\end{array}\right)\left(\begin{array}{cc}
1 & 0 \\
0 & -1
\end{array}\right)=0,
\end{equation}
 which is zero because we have a $50\%$ chance of getting either projection on the $Z$ axis. However, when we do the same for the $X$ axis the result is not the same,
 \begin{equation}
\left\langle s_{x}\right\rangle=\operatorname{Tr}\left(\hat{\rho} s_{x}\right)=\frac{\hslash}{2} \operatorname{Tr}\left(\begin{array}{cc}
1 / 2 & 1 / 2 \\
1 / 2 & 1 / 2
\end{array}\right)\left(\begin{array}{ll}
0 & 1 \\
1 & 0
\end{array}\right)=\frac{\hslash}{2},
\end{equation}
 meaning that we only have one option for the projection in this axis. Therefore this state does not truly represent the atoms coming out of the oven.\newline
 
 Before continuing we must realize that the $\ket{\psi}$ state is actually a pure state, since $\rho^2 = \rho$ and $\mu=\operatorname{Tr}(\rho^2) = 1$, since we were dealing a projector operator all along. \newline
 
 To truly encode all the properties of the silver atoms coming out of the oven we must use a mixed state, in the form of the density operator
 \begin{equation}
\hat{\rho}_{oven}=\frac{1}{2}|S_z, + \rangle\langle S_z, +|+\frac{1}{2}| S_z, - \rangle\langle S_z, -|,
\end{equation}
which can also be represented as
\begin{equation}
\hat{\rho}_{oven}=\frac{1}{2}\left(\begin{array}{l}
1 \\
0
\end{array}\right)\left(\begin{array}{ll}
1 & 0
\end{array}\right)+\frac{1}{2}\left(\begin{array}{l}
0 \\
1
\end{array}\right)\left(\begin{array}{ll}
0 & 1
\end{array}\right)=\frac{1}{2}\left(\begin{array}{ll}
1 & 0 \\
0 & 1
\end{array}\right).
\end{equation}
Notice that it is indeed mixed because $\hat{\rho}^2_{oven} = \frac{1}{4} \mathbb{I} \neq \hat{\rho}_{oven}$, and that $\mu_{oven}=1/2$.\newline

Now, every projection of spin in every direction has a $50\%$ chance of being measured:

\begin{equation}
\left\langle\boldsymbol{s}_{x}\right\rangle_{oven} = \left\langle\boldsymbol{s}_{y}\right\rangle_{oven} = \left\langle\boldsymbol{s}_{z}\right\rangle_{oven} = 0.
\end{equation}

In this sense, we encode the complete randomness of the spin within this mixed state.

\subsection{Two qubits system}

With the help of what we learned with the Stern-Gerlach, which is described with a single qubit, we can now analyze a more complex example where we consider a Hilbert space spanned by 4 possible states
$|00\rangle, \quad|01\rangle, \quad|10\rangle, \quad|11\rangle$,
where the first qubit refers to the subsystem $A$ and the second to the subsystem $B$, i.e.
\begin{equation}\label{2 qubits base}
|i j\rangle \equiv|i\rangle_{A}|j\rangle_{B} \equiv|i\rangle_{A} \otimes|j\rangle_{B}
\end{equation}
Now let us suppose that the system is in the pure state
\begin{equation}
|\psi\rangle=\frac{1}{\sqrt{2}}(|00\rangle+|11\rangle)
\end{equation}
so our density operator will be
\begin{align}
    \rho &\nonumber =\left| \psi \rangle \langle \psi\right|
    \\& = \unmed ( \ket{00}\bra{00} + \ket{00}\bra{11} + \ket{11}\bra{00} + \ket{11}\bra{11})
\end{align}
which can be represented as a $4 \mathrm{x} 4$ matrix following the same procedure as before, first we take
\begin{equation}
    \ket{0} = \begin{pmatrix} 0  \\ 1 \end{pmatrix},
\end{equation}
\begin{equation}
    \ket{1} = \begin{pmatrix} 1  \\ 0\end{pmatrix},
\end{equation}
then by following \eqref{2 qubits base} we get
\begin{equation}
    \ket{00} = \ket{0}_A \otimes \ket{0}_B = \begin{pmatrix} 0  \\ 1 \end{pmatrix}\otimes  \begin{pmatrix} 0  \\ 1 \end{pmatrix} =\begin{pmatrix} 0 \begin{pmatrix} 0  \\ 1 \end{pmatrix}  \\ 1 \begin{pmatrix} 0  \\ 1 \end{pmatrix} \end{pmatrix} 
    = \begin{pmatrix} 0 \\ 0 \\ 0 \\ 1 \end{pmatrix},
\end{equation}
\begin{equation}
    \ket{11} =  \ket{1}_A \otimes \ket{1}_B = \begin{pmatrix} 1  \\ 0 \end{pmatrix}\otimes  \begin{pmatrix} 1  \\ 0 \end{pmatrix} =\begin{pmatrix} 1 \begin{pmatrix} 1  \\ 0 \end{pmatrix}  \\ 0 \begin{pmatrix} 1  \\ 0 \end{pmatrix} \end{pmatrix}
    = \begin{pmatrix} 1 \\ 0 \\ 0 \\ 0 \end{pmatrix},
\end{equation}
\begin{equation}
    \ket{01} =  \ket{0}_A \otimes \ket{1}_B = \begin{pmatrix} 0  \\ 1 \end{pmatrix}\otimes  \begin{pmatrix} 1  \\ 0 \end{pmatrix} =\begin{pmatrix} 0 \begin{pmatrix} 1  \\ 0 \end{pmatrix}  \\ 1 \begin{pmatrix} 1  \\ 0 \end{pmatrix} \end{pmatrix} = \begin{pmatrix} 0 \\ 0 \\ 1 \\ 0 \end{pmatrix},
\end{equation}
\begin{equation}
    \ket{ 10} =  \ket{ 1}_A \otimes \ket{0}_B = \begin{pmatrix} 1  \\ 0 \end{pmatrix}\otimes  \begin{pmatrix} 0  \\ 1 \end{pmatrix} =\begin{pmatrix} 1 \begin{pmatrix} 0  \\ 1 \end{pmatrix}  \\ 0 \begin{pmatrix} 0  \\ 1 \end{pmatrix} \end{pmatrix} = \begin{pmatrix} 0 \\ 1 \\ 0 \\ 0 \end{pmatrix}.
\end{equation}
Is important to recognize that our pure state $\ket{\psi}$ is defined only with the states $\ket{00}$ and $\ket{11}$, and not with $\ket{01}$ nor $\ket{10}$.\newline

With these vectors we can now construct our density matrix:

\begin{align}
  \rho  = & \unmed \Bigg[ \begin{pmatrix} 0 \\ 0 \\ 0 \\ 1 \end{pmatrix} \begin{pmatrix} 0 & 0 & 0 & 1 \end{pmatrix} +  \begin{pmatrix} 0 \\ 0 \\ 0 \\ 1 \end{pmatrix} \begin{pmatrix} 1 & 0 & 0 & 0 \end{pmatrix} 
  \\& + \begin{pmatrix} 1 \\ 0 \\ 0 \\ 0 \end{pmatrix} \begin{pmatrix} 0 & 0 & 0 & 1 \end{pmatrix} + \begin{pmatrix} 1 \\ 0 \\ 0 \\ 0 \end{pmatrix} \begin{pmatrix} 1 & 0 & 0 & 0 \end{pmatrix} \Bigg]
  \\  = & \unmed \Bigg[ \begin{pmatrix} 0 & 0 & 0 & 0 \\ 0 & 0 & 0 & 0 \\ 0 & 0 & 0 & 0 \\ 0 & 0 & 0 & 1 \end{pmatrix}  + \begin{pmatrix} 0 & 0 & 0 & 0 \\ 0 & 0 & 0 & 0 \\ 0 & 0 & 0 & 0 \\ 1 & 0 & 0 & 0 \end{pmatrix}
  \\& + \begin{pmatrix} 0 & 0 & 0 & 1 \\ 0 & 0 & 0 & 0 \\ 0 & 0 & 0 & 0 \\ 0 & 0 & 0 & 0 \end{pmatrix} + \begin{pmatrix} 1 & 0 & 0 & 0 \\ 0 & 0 & 0 & 0 \\ 0 & 0 & 0 & 0 \\ 0 & 0 & 0 & 0 \end{pmatrix} \Bigg]
\end{align}
and therefore our density matrix in this representation is
\begin{equation}
    \rho = \unmed  \begin{pmatrix} 1 & 0 & 0 & 1 \\ 0 & 0 & 0 & 0 \\ 0 & 0 & 0 & 0 \\ 1 & 0 & 0 & 1 \end{pmatrix}.
\end{equation}
It turns out that for this particular case $\rho = \rho^2$ and from here it is clear that we are dealing with a pure state since the purity 
\begin{equation}
     \mu(\rho)=\operatorname{tr}\rho^2 = \operatorname{tr}  \begin{pmatrix} \unmed & 0 & 0 & \unmed \\ 0 & 0 & 0 & 0 \\ 0 & 0 & 0 & 0 \\ \unmed & 0 & 0 & \unmed \end{pmatrix} = \unmed + \unmed = 1.
\end{equation}

To study the susbsystem $A$ independently, we must first obtain its reduced density metric utilizing the partial trace over the subsystem $B$:
\begin{align}
\rho_{A}=& \operatorname{tr}_{B} \rho 
\\=&\nonumber
\frac{1}{2}\quad_{B}\langle 0|\Big( \ket{00}\bra{00} + \ket{00}\bra{11} + \ket{11}\bra{00} + \ket{11}\bra{11}\Big) |0\rangle_{B}
\\ &
+\frac{1}{2}\quad_{B}\langle 1|  \Big( \ket{00}\bra{00} + \ket{00}\bra{11} + \ket{11}\bra{00} + \ket{11}\bra{11}\Big) |1\rangle_{B}
\\=&
\frac{1}{2}\quad_{B}\langle 0|\Big( \ket{00}\bra{00}\Big) |0\rangle_{B}
+\frac{1}{2}\quad_{B}\langle 1|  \Big(\ket{11}\bra{11}\Big) |1\rangle_{B}
\\=& \frac{1}{2}\left(|0\rangle_{A A}\langle 0|+| 1\rangle_{A A}\langle 1|\right) 
\\ = & \unmed \mathbb{I}_{2 \times 2} .
\end{align}

The partial trace can be easily understood in braket notation, but it is a little bit more complicated in terms of matrices, for a general two qubit system we will have
\begin{align}
\rho_A &=\operatorname{tr}_{B}\left(\begin{array}{llll}
\rho_{11} & \rho_{12} & \rho_{13} & \rho_{14} \\
\rho_{21} & \rho_{22} & \rho_{23} & \rho_{24} \\
\rho_{31} & \rho_{32} & \rho_{33} & \rho_{34} \\
\rho_{41} & \rho_{42} & \rho_{43} & \rho_{44}
\end{array}\right)
\\&=
\left(\begin{array}{ll}
\operatorname{tr}\left(\begin{array}{ll}
\rho_{11} & \rho_{12} \\
\rho_{21} & \rho_{22}
\end{array}\right) & \operatorname{tr}\left(\begin{array}{ll}
\rho_{13} & \rho_{14} \\
\rho_{23} & \rho_{24}
\end{array}\right) \\
\operatorname{tr}\left(\begin{array}{ll}
\rho_{31} & \rho_{32} \\
\rho_{41} & \rho_{42}
\end{array}\right) & \operatorname{tr}\left(\begin{array}{ll}
\rho_{33} & \rho_{34} \\
\rho_{43} & \rho_{44}
\end{array}\right)
\end{array}\right)
\\&
=\left(\begin{array}{ll}
\rho_{11}+\rho_{22} & \rho_{13}+\rho_{24} \\
\rho_{31}+\rho_{42} & \rho_{33}+\rho_{44}
\end{array}\right)
\end{align}
and if we wanted to study the subsystem $B$, we would do the partial trace over $A$, which is
\begin{align}
\rho_B &=\operatorname{tr}_{A}\left(\begin{array}{llll}
\rho_{11} & \rho_{12} & \rho_{13} & \rho_{14} \\
\rho_{21} & \rho_{22} & \rho_{23} & \rho_{24} \\
\rho_{31} & \rho_{32} & \rho_{33} & \rho_{34} \\
\rho_{41} & \rho_{42} & \rho_{43} & \rho_{44}
\end{array}\right)
\\&=
\left(\begin{array}{ll}
\operatorname{tr}\left(\begin{array}{ll}
\rho_{11} & \rho_{13} \\
\rho_{31} & \rho_{33}
\end{array}\right) & \operatorname{tr}\left(\begin{array}{ll}
\rho_{12} & \rho_{14} \\
\rho_{32} & \rho_{34}
\end{array}\right) \\
\operatorname{tr}\left(\begin{array}{ll}
\rho_{21} & \rho_{23} \\
\rho_{41} & \rho_{43}
\end{array}\right) & \operatorname{tr}\left(\begin{array}{ll}
\rho_{22} & \rho_{24} \\
\rho_{42} & \rho_{44}
\end{array}\right)
\end{array}\right)
\\&
=\left(\begin{array}{ll}
\rho_{11}+\rho_{33} & \rho_{12}+\rho_{34} \\
\rho_{21}+\rho_{43} & \rho_{22}+\rho_{44}
\end{array}\right).
\end{align}
For our particular case, we have
\begin{align}
\rho_A &=\operatorname{tr}_{B} \unmed  \begin{pmatrix} 1 & 0 & 0 & 1 \\ 0 & 0 & 0 & 0 \\ 0 & 0 & 0 & 0 \\ 1 & 0 & 0 & 1 \end{pmatrix}
\\&= \unmed
\left(\begin{array}{ll}
\operatorname{tr}\left(\begin{array}{ll}
1 & 0 \\
0 & 0
\end{array}\right) & \operatorname{tr}\left(\begin{array}{ll}
0 & 1 \\
0 & 0
\end{array}\right) \\
\operatorname{tr}\left(\begin{array}{ll}
0 & 0 \\
1 & 0
\end{array}\right) & \operatorname{tr}\left(\begin{array}{ll}
0 & 0 \\
0 & 1
\end{array}\right)
\end{array}\right)
\\&
=\unmed\left(\begin{array}{ll}
1 & 0 \\
0 & 1
\end{array}\right),
\end{align}
which is exactly the same result that we obtained with the braket notation.\newline

Now, with this result we can calculate the von Neumann entropy of subsystem $A$, which is
\begin{equation}
\begin{aligned}
S_{A} &=-\operatorname{tr} \rho_{A} \log \rho_{A} \\
&=-2 \times \frac{1}{4} \log \frac{1}{4} \\
&=\log 2
\end{aligned}
\end{equation}
Since $\rho_{A}$ is proportional to the identity matrix of a 2-state system, it means that $\rho_{A}$ is maximally mixed, and that the initial state $|\psi\rangle$ is maximally entangled.\newline

From this particular example we learn what the entanglement entropy actually quantifies, it counts the number of entangled qubits between the subsystems $A$ and $B$. If we had $k$ qubits in each subystem $A$ and $k$ qubits in subystem $B$, then in a maximally entangled state our entropy would be $S_{A}=k \log 2$.  In this case we only have two qubits, one in each subsystem, so the result is only $S_{A} =\log 2$. This could also be understood in terms of states because $k$ qubits have $2^{k}$ states, so $e^{S_{A}}$ would count for us the number of entangled states \cite{Lectures Quantum Gravity}.

\subsection{Two coupled harmonic oscillators}

Let us now turn our attention to systems constructed with harmonic oscillators, which have been used to model circuit complexity within quantum field theories \cite{Simetricos Jefferson}, describe solid state physics, and even study the entropy of black holes utilizing the now familiar concepts of von Neumann entropy and reduced density matrices \cite{Srednicki, Bombelli}. First, we will study the simplest system of this type, consisting of just two coupled harmonic oscillators, and then we will generalize what we learn from it so we can solve the much harder system of N coupled harmonic oscillators.\newline

This quantum system is defined by the Hamiltonian

\begin{equation}
H=\frac{1}{2}\left[p_{1}^{2}+p_{2}^{2}+k_{0}\left(q_{1}^{2}+q_{2}^{2}\right)+k_{1}\left(q_{1}-q_{2}\right)^{2}\right]
\end{equation}
and to find its wave function we must first "decouple" the oscillators via the \textit{canonical transformation}:

\begin{equation}
q_{+}=\frac{1}{\sqrt{2}}\left(q_{1}+q_{2}\right),
\end{equation}
\begin{equation}
q_{-}=\frac{1}{\sqrt{2}}\left(q_{1}-q_{2}\right),
\end{equation}
\begin{equation}
p_{+}=\frac{1}{\sqrt{2}}\left(p_{1}+p_{2}\right),
\end{equation}
\begin{equation}
p_{-}=\frac{1}{\sqrt{2}}\left(p_{1}-p_{2}\right),
\end{equation}
which leaves us the transformed Hamiltonian

\begin{equation}
H=H_+ + H_-=\frac{1}{2}\left(p_{+}^{2}+\omega_{+}^{2} q_{+}^{2}+p_{-}^{2} +\omega_{-}^{2} q_{-}^{2}\right)
\end{equation}
with \begin{equation}
\omega_{+}=k_{0}^{1 / 2}
\end{equation}
and
\begin{equation}
\omega_-=\left(k_{0}+2 k_{1}\right)^{1 / 2}.
\end{equation}
This version of the Hamiltonian can be solved analytically since the Schrödinger equation now reads
\begin{equation}
    H\ket{\psi} = (H_+ + H_-)\ket{\psi} = (E_+ + E_-) \ket{\psi},
\end{equation}
so the solution of our general problem will be the product of the solutions of each separate subsystem, in terms of $q_+$ and $q_-$ this is

\begin{align}
\psi_{n,m}(q_+,q_-)&\label{2 psi OASA}=\Big( \frac{\omega_+}{\hslash}\Big)^{1/4}\chi_n\Big(q_+\sqrt{\frac{\omega_+}{\hslash}}\Big)\Big(\frac{\omega_-}{\hslash}\Big)^{1/4}\chi_m\Big(q_-\sqrt{\frac{\omega_-}{\hslash}}\Big),
\end{align}
where $\chi_n(x)$ is the \textit{Hermite function} defined by
\begin{equation}
\chi_{n}(x)=\frac{e^{-x^{2} / 2}}{\sqrt{2^{n} n ! \sqrt{\pi}}} H_{n}(x)
\end{equation}
with $H_{n}(x)$ the Hermite polynomials. \newline

If we write the wave function utilizing our original parameters and variables it takes the form
\begin{equation}
\psi_{n,m}=\Big( \frac{k_0^2+2k_0k_1}{\hslash^4}\Big)^{1/8}\chi_n\Big((q_1+q_2)\Big[\frac{k_0}{(2\hslash)^2}\Big]^{\frac{1}{4}}\Big)\chi_m\Big((q_1-q_2)\Big[\frac{k_0+2k_1}{(2\hslash)^2}\Big]^{\frac{1}{4}}\Big)
\end{equation}
and the ground state is then described by
\begin{equation}
\psi_{0}\left(q_{+}, q_{-}\right)=\frac{\left(\omega_+\omega_{-}\right)^{1 / 4} }{(\pi \hslash)^{1/2}}\exp \left[-\frac{1}{2 \hslash}\left(\omega_{+} q^{2}_+ +\omega_-q^{2}_-\right) \right],
\end{equation}
or equivalently

\begin{equation}\label{Jefferson base x1x2}
  \psi_{0}\left(q_{1}, q_{2}\right)=\frac{\left(\omega_+\omega_{-}\right)^{1 / 4} }{(\pi \hslash)^{1/2}} \exp \left[-\frac{1}{4\hslash}\left(\omega_{+} \left(q_{1}+q_{2}\right)^{2} +\omega_-\left(q_{1}-q_{2}\right)^{2}\right) \right] ,
\end{equation}
which is a Gaussian state, so we have two possible paths to calculate the entropy of this system, via the wave function or the quantum covariance matrix. We will do both.

\subsubsection{Using the wave function}

Since
\begin{equation}
    \psi(q_1,q_2)= \braket{q_1,q_2}{\psi}
\end{equation}
we can construct the density matrix of this state in the position basis as
\begin{align}
\rho_0 &= \braket{q_1,q_2}{\psi_0}\braket{\psi_0}{q_1^\prime, q_2^\prime}
\\& = \psi_0(q_1,q_2) ( \psi_0(q_1^\prime,q_2^\prime) )^* = \psi_0(q_1,q_2)  \psi_0(q_1^\prime,q_2^\prime) 
\\&= \frac{\left(\omega_+\omega_{-}\right)^{\unmed}}{\pi\hslash}\exp \left[-\frac{1}{4\hslash}\left(\omega_{+}[\left(q_{1}+q_{2}\right)^{2} + \left(q^\prime_{1}+q^\prime_{2}\right)^{2}] +\omega_-[\left(q_{1}-q_{2}\right)^{2} + \left(q^\prime_{1}-q^\prime_{2}\right)^{2}]\right) \right]  
\end{align}
which given the normalization of the wave function it follows that the state is pure, i.e. $\mu(\rho_0)=\operatorname{Tr}(\rho^2)=1$.\newline

\begin{figure}[h!]
    \centering
    \includegraphics[width=0.75\textwidth]{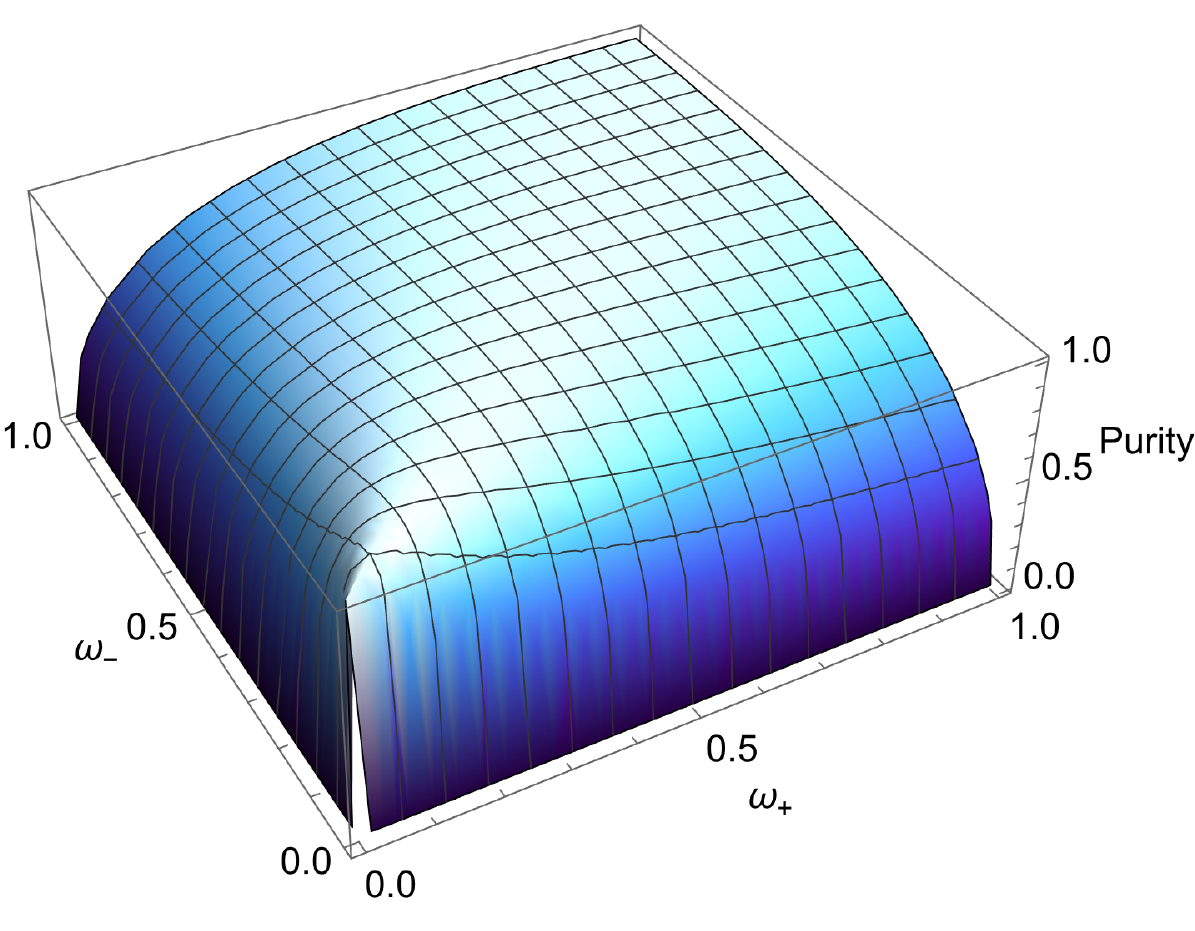}
    \caption{3D plot of the purity of a reduced density matrix of any of the oscillators with respect to $\omega_+$ and $\omega_-$. }
    \label{purity3d}
\end{figure}

\begin{figure}[h!]
    \centering
    \includegraphics[width=0.75\textwidth]{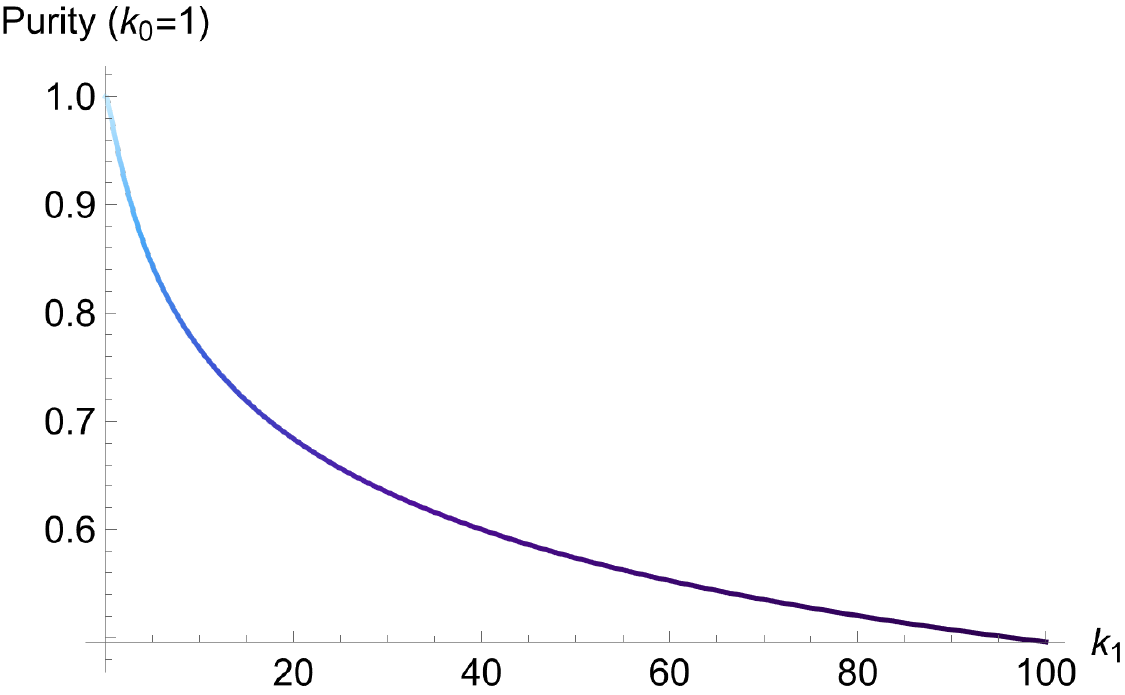}
    \caption{2D plot of the purity of a reduced density matrix of any of the oscillators with respect to $k_1$ while setting $k_0 = 1$. }
    \label{purity2d}
\end{figure}

To study the entanglement between the particle $1$ and $2$ we must first obtain the reduced density matrices for each subsystem, meaning that in order to obtain the reduced density matrix of the particle $1$ we must do trace with respect to $2$, and vice versa for the reduced density matrix of $2$. In terms of these continuous variables it means to integrate (the sum of the trace) with respect of the same variable in both wave functions (over the diagonal, same indices). For example, the reduced density matrix of $2$ is 

\begin{align}
\rho_{2}\left(q_{2}, q_{2}^{\prime}\right)=&\int_{-\infty}^{+\infty} d q_{1} \psi_{0}\left(q_{1}, q_{2}\right) \psi_{0}^{*}\left(q_{1}, q_{2}^{\prime}\right)
\\=&
\int_{-\infty}^{+\infty} d q_{1}  \frac{\left(\omega_+\omega_{-}\right)^{1 / 2}}{\pi\hslash}\exp \Big[-\frac{1}{4\hslash}\Big(\omega_{+}[\left(q_{1}+q_{2}\right)^{2} + \left(q_{1}+q^\prime_{2}\right)^{2}] \\&+\omega_-[\left(q_{1}-q_{2}\right)^{2} + \left(q_{1}-q^\prime_{2}\right)^{2}]\Big) \Big] 
\\=&\label{reducida rho2 gammabeta}
\left( \frac{ 2 \omega_{+}\omega_{-} }{\pi \hslash \left(\omega_{+}+\omega_{-}\right)}\right)^{1 / 2} \exp \left[-\frac{\gamma}{ 2} \left(q_{2}^{2}+q_{2}^{\prime 2}\right)    +\beta q_{2} q_{2}^{\prime}\right],
\end{align}
with 
\begin{equation}\label{mdens red beta}
    \beta=\frac{\left(\omega_{+}-\omega_{-}\right)^{2}}{4\hslash \left(\omega_{+}+\omega_{-}\right)} 
\end{equation}
and
\begin{equation}\label{mdens red gamma}
    \gamma= \frac{ 2 \omega_{+}\omega_{-} }{\hslash\left(\omega_{+}+\omega_{-}\right)} + \beta,
\end{equation}
or in terms of our original parameters 
\begin{align}
    \rho_2(q_2,q_2^\prime)=&\nonumber \left(\frac{2 \sqrt{k_0 (k_0+2 k_1)}}{ (\pi  \hslash)    \sqrt{k_0} + \sqrt{k_0+2 k_1}}\right)^{1/2}\times
    \\&\exp\left[-\frac{\sqrt{k_0 (k_0+2 k_1)} \left(3 {q^\prime_{2}}^2+2 q^\prime_{2} q^\prime_{2}+3 {q^\prime_{2}}^2\right)+k_0 (q^\prime_{2}-q^\prime_{2})^2+k_1 (q^\prime_{2}-q^\prime_{2})^2}{4 \hslash  \left(\sqrt{k_0+2 k_1}+\sqrt{k_0}\right)}\right]
\end{align}
and this reduced density matrix has purity
\begin{equation}\label{Jefferson purity densmat}
    \mu_2(\rho_2)= \frac{2 (k_0 (k_0+2 k_1))^{1/4}}{\sqrt{k_0+2 k_1}+\sqrt{k_0}} =
    \frac{2 \sqrt{\omega_+ \omega_-}}{\omega_+ + \omega_-},
\end{equation}
which is plotted on Figure \ref{purity3d}  \footnote{Notice that not every point is permitted on the graph, since $\omega_{+}=k_{0}^{1 / 2}$ and $\omega_-=\left(k_{0}+2 k_{1}\right)^{1 / 2}$ so if $w_+=1$ then $w_-$ must be at least $1$ otherwise we are allowing $k_1$ to get negative values} and can be understood in terms of the coupling constant with the help of Figure  \ref{purity2d}, notice how the purity is decreases as the coupling constant increases its strength.

 Now, we need the eigenvalues $p_{n}$ of $\rho_{2}\left(q_2, q_2^{\prime}\right)$ :
\begin{equation}\label{eigenvalues entropy}
\int_{-\infty}^{+\infty} d q^{\prime}_2 \rho_{2}\left(q_2, q_2^{\prime}\right) f_{n}\left(q_2^{\prime}\right)=p_{n} f_{n}(q_2)
\end{equation}
since we can construct the entropy in terms of them as $S=-\sum_{n} p_{n} \ln p_{n}$. The solution of \eqref{eigenvalues entropy} is found by noticing that once we carry out the integral with respect to $q_2^\prime$, we will be left with a function solely in terms of $q_2$; so if we multiply \eqref{eigenvalues entropy} by  $f_{n}^{-1}(q_2)$, on the right hand side only remains our eigenvalue, and on the left side this factor would need to cancel out every term dependent of $q_2$, including the one emergent from the integral, in order for our eigenvalue to be independent of both $q_2$ and $q_2^\prime$.\newline

Taking this into consideration and given that our system is composed by harmonic oscillators, the function that is natural to generate the factors needed will be the Hermite polynomials multiplied by an exponential to cancel out the term $\exp [- \frac{\gamma}{2} q_2^2]$ that can be extracted from the integral. Therefore we propose
\begin{equation}
f_{n}(q)=H_{n}\left(\alpha^{a} x\right) \exp \left(-\frac{\alpha q^{2}} { 2}\right)
\end{equation}
where we need to find the exact values of $\alpha$ and $a$.\newline

To do so, we begin by analyzing the case of $n=0$, here
\begin{equation}
\int_{-\infty}^{+\infty} d q^{\prime}_2 \left( \frac{ 2 \omega_{+}\omega_{-} }{\pi \hslash \left(\omega_{+}+\omega_{-}\right)}\right)^{1 / 2} \exp \left[-\frac{\gamma}{ 2} \left(q_{2}^{2}+q_{2}^{\prime 2}\right)    +\beta q_{2} q_{2}^{\prime}\right] f_{0}\left(q_2^{\prime}\right)=p_{0} f_{0}(q_2),
\end{equation}
and once we extract all the terms independent of $q_2^\prime$, the integral to solve will be
\begin{equation}
    \int_{-\infty }^{\infty } \exp [\beta  (q_2 q_2^{\prime})] \exp \left[\frac{1}{2} \left(-{q_2^\prime}^2\right) (\alpha +\gamma )\right] \, dq_2^{\prime} = \frac{\sqrt{2 \pi }}{\sqrt{\alpha +\gamma }}  e^{\frac{\beta ^2 q_2^2}{2 (\alpha +\gamma )}},
\end{equation}
then \eqref{eigenvalues entropy} for $n=0$ will yield the eigenvalue
\begin{equation}
p_{0} =  \frac{\sqrt{2 \pi } }{\sqrt{\alpha +\gamma }} \exp \left[\frac{1}{2} (-\gamma ) q_2^2\right] \exp \left[\frac{\beta ^2 q_2^2}{2 (\alpha +\gamma )}\right]  \exp \left[\frac{\alpha  q_2^2}{2}\right].
\end{equation}
Since we want it to be a constant, we need the argument of the exponential to be zero, i.e.
\begin{equation}
    \frac{\alpha ^2+\beta ^2-\gamma ^2}{2 (\alpha +\gamma )}=0,
\end{equation}
from this equation we find that our $\alpha$ must be
\begin{equation}
\alpha=\frac{\left(\omega_{+} \omega_{-}\right)^{1 / 2}}{\hslash}=\left(\gamma^{2}-\beta^{2}\right)^{1 / 2},
\end{equation}
and our eigenvalue for this case is then
\begin{equation}
p_{0}=(1-\xi)  
\end{equation}
with
\begin{equation}\xi= \frac{(\omega_+ -\omega_- )^2}{\omega_+ ^2+6 \omega_+  \omega_- +4 \omega_+  \sqrt{\omega_+  \omega_- }+4 \omega_-  \sqrt{\omega_+  \omega_- }+\omega_- ^2}=\frac{\beta}{\gamma+\alpha}.  \end{equation} 

Solving for $n=1$ we find that $a=1/2$ and with it our general eigenfunction turns our to be
\begin{equation}
f_{n}(x)=H_{n}\left( \sqrt{\alpha} q\right) \exp \left(-\frac{\alpha q^{2}} { 2}\right)
\end{equation}
that produces the $n$-th eigenvalue
\begin{equation}
p_{n}=(1-\xi)\xi^n. 
\end{equation}

\begin{figure}[h!]
    \centering
    \includegraphics[width=0.75\textwidth]{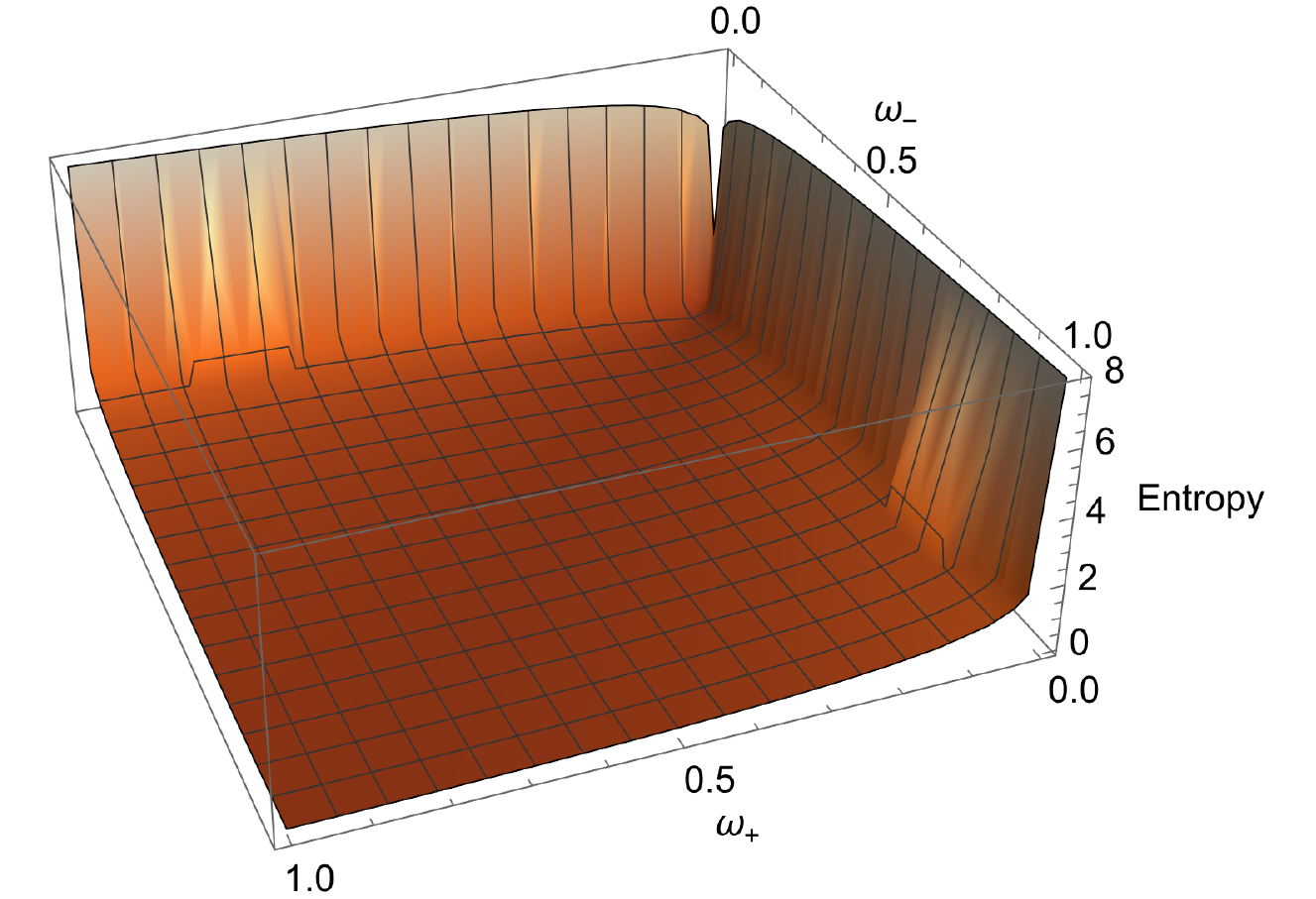}
    \caption{3D plot of the entropy of a reduced density matrix of any of the oscillators with respect to $\omega_+$ and $\omega_-$. }
    \label{entropy3d}
\end{figure}

\begin{figure}[h!]
    \centering
    \includegraphics[width=0.75\textwidth]{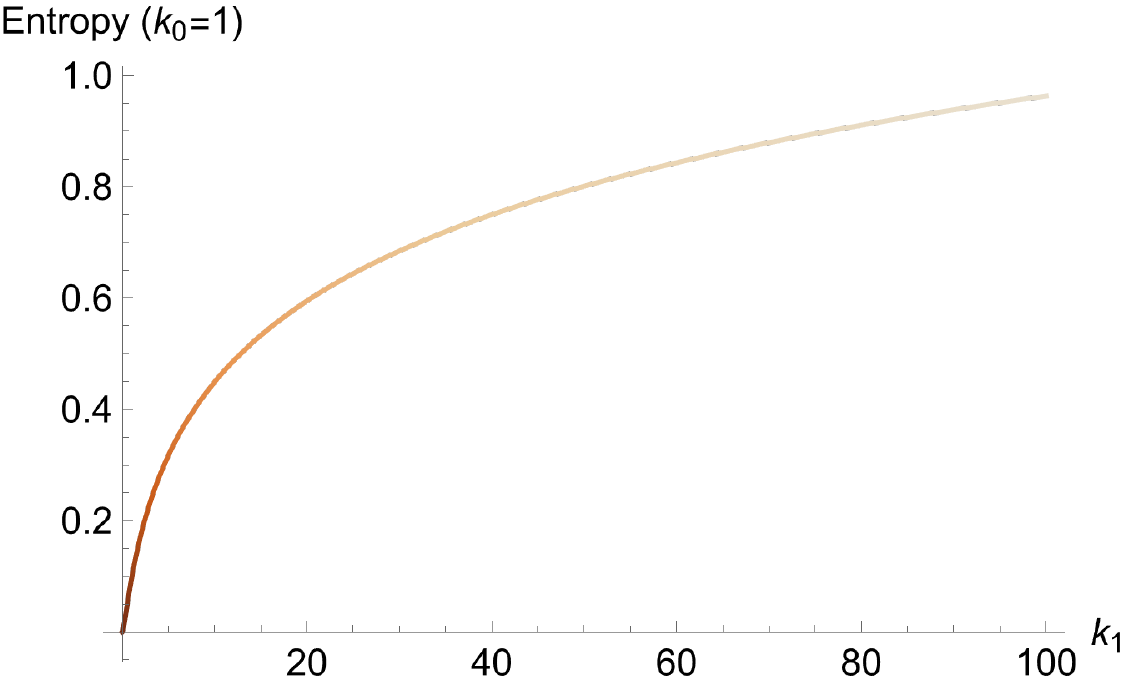}
    \caption{2D plot of the entropy of a reduced density matrix of any of the oscillators with respect to $k_1$ while setting $k_0 = 1$. }
    \label{entropy2d}
\end{figure}

Then, in terms of these eigenvalues the entropy is
\begin{align}
S &=-\sum_{n=0}^{\infty} p_{n} \ln p_{n} \\
&=-\sum_{n=0}^{\infty}\left(1-\xi\right) \xi^{n} \ln \left[\left(1-\xi\right) \xi^{n}\right] \\
&=-\left(1-\xi\right) \sum_{n=0}^{\infty} \xi^{n}\left[\ln \left(1-\xi\right)+\ln \xi^{n}\right] \\
&\label{entropia eigenvalores suma}=-\left(1-\xi\right)\left[\ln \left(1-\xi\right) \sum_{n=0}^{\infty} \xi^{n}+\ln \xi \sum_{n=0}^{\infty} n \xi^{n}\right].
\end{align}
Since $0<\xi<1$ each of these sums can be expressed as

\begin{equation}
\sum_{n=0}^{\infty} \xi^{n} =\frac{1}{1-\xi},
\end{equation}
\begin{equation}
\sum_{n=0}^{\infty} n \xi^{\mu_{i}} =\frac{\xi}{\left(1-\xi\right)^{2}},
\end{equation}
and substituting these results in \eqref{entropia eigenvalores suma}
\begin{align}
S&=-\left(1-\xi\right)\left[\ln \left(1-\xi\right) \frac{1}{1-\xi}+\ln \xi \frac{\xi}{\left(1-\xi\right)^{2}}\right],
\end{align}
 we get our final expression for the entropy of the system:
\begin{equation}\label{Jefferson entropy densmat}
S(\xi)=-\ln (1-\xi)-\frac{\xi}{1-\xi} \ln \xi,
\end{equation}
Which can be understood with the help of Figures \ref{entropy3d} and \ref{entropy2d}, where we recognize that the entropy only depends on the proportion between our initial parameters $k_0$ and $k_1$, meaning that the coupling strength among the oscillators directly affects the entropy of the subsystems.


It should be noted that given the symmetry of $q_1$ and $q_2$ in our Hamiltonian, the results for $\rho_2$ are the same as those of $\rho_1$ interchanging $q_2 \leftrightarrow q_1$.\newline

\subsubsection{Using the quantum covariance matrix}

We will now see how to arrive to the same results using the quantum covariance matrix defined by \eqref{qcov matrix}, which given that our state is Gaussian, contains all the relevant information of the purity in \eqref{purity gaussian}, and entropy in \eqref{von Neumann gaussian} and \eqref{vN symplectic}. We begin by obtaining the quantum covariance matrix of our system, keeping in mind that we are working with the ground state:

\begin{equation}
    \sigma=\begin{pmatrix}
\sigma_{q_1 q_1} & \sigma_{q_2 q_1} & \sigma_{p_1 q_1} & \sigma_{p_2 q_1} \\
\sigma_{q_2 q_1} & \sigma_{q_2 q_2} & \sigma_{p_1 q_2} & \sigma_{p_2 q_2} \\
\sigma_{p_1 q_1} & \sigma_{p_1 q_2} & \sigma_{p_1 p_1} & \sigma_{p_2 p_1} \\
\sigma_{p_2 q_1} & \sigma_{p_2 q_2} & \sigma_{p_2 p_1} & \sigma_{p_2 p_2} 
\end{pmatrix},
\end{equation}
where every entry is given by the expected values of \eqref{qcov matrix}, so all the \underline{non null} integrals that we need in order to build the quantum covariance matrix are:

\begin{align}
    \langle q_1^2 \rangle &= \int _{-\infty }^{\infty }\int _{-\infty }^{\infty }\psi_0 q_1^2 \psi_0 dq_1 dq_2 
    \\&= \frac{ \hslash }{4} \left(\frac{1}{\sqrt{k_0}}+\frac{1}{\sqrt{k_0+2 k_1}}\right) = \frac{ \hslash }{4} \left(\frac{1}{\omega_+}+\frac{1}{\omega_-}\right) ,
\end{align}

\begin{align}
    \langle q_2^2 \rangle &= \int _{-\infty }^{\infty }\int _{-\infty }^{\infty }\psi_0 q_2^2 \psi_0 dq_1 dq_2 
    \\&=  \frac{ \hslash }{4} \left(\frac{1}{\sqrt{k_0}}+\frac{1}{\sqrt{k_0+2 k_1}}\right)= \frac{ \hslash }{4} \left(\frac{1}{\omega_+}+\frac{1}{\omega_-}\right) ,
\end{align}

\begin{align}
    \langle p_1^2 \rangle =& -\hslash^2 \int _{-\infty }^{\infty }\int _{-\infty }^{\infty }\psi_0  \frac{\partial^2 \psi_0}{\partial q_1^2}  dq_1 dq_2 
    \\&= \frac{\hslash}{4} (\sqrt{k_0} + \sqrt{k_0 + 2 k_1})  = \frac{\hslash}{4} (\omega_+ + \omega_-) ,
\end{align}

\begin{align}
    \langle p_2^2 \rangle =& -\hslash^2 \int _{-\infty }^{\infty }\int _{-\infty }^{\infty }\psi_0  \frac{\partial^2 \psi_0}{\partial q_2^2}  dq_1 dq_2 
    \\&= \frac{\hslash}{4} (\sqrt{k_0} + \sqrt{k_0 + 2 k_1})  = \frac{\hslash}{4} (\omega_+ + \omega_-) ,
\end{align}

\begin{align}
    \unmed \langle q_1 q_2 + q_2 q_1 \rangle &= \int _{-\infty }^{\infty }\int _{-\infty }^{\infty }\psi_0 q_1 q_2 \psi_0 dq_1 dq_2 
    \\&= \frac{\hslash }{4}  \left(\frac{1}{\sqrt{k_0}}- \frac{1}{\sqrt{k_0+2 k_1}}\right)= \frac{\hslash }{4}  \left(\frac{1}{\omega_+}-\frac{1}{\omega_-}\right),
\end{align}

\begin{align}
    \unmed\langle p_1 p_2 + p_2 p_1 \rangle &=  \int _{-\infty }^{\infty }\int _{-\infty }^{\infty }\psi_0 \frac{\partial^2 \psi_0}{\partial q_1 \partial q_2}  dq_1 dq_2 
    \\&= \frac{\hslash}{4} (\sqrt{k_0} - \sqrt{k_0 + 2 k_1})  = \frac{\hslash}{4} (\omega_+ - \omega_-) ,
\end{align}
with $\psi_0$ given by \eqref{Jefferson base x1x2}. Therefore, the quantum covariance matrix takes the form
\begin{align}
    \sigma&=
\frac{\hslash}{4}\begin{pmatrix}
\left(\frac{1}{\omega_+}+\frac{1}{\omega_-}\right) & \left(\frac{1}{\omega_+}-\frac{1}{\omega_-}\right) & 0 & 0 \\
\left(\frac{1}{\omega_+}-\frac{1}{\omega_-}\right)& \left(\frac{1}{\omega_+}+\frac{1}{\omega_-}\right) & 0 & 0 \\
0 & 0 & (\omega_+ + \omega_-)  & (\omega_+ - \omega_-)  \\
0 & 0 & (\omega_+ - \omega_-)  & (\omega_+ + \omega_-)  
\end{pmatrix}
\end{align}
which has a determinant $\operatorname{det}(\sigma) = \frac{\hslash^4}{16}$.\newline

In order to calculate both the purity and entropy of the reduced subsystems, we must first follow \eqref{qcov matrix} once again to produce the reduced quantum covariance matrices for each of our oscillators:

\begin{equation}
    \sigma_1 = \sigma_2 =
\frac{\hslash}{4}\begin{pmatrix}
\left(\frac{1}{\omega_+}+\frac{1}{\omega_-}\right) &   0 \\
0 & (\omega_+ + \omega_-),  
\end{pmatrix}
\end{equation}
in this particular case they turned out to be the same given the symmetry between them in our Hamiltonian, so any result that we obtain for one, applies for the other. Now, the determinant of this matrix is
\begin{equation}
\operatorname{det}(\sigma_{1})=\operatorname{det}(\sigma_{2})=\sqrt{\left(\frac{\hslash}{4}\right)^{2}\left(\frac{1}{\omega_{+}}+\frac{1}{\omega_{-}}\right)\left(\omega_{+}+\omega_{-}\right)}=\frac{\hslash}{4} \frac{\omega_{+}+\omega_{-}}{\sqrt{\omega_{+} \omega_{-}}},
\end{equation}
and with it we can use \eqref{purity gaussian} to obtain the purity of our reduced subsystems, which yields
\begin{equation}
\mu(1)=\mu(2)=\frac{\hslash}{2} \frac{1}{\sqrt{\frac{\hslash^{2}}{16}\left(\omega_{+}+\omega_{-}\right)\left(\frac{1}{\omega_{+}}+\frac{1}{\omega_{-}}\right)}}=\frac{2 \sqrt{\omega_{+} \omega_{-}}}{\omega_{+}+\omega_{-}},
\end{equation}
and it is exactly what we got in \eqref{Jefferson purity densmat} using the density matrix.\newline

Finally for the entropy, we calculate the symplectic eigenvalue of either $\sigma_1 /\hslash$ or $\sigma_2 /\hslash $, with the help of \eqref{Omega prima}, which turns out to be
\begin{equation}
    \nu= \frac{\omega_+ + \omega_-}{4 \sqrt{\omega_+ \omega_- }}
\end{equation}
and using \eqref{vN symplectic} we get the entropy
\begin{align}
S_1\left(\nu\right)=S_2\left(\nu\right)= &\left(\left(\frac{\omega_+ + \omega_-}{4 \sqrt{\omega_+ \omega_- }}\right)+\frac{1}{2}\right) \ln \left(\left(\frac{\omega_+ + \omega_-}{4 \sqrt{\omega_+ \omega_- }}\right)+\frac{1}{2}\right)
\\&-\left(\left(\frac{\omega_+ + \omega_-}{4 \sqrt{\omega_+ \omega_- }}\right)-\frac{1}{2}\right) \ln \left(\left(\frac{\omega_+ + \omega_-}{4 \sqrt{\omega_+ \omega_- }}\right)-\frac{1}{2}\right),
\end{align}
which is an alternative form of \eqref{Jefferson entropy densmat}.\newline
    
With this particular example we can see that for a Gaussian state, we have the liberty of choosing between the standard way of calculating the purity and entropy in terms of the density matrix, or take the perhaps more approachable path of the quantum covariance matrix. Our decision in reality depends on what type of integrals have to be calculated and if the entropy eigenvalue equation like \eqref{eigenvalues entropy} is solvable.

\subsection{N coupled harmonic oscillators}

Let us now consider the most general case of coupling between $N$ harmonic oscillators, described by
\begin{equation}
H=\frac{1}{2}\left(\vec{p}^{\top} \cdot \vec{p}+\vec{q}^{\top} \cdot K \cdot \vec{q}\right)
\end{equation}
where 
\begin{equation}
    \vec{q} =\left(\begin{array}{c}
q_{1} \\
\vdots \\
q_{N}
\end{array}\right) ,
\end{equation}
\begin{equation}
    \vec{p} =\left(\begin{array}{c}
p_{1} \\
\vdots \\
p_{N}
\end{array}\right),
\end{equation}
and the matrix $K$ contains all the information about the coupling between oscillators, for example the entry $K_{(1,2)}$ is the coupling constant between the oscillators $q_1$ and $q_2$. It should be noted that $K$ is a real symmetric matrix with positive eigenvalues.\newline

This system has been specially useful to understand the entropy of black holes \cite{Srednicki, Bombelli, entropy black holes}, but given its complexity the noteworthy results are regularly just stated without much explanation. Here we will study it in detail as a final example utilizing a combination of the techniques used in the case of the two coupled harmonic oscillators.\newline

To obtain the wave function of this system we will follow the same procedure that we used in the previous example, first we will decouple the oscillators with a canonical transformation and then the solution will be the product of the individual decoupled wave functions. What this canonical transformation does is diagonalize our matrix $K$ such that the elements of this new diagonal matrix $W$ are the frequencies of the normal modes (our decoupled oscillators) i.e.
\begin{equation}
    W =\operatorname{diag}\left(\omega_{1}, \ldots \omega_{N}\right).
\end{equation}
 We do so utilizing orthogonal matrices $U \cdot U^{\top} =\mathbb{1}$ such that we can write
\begin{equation}
K =U^{\top} \cdot W^{2} \cdot U, \\
\end{equation}
meaning that our normal modes coordinates are
\begin{equation}
\vec{Q} =U \cdot \vec{q}, 
\end{equation}
and
\begin{equation}
\vec{P} =U \cdot \vec{p}.
\end{equation}
Now, the general wave function of our global system will be the product of the wave functions of our normal modes
\begin{equation}
\psi_{n_{1} \ldots n_{N}} =\prod_{a=1}^{N} \psi_{n_{a}} 
\end{equation}
where each individual solution is
\begin{equation}
\psi_{n_{a}} =\left(\frac{\omega_{a}}{\hslash}\right)^{\frac{1}{4}} \chi_{n_{a}}(\xi)
\end{equation}
with
\begin{equation}
\xi_{a}=Q_{a} \sqrt{\frac{\omega_{a}}{\hslash}}.
\end{equation}

The ground state will be then characterized by the wave function
\begin{equation}
\psi_{0} =\frac{[\operatorname{det} W]^{\frac{1}{4}}}{(\pi \hslash)^{\frac{N}{4}}} \exp \left(-\frac{1}{2 \hslash} \vec{Q}^{\top} W \vec{Q}\right),
\end{equation}
that in terms of our original variables $q_1,\dots,q_N$ the exponent is
\begin{equation}
-\frac{1}{2 \hslash} \vec{Q}^{\top} W \vec{Q}=\frac{1}{2 \hslash} \vec{q}^{\top} U^{\top} W U \vec{q}.
\end{equation}

At this point we can already calculate the purity using the quantum covariance matrix, for example with the subsystems defined such that the one consist of the first $n$ oscillators and another of the last $N-n$, by obtaining the reduced matrices
\begin{equation}
\hat{\sigma}_{p_{a} p_{b}}=\frac{1}{2}\left\langle\hat{p}_{a} \hat{p}_{b}+\hat{p}_{b} \hat{p}_{a}\right\rangle-\left\langle\hat{p}_{a}\right\rangle\left\langle\hat{p}_{b}\right\rangle,
\end{equation}
\begin{equation}
\hat{\sigma}_{q_{a} q_{b}}=\frac{1}{2}\left\langle\hat{q}_{a} \hat{q}_{b}+\hat{q}_{b} \hat{q}_{a}\right\rangle-\left\langle\hat{q}_{a}\right\rangle\left\langle\hat{q}_{b}\right\rangle
\end{equation}
that in terms of our matrices $U$ and $W$ are
\begin{equation}\label{N osc sigma pp}
\left(\hat{\sigma}_{p_{a} p_{b}}\right)=\frac{\hslash}{2} U \cdot W \cdot U^{\top}=\left(\begin{array}{cc}
A & B \\
B^{\top} & C
\end{array}\right)
\end{equation}
\begin{equation}
\left(\hat{\sigma}_{q_{a} q_{b}}\right)=\frac{\hslash}{2} U \cdot W^{-1} \cdot U^{\top}=\left(\begin{array}{cc}
E & F \\
F^{\top} & G
\end{array}\right)
\end{equation}
where they satisfy
\begin{equation}
\left(\hat{\sigma}_{p_{a} p_{b}}\right)\left(\hat{\sigma}_{q_{a} q_{b}}\right)=\left(\frac{\hslash}{2}\right)^{2} \mathbb{1}_{n \times n}.
\end{equation}
With this and \eqref{purity gaussian}, our purity for the subsystem consisting of the first $n$ oscillators will be
\begin{align}
\mu(1, \ldots, n)&=\int_{-\infty}^{\infty} d q_{1} . . d q_{n} d q_{1}^{\prime} \ldots d q_{n}^{\prime} \rho\left(q_{1}, . ., q_{n} \mid q_{1}^{\prime}, \ldots, q_{n}^{\prime}\right) \rho\left(q_{1}^{\prime}, . ., q_{n}^{\prime} \mid q_{1}, \ldots, q_{n}\right)
\\&
=\left(\frac{\hslash}{2}\right)^{n} \frac{1}{\sqrt{\operatorname{det} A \operatorname{det} E}}
\end{align}
and for the one with the lasts $N-n$,
\begin{equation}
\mu( n+1, \ldots, N)=\left(\frac{\hslash}{2}\right)^{N-n} \frac{1}{\sqrt{\operatorname{det} C \operatorname{det} F}}.
\end{equation}

We should remark that it only makes sense to obtain the purity and entropy for interacting sections of the whole system, otherwise the entanglement would automatically be zero. For example, if we were dealing with a chain of $40$ oscillators were each only interacts with its nearest neighbors (in the same fashion as our previous example), the subsystems must be constructed between blocks of contiguous oscillators, i.e. one block could be formed by the oscillators $\{1,\dots,10\}$ and another one by the $\{20,\dots, 40\}$ but it would not be of much interest to construct one block utilizing the oscillators $\{1,3,5,10,13\}$ specifically.\newline

For the entropy we will show how to obtain it via the eigenvalue equation. For this we must get our density matrix, noticing first that we can write the ground state wave function in terms of the reduced quantum covariance matrix for the momenta \eqref{N osc sigma pp} as 
\begin{equation}
\psi_{0}(\vec{q})=\frac{[\operatorname{det} W]^{\frac{1}{4}}}{(\pi \hslash)^{\frac{N}{4}}} \exp \left(-\frac{1}{\hslash^{2}} \vec{q}^{\top} \sigma^{(p)} \vec{q}\right).
\end{equation}
and with it we can construct our density matrix in the same way as before:
\begin{equation}
\begin{aligned}
\rho\left(\vec{q}, \vec{q}^{\prime}\right) &=\psi_{0}(q) \psi_{0}\left(q^{\prime}\right) \\
&=\frac{[\operatorname{det} W]^{\frac{1}{2}}}{(\pi \hslash)^{\frac{N}{2}}} \exp \left[-\frac{1}{\hslash^{2}}\left(\vec{q}^{\top} \sigma^{(p)} \vec{q}+\vec{q}^{\prime \top} \sigma^{(p)} \vec{q}^{\prime}\right)\right]
\end{aligned}
\end{equation}

To obtain the reduced density matrix of the first $n$ oscillators, collectively denoted from here on out as  $\chi=\left(q_{1}, \ldots, q_{n}\right)$, we take the trace with respect to the last $N-n$, i.e.
\begin{align}
\rho_{\{n\}}\left(\chi, \chi^{\prime}\right)=& \int_{-\infty}^{\infty} \mathrm{d} q_{n+1} \ldots \mathrm{d} q_{N} \rho\left(\chi, q_{n+1}, \dots, q_N, \chi^{\prime}, q_{n+1}, \dots, q_N\right) \\
=&\nonumber \frac{\left(\hslash \sqrt{\frac{\pi}{2}}\right)^{N-n}}{(\pi \hslash)^{\frac{N}{2}}}\left(\frac{\operatorname{det} W}{\operatorname{det} C}\right)^{\frac{1}{2}} \exp \left\{-\frac{1}{\hslash^{2}}\left[\vec{\chi}^{\top} \cdot\left(A-\frac{M}{2}\right) \cdot \vec{\chi}\right.\right.\\
&\label{mdens red n} \left.\left.+\vec{\chi}^{\prime \top} \cdot\left(A-\frac{M}{2}\right) \cdot \vec{\chi}^{\prime}-\vec{\chi}^{\top} \cdot M \cdot \vec{\chi}^{\prime}\right]\right\}
\end{align}
where the subindex $\{n\}$ denotes the set of the first $n$ oscillators and the $A$ and $M$ matrices come from $\sigma^{(p)}$  defining
\begin{equation}
M =B \cdot C^{-1} B^{\top}.
\end{equation}
We must realize that we are doing this in order to obtain an equation analogous to \eqref{reducida rho2 gammabeta}, and also that $A=(A)_{n \times n}$ and $C =(C)_{(N-n) \times(N-n)}$. \newline

The entropy of the first $n$ oscillators will be once again given by
\begin{equation}
S_{\{n\}}=-\operatorname{Tr}\left(\rho_{\{n\}} \ln \rho_{\{n\}}\right)=-\sum_{\mu} p_{\mu} \ln p_{\mu},
\end{equation}
where $p_{\mu}$ are the eigenvalues of $\rho_{n}$, solutions of
\begin{equation}
\int_{-\infty}^{\infty} \mathrm{d} q_{1}^{\prime} \ldots \mathrm{d} q_{n}^{\prime} \rho_{\{n\}}\left(\chi, \chi^{\prime}\right) f_{\mu}\left(\chi^{\prime}\right)=p_{\mu} f_{\mu}(\chi).
\end{equation}
To find a manageable solution of this equation, we must first notice that the matrix $A-\frac{M}{2}$ is symmetric, therefore we can diagonalize it in the same way as we did with $K$, this is
\begin{equation}\label{N osc lol}
A-\frac{M}{2}=\lambda^{\top} \Theta \lambda
\end{equation}
with $\lambda^{\top} \lambda =\mathbb{1}$, and $
\Theta =\operatorname{diag}\left(\theta_{1}, \ldots, \theta_{n}\right)$. Then, the product of matrices within the exponent of \eqref{mdens red n} can be rewritten as
\begin{equation}
\begin{aligned}
\vec{\chi}^{\top} \cdot\left(A-\frac{M}{2}\right) \cdot \vec{\chi} &=\vec{\chi}^{\top} \lambda^{\top} \Theta \lambda \vec{\chi}=\left(\vec{\chi}^{\top} \lambda^{\top} \Theta^{\frac{1}{2}}\right)\left(\Theta^{\frac{1}{2}} \lambda \vec{\chi}\right) \\
&=\frac{Z^{\top} Z}{2}=\sum_{i=1}^{n} \frac{Z_{i}^{2}}{2}
\end{aligned}
\end{equation}
where $\frac{\vec{Z}}{\sqrt{2}}=\Theta^{\frac{1}{2}} \lambda \vec{\chi}$ and $\Theta^{\frac{1}{2}}=\operatorname{diag}\left(\theta_{1}^{\frac{1}{2}}, \ldots, \theta_{n}^{\frac{1}{2}}\right)$ since $\Theta$ is a diagonal matrix. \newline
Now we do the same for the other products in the exponent of \eqref{mdens red n}:
\begin{equation}
\vec{\chi}^{\prime \top} \cdot\left(A-\frac{M}{2}\right) \cdot \vec{\chi}^{\prime}=\sum_{i=1}^{n} \frac{Z_{i}^{\prime 2}}{2}
\end{equation}
where we just added the prime in $\frac{\vec{Z}^{\prime}}{\sqrt{2}}=\Theta^{\frac{1}{2}} \lambda \overrightarrow{\chi^{\prime}}$. Using this, we can write \begin{equation}\vec{\chi}=\frac{\lambda^{\top} \Theta^{-\frac{1}{2}} \vec{Z}}{\sqrt{2}},\end{equation} 
\begin{equation}\vec{\chi}^{\prime}=\frac{\lambda^{\top} \Theta^{-\frac{1}{2}} \vec{Z}^{\prime}}{\sqrt{2}},\end{equation}
then we have
\begin{equation}\label{n osc znz}
\vec{\chi}^{\top} \cdot M \cdot \vec{\chi}^{\prime}=\vec{Z}^{\top} \cdot V \cdot \vec{Z}^{\prime},
\end{equation}
with $V=\frac{\Theta^{\frac{1}{2}} \lambda M \lambda^{\top} \Theta^{-\frac{1}{2}}}{2} .$ Using \eqref{n osc znz} we are able rewrite our density matrix in a much simpler way:
\begin{equation}
\rho_{n}\left(\chi, \chi^{\prime}\right)=\beta \exp \left\{-\frac{1}{h^{2}} \left[ \sum_{i=1}^{n} \frac{1}{2}\left(Z_{i}^{2}+Z_{i}^{\prime 2}\right)-\vec{Z}^{\top} \cdot V \cdot \vec{Z}^{\prime}\right]\right\}
\end{equation}
with
\begin{equation}
\beta=\frac{\left(\hslash \sqrt{\frac{\pi}{2}}\right)^{N-n}}{(\pi \hslash)^{\frac{N}{2}}}\left(\frac{\operatorname{det} W}{\operatorname{det} C}\right)^{\frac{1}{2}}.
\end{equation}
But it can be simplified even further if we notice that the matrix $V$ is also symmetric, so we diagonalize it in the same way as the others
\begin{equation}
V=\eta^{\top} \varphi \eta
\end{equation}
where once again $\eta^{\top} \eta =\mathbb{1}$ and $\varphi=\operatorname{diag}\left(\varphi_{1}, \ldots, \varphi_{n}\right)$. Then we can combine the $\eta$ and $\vec{Z}$  using
\begin{equation}
\begin{aligned}
\vec{Y} &=\eta \vec{Z} \\
\vec{Y}^{\prime} &=\eta \vec{Z}^{\prime}
\end{aligned}
\end{equation}
for a more compact notation, 
\begin{equation}
\vec{Z}^{\top} V \vec{Z}^{\prime}=\vec{Z}^{\top} \eta^{\top} \varphi \eta \vec{Z}^{\prime}=\vec{Y}^{\top} \varphi \vec{Y}^{\prime}=\sum_{i=1}^{n} \varphi_{i} Y_{i} Y_{i}^{\prime}
\end{equation}
and for the squared terms
\begin{equation}
\sum_{i=1}^{n} Z_{i}^{2}=\vec{Z}^{\top} \vec{Z}=\vec{Y}^{\top} \eta \eta^{\top} \vec{Y}=\vec{Y}^{\top} \vec{Y}=\sum_{i=1}^{n} Y_{i}^{2},
\end{equation}
thus our density matrix becomes
\begin{equation}
\begin{aligned}
\rho_{n}\left(\chi, \chi^{\prime}\right) &=\beta \exp \left[-\frac{1}{\hslash^{2}} \sum_{i=1}^{n}\left(\frac{Y_{i}^{2}}{2}+\frac{Y_{i}^{\prime 2}}{2}-\varphi_{i} Y_{i} Y_{i}^{\prime}\right)\right] \\
&=\beta \prod_{i=1}^{n} \exp \left[-\frac{1}{\hslash^{2}}\left(\frac{Y_{i}^{2}}{2}+\frac{Y_{i}^{\prime 2}}{2}-\varphi_{i} Y_{i} Y_{i}^{\prime}\right)\right].
\end{aligned}
\end{equation}
Now, to solve the eigenvalue equation
\begin{equation}
\beta \int_{-\infty}^{\infty} \mathrm{d} q_{1}^{\prime} \ldots \mathrm{d} q_{n}^{\prime} \prod_{i=1}^{n} \exp \left[-\frac{1}{\hslash^{2}}\left(\frac{Y_{i}^{2}}{2}+\frac{Y_{i}^{\prime 2}}{2}-\varphi_{i} Y_{i} Y_{i}^{\prime}\right)\right] f_{\mu}\left(\chi^{\prime}\right)=p_{\mu} f_{\mu}(\chi),
\end{equation}
we must change the variables of integration from  $\left(q_{1}^{\prime}, \ldots, q_{n}^{\prime}\right) \rightarrow\left(Y_{1}, \ldots, Y_{n}\right)$, so we need the Jacobian of this transformation
\begin{equation}
J\left(Y^{\prime}\right) =\left(\begin{array}{ccc}
\frac{\partial q_{1}}{\partial  Y_{1}} & \cdots & \frac{\partial q_{1}}{\partial Y_{n}} \\
\vdots & \ddots & \vdots \\
\frac{\partial q_{n}}{\partial Y_{1}} & \cdots & \frac{\partial q_{n}}{\partial Y_{n}}
\end{array}\right)=\frac{1}{\sqrt{2}} \lambda^{\top} \Theta^{-\frac{1}{2}} \eta^{\top}, \end{equation}
then
\begin{equation}
     \operatorname{det} (J) =\frac{1}{2^{\frac{n}{2}}} \operatorname{det}\left(\lambda^{\top}\right) \operatorname{det}\left(\Theta^{-\frac{1}{2}}\right)\operatorname{det}\left(\eta^{\top}\right)
\end{equation}
and remembering that both $\lambda$ and $\eta$ are ortogonal, i.e.  $\operatorname{det}\left(\lambda^{\top}\right)=\operatorname{det}\left(\eta^{\top}\right)=1$,
\begin{equation}
    \operatorname{det}(J) =\frac{1}{2^{\frac{n}{2}}} \operatorname{det}\left(\Theta^{-\frac{1}{2}}\right),
\end{equation}
but from \eqref{N osc lol}
\begin{equation}
 \operatorname{det}\left(A-\frac{M}{2}\right)=\operatorname{det} (\Theta)
\end{equation}
then since $\Theta$ is diagonal,
\begin{equation}
\operatorname{det} (J)=\frac{1}{2^{\frac{n}{2}}}\left(\operatorname{det}\left( A-\frac{M}{2}\right)\right)^{-\frac{1}{2}}.
\end{equation}
With this result at hand we can perform our desired change of variables from $\left(q_{1}^{\prime}, \ldots, q_{n}^{\prime}\right) \rightarrow\left(Y_{1}, \ldots, Y_{n}\right)$, leaving us the simpler eigenvalue equation
\begin{equation}
\beta \operatorname{det}(J) \int_{-\infty}^{\infty} \mathrm{d} Y_{1}^{\prime} \ldots \mathrm{d} Y_{n}^{\prime} \prod_{i=1}^{n} \exp \left[-\frac{1}{\hslash^{2}}\left(\frac{Y_{i}^{2}}{2}+\frac{Y_{i}^{\prime 2}}{2}-\varphi_{i} Y_{i} Y_{i}^{\prime}\right)\right] f_{\mu}\left(Y^{\prime}\right)=p_{\mu} f_{\mu}(Y)
\end{equation}
If we take $f_{\mu}\left(Y^{\prime}\right)=\prod_{i=1}^{n} f_{n}\left(Y_{i}^{\prime}\right)$, we have
\begin{equation} \label{157}
\beta \operatorname{det}(J) \prod_{i=1}^{n} \int_{-\infty}^{\infty} \mathrm{d} Y_{i}^{\prime} \exp \left[-\frac{1}{\hslash^{2}}\left(\frac{Y_{i}^{2}}{2}+\frac{Y_{i}^{\prime 2}}{2}-\varphi_{i} Y_{i} Y_{i}^{\prime}\right)\right] f_{\mu_{i}}\left(Y^{\prime}_i\right)=p_{\mu} \prod_{i=1}^{n} f_{\mu_{i}}\left(Y_{i}\right)
\end{equation}
which can be solved in the same way as our past example since it is essentially the product of eigenvalue equations corresponding to individual oscillators, or speaking formally, each correlates to a normal mode. Following this approach our general solution will be in terms of the parameters of the individual solutions, specifically
\begin{equation}
f_{\mu_{i}} =H_{\mu_{i}}\left(\alpha_{i}^{\frac{1}{2}} Y_{i}\right) \exp \left(-\frac{\alpha_{i} Y_{i}^{2}}{2}\right) 
\end{equation}
\begin{equation}
    \bar{p}_{\mu_{i}} =\left(1-\xi_{i}\right) \xi_{i}^{\mu_{i}}
\end{equation}
\begin{equation}
    \xi_{i} =\frac{\varphi_{i}}{1+\left(1-\varphi_{i}^{2}\right)^{1/2}}
\end{equation}
\begin{equation}
    \alpha_{i} =\left(\left(\frac{1}{\hslash^{2}}\right)^{2}-\left(\frac{\varphi_{i}}{\hslash ^{2}}\right)^{2}\right)^{\frac{1}{2}} 
\end{equation}
where we should remark that $\bar{p}_{\mu_{i}}$ is the eigenvalue for an individual eigenvalue equation and $p_\mu$ is the general eigenvalue.\newline

Using these results in \eqref{157} gives us
\begin{equation}
\beta \operatorname{det}(J) \prod_{i=1}^{n}\left(\frac{\hslash \sqrt{\pi}}{\sqrt{1-\varphi_{i}}} \bar{p}_{\mu_{i}} f_{\mu_{i}}(Y)\right)=p_{\mu} \prod_{i=1}^{n} f_{\mu_{i}}\left(Y_{i}\right),
 \end{equation}
 which implies that our general eigenvalue is
 \begin{equation}
      p_{\mu} = \beta \operatorname{det}(J) \prod_{i=1}^{n}\left(\frac{\hslash \sqrt{\pi}}{\sqrt{1-\varphi_{i}}}\right) \prod_{j=1}^{n}\left(\bar{p}_{\mu_{j}}\right) .
 \end{equation}
When we plug in it the values for $\beta$ and $\operatorname{det}(J)$ we get
\begin{equation}\label{161}
p_{\mu}=2^{-\frac{N}{2}} \hslash^{\frac{N}{2}}\left[\frac{\operatorname{det} W}{\operatorname{det} C} \operatorname{det}\left(A-\frac{M}{2}\right)^{-1} \prod_{i=1}^{n} \frac{1}{\left(1-\varphi_{i}\right)}\right]^{1/2} \prod_{j=1}^{n} \bar{p}_{\mu_{j}} .
\end{equation}
which can be simplified taking the determinant of \eqref{N osc sigma pp}, 
\begin{equation}
\operatorname{det}\left(\sigma^{(p)}\right)=\left(\frac{\hslash}{2}\right)^{N} \operatorname{det} W=\operatorname{det}\left(A-B C^{-1} B^{\top}\right) \operatorname{det} C = \operatorname{det}(A-M) \operatorname{det} C 
\end{equation}
then the ratio of the determinants within our eigenvalue is
\begin{equation}
    \frac{\operatorname{det} W}{\operatorname{det} C}=\left(\frac{2}{\hslash}\right)^{N} \operatorname{det}(A-M).
\end{equation}
Substituting this in \eqref{161} we obtain
\begin{equation}
p_{\mu}=\left[\frac{\operatorname{det}(A-M) \operatorname{det}\left(A-\frac{M}{2}\right)^{-1}}{\prod_{i=1}^{n}\left(1-\varphi_{i}\right)}\right]^{\frac{1}{2}} \prod_{j=1}^{n} \bar{p}_{\mu_{j}}
\end{equation}
and to simplify even further we make use that it is possible to change
\begin{equation}
\prod_{i=1}^{n}\left(1-\varphi_{i}\right)=\operatorname{det}\left(\mathbb{1}_{n}-V\right)
\end{equation}
and we can verify numerically that
\begin{equation}
\left[\frac{\operatorname{det}(A-M)}{\operatorname{det}\left(A-\frac{M}{2}\right) \operatorname{det}\left(\mathbb{1}_{n}-V\right)}\right]^{\frac{1}{2}}=1,
\end{equation}
therefore, we finally get a simple form for our general eigenvalue
\begin{equation}
\begin{aligned}
p_{\mu} &=\prod_{i=1}^{n} \bar{p}_{\mu_{i}} \\
 &= \prod_{i=1}^{n} \left(1-\xi_{i}\right) \xi_{i}^{\mu_{i}} \\
 &=\prod_{i=1}^{n} \left(1-\frac{\varphi_{i}}{1+\left(1-\varphi_{i}^{2}\right)^{\frac{1}{2}}}\right)\left( \frac{\varphi_{i}}{1+\left(1-\varphi_{i}^{2}\right)^{\frac{1}{2}}}\right)^{\mu_{i}}.
\end{aligned}
\end{equation}
Using this result we can calculate the entropy
\begin{equation}
\begin{aligned}
S_{n} &=-\sum_{\mu=0}^{\infty} p_{\mu} \ln p_{\mu} \\
&=-\sum_{\left\{\mu_{i}, \mu_{j}=0\right\}}^{\infty}\left(\prod_{i=1}^{n} \bar{p}_{\mu_{i}}\right) \ln \left(\prod_{j=1}^{n} \bar{p}_{\mu_{j}}\right) \\
&=-\sum_{\left\{\mu_{i}, \mu_{j}=0\right\}}^{\infty}\left(\prod_{i=1}^{n} \bar{p}_{\mu i}\right) \sum_{j=1}^{n} \ln \bar{p}_{\mu j} \\
&=\sum_{i=1}^{n}\left(-\sum_{\mu_{i}=0}^{\infty} \bar{p}_{\mu_{i}} \ln \left(\bar{p}_{\mu_{i}}\right)\right) \\
&= \sum_{i=1}^{n} S_{i}
\end{aligned}
\end{equation}
Notice that $S_{i}$ corresponds to the entropy of the i-th oscillator, so if we use the results of the previous example, 

\begin{equation}\nonumber
S_{i}=-\ln \left(1-\xi_{i}\right)-\frac{\xi_{i}}{1-\xi_{i}} \ln \xi_{i}
\end{equation}
we arrive at the final expression for the correspondent entropy  to the subsystem composed by the first $n$ oscillators:
\begin{equation}
S_{n} =\sum_{i=1}^{n} S_i = \sum_{i=1}^{n} \left(-\ln \left(1-\xi_{i}\right)-\frac{\xi_{i}}{1-\xi_{i}} \ln \xi_{i}\right).
\end{equation}

From these examples we learned how to explicitly calculate the purity and entropy; and if we are dealing with a Gaussian state, that it is possible to obtain the same results with either the quantum covariance matrix or the density matrix.\newline

In the final chapters of this thesis, inspired by Berry \cite{classical Berry} and Hannay \cite{classical Hannay}, we will inquire if there is a classical analog of the quantum covariance matrix and investigate if we can use it to obtain classical analogs of both the purity and entropy.

\part{Classical aspect of the quantum information}

\chapter{Classical analogs of the Quantum Metric Tensor and Berry's phase}

In the next chapter we will propose a new way to calculate the quantum covariance matrix, purity and the entropy using a purely classical approach, specifically we will be able to do so when working with Gaussian states. This type of techniques were first introduced by Berry for the Wigner functions \cite{classical Berry}, and were later popularized by Hannay with a classical analog of the Berry's phase \cite{classical Hannay}. In this chapter we will explain their ideas and briefly review the Action-Angle variables since they will be our main work tools.\newline

Before getting into it, we must first discuss why even if we can produce a classical mathematical apparatus which reproduces the results that are obtained with quantum mechanics, certain properties of nature are intrinsically a quantum phenomenon (entanglement for example) and can not be replicated using a classical system.

\section{Why is entanglement only a quantum effect and not a classical one?}

For the last century, we as physicist have been trying to attain a clear understanding of what really differentiates a quantum theory from a classical one. We may think, for example, that the concept of superposition is perhaps the main component of a quantum theory, however it also appears in classical wave mechanics, as in Young’s
double-slit experiment. So why is it any different when it appears in quantum mechanics? \newline

This becomes apparent, and is rather radical, when we repeat the double-slit experiment using a source that emits a single photon at a time, or in reality, when we interfere any single quantum object. It has been observed in the laboratory \cite{interferencia fotones, interferencia electrones, interferencia neutrones, interferencia atomos} that when we send a particle  at a the time through the slits, the same interference pattern emerges, but since we sent them one by one, they could not interfere with any other particle. This is the main difference of superposition in quantum mechanics, here even single particles can interfere, where in classical mechanics only waves can do so. \newline

Interference of matter can be thoroughly described by the wave function of quantum mechanics, but there have been attempts to explain it, and all the philosophical questions that it arises, by using what are known as hidden variables. These are some sort of unknown physical parameters that completely determine the path of the single particle, which under this formalism behaves just as any classical object and the randomness of its trajectory is a result of these variables interacting with (or being part of) the body. However, it was shown by J. Bell \cite{Bell} that there cannot exist a theory of local hidden variables that successfully reproduces all the predictions of quantum mechanics, and therefore that is impossible to encompass the whole of Nature within a classical framework. \newline

Nowadays, specially in the study of optical waves, the superpositions between different degrees of freedom of a physical system has been sometimes called "classical entanglement" when the phenomena at hand can be completely understood with classical theories, like classical electrodynamics. In these cases we believe that there is no need to utilize "quantum labels" since they are not needed, and that doing so could be detrimental since it can lead to confusion \cite{entanglement clas vs quant}. \newline  

Entanglement is one of the most important implications of  quantum physics, and there are classical analogs for it \cite{classical entanglement}, but they do not imply that classical systems can experiment entanglement in the full sense of the word, specifically since the spatial separation condition is not present in this classical counterpart. We must always keep in mind that the classical analogs are exactly that, analogies, not a new way to circumvent quantum implications, nor get them when they are not present in the physical system. But these classical analogs have value of their own since they helps us calculate certain quantities and specially teach us how to exactly discern what truly makes something a quantum property.\newline

In this thesis we will give a classical analog of the quantum covariance matrix, the purity and the entropy, but we want to make clear that it must be treated as a mathematical apparatus that will facilitate calculations for Gaussian states, and is not meant to be understood as a "classical way to view quantum mechanics" nor viewed as an equivalent procedure to the quantum ones explained in the previous chapter.\newline

Before we begin explaining the most famous classical analog of a quantum property, Hannay's angle, let us briefly review the action-angle variables since most classical analogs are in terms of them.

\section{Hamilton-Jacobi equation and the Action-Angle variables}

The Hamilton–Jacobi equation is a formulation of classical mechanics, equivalent to the Newton's laws of motion, Lagrangian mechanics and Hamiltonian mechanics. The principal advantage of using this formalism instead of the others is that it allows us to identify conserved quantities within our systems of study even when we have not solved it completely. It is also the only formulation in which we can represent the motion of a particle in terms of a wave \cite{Goldstein} and thus became an inspiration for Schrödinger to postulate his groundbreaking equation \cite{Schrodinger}.\newline

The \textit{Hamilton-Jacobi} equation can be understood as a canonical transformation that takes us from our original variables to new ones such that the equations of motions become zero, meaning that our transformed Hamiltonian denoted by $K$ does not carry any time evolution and thus is identically zero. This peculiar transformation is explicitly written as

\begin{equation}\label{Hamilton-Jacobi}
K = H\left(q_{1}, \ldots, q_{n} ; \frac{\partial S}{\partial q_{1}}, \ldots, \frac{\partial S}{\partial q_{n}} ; t\right)+\frac{\partial S}{\partial t}=0
\end{equation}
where $H$ is our original Hamiltonian and $S$ in this context is called \textit{"Hamilton's principal function"}, which as the notation implies will turn out to be the\textit{ action} of our system
\begin{equation}
    S=S\left(q_{1}, \ldots, q_{n} ; \alpha_{1}, \ldots, \alpha_{n+1} ; t\right)=\int dt L.
\end{equation}
We must notice that our Hamiltonian is written in terms of the action, substituting the momentum as
\begin{equation}\label{momentum H-J}
    p_i = \frac{\partial S}{\partial q_i}
\end{equation}
and that $S$ itself is dependent of $\alpha_{1}, \ldots, \alpha_{n+1}$ which are $n+1$  independent constants of integration (the final constant $\alpha_{n+1}$ is an additive constant) that in the most common cases will be related to the energy.\newline

To solve a system in this formulation of classical mechanics we must first find the value of $S$ that is obtained with the differential equation that emerges once we plug in our Hamiltonian in \eqref{Hamilton-Jacobi}, and then we get its derivatives
\begin{equation}
\frac{\partial S(q, \alpha, t)}{\partial \alpha_{i}}=\beta_{i}
\end{equation}
\begin{equation}
\frac{\partial S(q, \alpha, t)}{\partial q_{i}}=p_{i}
\end{equation}
since with them we can write
\begin{equation}
q_{j}=q_{j}(\alpha, \beta, t),
\end{equation}
and completely solve our system \cite{Goldstein}.\newline

For simplicity, from this point onward we will focus on the case of one degree of freedom.\newline

If our Hamiltonian does not depend explicitly on time, all time dependence  of \eqref{Hamilton-Jacobi} concentrates on the partial of the action
\begin{equation}
H\left(q, \frac{\partial S}{\partial q}\right)+\frac{\partial S}{\partial t}=0
\end{equation}
so we can hypothesize that our solution will be of the form
\begin{equation}
S(q, t)=W(q)+ T(t),
\end{equation}
where we have separated the coordinates and time dependence of the action, $W$ is called the \textit{ Hamilton's characteristic function} (the action $S$ was the principal function). Then \eqref{Hamilton-Jacobi} can be written as
\begin{equation}
\mathrm{H}\left(\mathrm{q}, \frac{\partial \mathrm{W}}{\partial \mathrm{q}}\right)=-\frac{\mathrm{dT}}{\mathrm{dt}},
\end{equation}
and since both sides of the equation depend on different variables, they must be equal to a constant, the energy $E$, given that our Hamiltonian is time independent,
\begin{equation}
H\left(q, \frac{\partial W}{\partial q}\right)=E=- \frac{d T}{d t}.
\end{equation}
So all that there is left to do is find $W$ in terms of $E$ and $q$, from which we can find $q$ itself from the partials of $W$ as stated above.\newline

For systems with a periodic behavior in phase space\footnote{Notice that we say "behavior in phase space" and not "motion", this is because motion characterizes the behavior of the particle in time and having a periodic behavior in phase space does not necessarily mean that the motion of the particle will be periodic.}, we at first might be more interested in finding the frequency of it, rather than the explicit description of the motion itself. When this is the case, there are a particular set of coordinates known as action-angle variables that when used alongside the Hamilton-Jacobi equation, they almost effortlessly give us the frequency of the motion while also having the advantage that make \eqref{Hamilton-Jacobi} completely separable.\newline

We can divide this periodic behavior in two important categories:
\begin{itemize}
    \item \textit{Oscillatory} - For this type of motion is common for the momentum to be related with its conjugate coordinate by a quadratic equation, so the trajectory forms a closed loop when plotted in phase space, meaning that the particle will go back and forth between turning points retracing its steps every half oscillation, one physical example of this will be the harmonic oscillator.

    \item \textit{Rotatory} -  As the name implies, this type of motion is obtained when the position coordinate returns to its original position without retracing its steps, a physical example would be any type of wheel. In this case, the plot in phase space is not closed and behaves similarly to a cosine function.
\end{itemize}
It should be noted that the two types of periodic behavior are bounded in phase space, and also that they both can be found in the same physical system where the one that is exhibited depends on how much energy the system possess. For example, a simple pendulum can display oscillatory motion when its energy is not enough to go over the top, but has a rotatory type of motion when it does \cite{Goldstein, Calkin} \newline

With these cases in mind, we can define our \textit{action-angle variables} as a canonical transformation, where the angle variable takes the role of the position, and the action variable of the momentum. To get them explicitly we use Hamilton's characteristic function in \eqref{momentum H-J} so we can write it as
\begin{equation}
\Delta W=\int p d q,
\end{equation}
since our system is considered to be periodic, we can take the closed integral so our system returns to its original state. We denote this change in $W$ by $2\pi$ as our \textit{action variable}
\begin{equation}
I=\frac{1}{2 \pi} \oint p d q,
\end{equation}
this name comes from the fact that they have the same units as the action $S$, i.e. angular momentum.\newline

The generalized coordinate conjugate to $I$, will be the \textit{angle variable} that we will denote by $\phi$, and is defined by the partial
\begin{equation}
\phi=\frac{\partial W}{\partial I},
\end{equation}
and it does not have any units, hence the "angle" name.\newline

These action-angle variables have the important characteristic that the Hamiltonian can be written in terms of the action variable alone $H=H(I)$, in this manner we have made the angle variable a cyclic or ignorable coordinate. This has important implications for the equations of motion, which for the action variable will be
\begin{equation}
\dot{I}=-\frac{\partial H(I)}{\partial \phi} = 0
\end{equation}
meaning that the action variable itself is a constant of motion (we constructed it in this way), and more importantly
\begin{equation}
\dot{\phi}=\frac{\partial H(J)}{\partial I}=\omega(I),
\end{equation}
where $\omega$ is the angular frequency of our periodic system, which is also a constant of motion since it only depends on $I$.\newline

Notice that we never completely solved our system, but nonetheless we found the frequency just by using these new variables \cite{Goldstein, Calkin}.

\section{Hannay's angle}

In section $2.2$ we explored the components of the Quantum Geometric Tensor, the real part being the Quantum Metric Tensor and the imaginary one turn out to be closely related to the Berry curvature, from which we can obtain Berry's phase. This was an extra phase that appeared in our wave function when we vary the parameters of a system adiabatically and in a cyclic form.\newline

For classical systems it was well established that adiabatic changes in the parameters result in conserved quantities associated to the action variable, however it had eluded physicist how these variations affect the angle variable and if there was any physical significance in these changes of a regularly cyclic coordinate.\newline

Hannay \cite{classical Hannay}, inspired by Berry \cite{Berry}, was the first to closely examine what occurs to the angle variable when we vary the parameters of a classical systems in a closed loop. He found that an anoholonomy also appears for classical systems within the angle variable.\newline

Consider a one-dimensional system, described by the Hamiltonian $\mathrm{H}(q, p, \lambda(t))$, where $\lambda$ are the parameters that are changed adiabatically and thus their dependence on time. We are interested in the equations of motion satisfied by the action-angle variables $(I,\phi)$ that can be obtained for each instant $t$ from the $(q, p)$ using a time dependent canonical transformation, with generating function $W(\mathrm{q}, \mathrm{I}, \lambda(t))$.\newline

These action-angle variables satisfy Hamilton's canonical equations with Hamiltonian
\begin{equation}
K(\phi, I, \mathbf{X}(t))=H(I, \lambda(t))+\left(\frac{\partial W(q, I, \lambda(t))}{\partial \lambda }\right)_{q, I} \cdot \frac{d \lambda(t)}{d t}
\end{equation}
where its important to note that $H(I, \lambda(t))$ is written in terms of only the action variable (see the previous section) and the parameters, and that the Hamiltonian $K$ for the action-angle variables not only depends on the parameters but also on on the rate of change $d \lambda(t) / dt$ of them. \newline

The partial of the generating function $W$ with respect to the parameters $\lambda$ can be expressed as
\begin{equation}
\left(\frac{\partial W}{\partial \lambda}\right)_{\phi, I}=\left(\frac{\partial W}{\partial q}\right)_{I, \lambda}\left(\frac{\partial q}{\partial \lambda}\right)_{\phi, I}+\left(\frac{\partial W}{\partial \lambda}\right)_{q, I}= p \left(\frac{\partial q}{\partial \lambda}\right)_{\phi, I} +\left(\frac{\partial W}{\partial \lambda}\right)_{q, I}
\end{equation}
and with this we can explicitly write the equations of motion for the action-angle variables 
\begin{equation}
\frac{d \phi}{d t}=\omega(I, \lambda)+\frac{\partial}{\partial I}\left(-p \frac{\partial q}{\partial \lambda}+\frac{\partial W}{\partial \lambda}\right) \cdot \frac{d \lambda}{d t}
\end{equation}
\begin{equation}
\frac{\mathrm{d} I}{\mathrm{dt}}=-\frac{\partial}{\partial \phi}\left(-p \frac{\partial q}{\partial \lambda}+\frac{\partial W}{\partial \lambda}\right) \cdot \frac{d\lambda}{d t} 
\end{equation}
where $\omega(\mathrm{I}, \lambda)=\partial \mathrm{H}_{0}(\mathrm{I}, \lambda) / \partial \mathrm{I}$ is the angular frequency.\newline

Up to this point we have made no assumptions on how the parameters change with time, so if we want them to change adiabatically we need to make the rate of change so slow that the would need several cycles (in phase space) in order to notice them, specifically
\begin{equation}
\frac{1}{\lambda}\frac{d\lambda}{dt} \ll \omega.
\end{equation}
Also since we need whole cycles to perceive any changes, we can approximate our results by averaging over them. For example, any function $f$ can be substituted by its average in the following way
\begin{equation}
\langle f \rangle=\frac{1}{2 \pi} \int_{0}^{2 \pi} f \mathrm{d} \phi.
\end{equation}

With this in mind, our averaged action variable equation of motion is
\begin{equation}
\left\langle\frac{\mathrm{d} I}{\mathrm{dt}}\right\rangle \approx 0 .
\end{equation}
since it was single-valued function of $\phi$ and the averaging returns a null value, implying that the action variable is an \textit{adiabatic invariant}, notice that it wasn't invariant under any rate of change in the parameters, it became invariant when we set the rate of change to an adiabatic one.\newline

For the angle variable on the other hand we get
\begin{equation}\label{averaged eom phi}
\frac{d \phi}{dt} \approx \omega(I, \lambda)+\frac{\partial \mathcal{A}(I, \lambda)}{\partial I} \cdot \frac{d \lambda}{dt}
\end{equation}
with the function $A$ carrying the averages as
\begin{equation}
\mathcal{A}(\mathrm{I}, \lambda)=-\langle p \frac{\partial q}{\partial \lambda}\rangle + \langle\frac{\partial W}{\partial \lambda}\rangle.
\end{equation}

We can integrate \eqref{averaged eom phi} with respect to time to obtain the change in the angle variable when going from a time $t_0$ to another one $t_f$,
\begin{equation}
\Delta \phi \approx \int_{t_{0}}^{t_f} \omega(I, \lambda(t)) d t+\frac{\partial}{\partial I} \int_{\lambda_{0}}^{\lambda_f} \mathcal{A}(I, \lambda) \cdot d \lambda .
\end{equation}
 where we used the fact that the action variable is constant. The first term is just how much the angle variable changed in the time that passed ($t_f - t_0$) with respect to the "zero" of the variable, the point that we take as our origin. However, the second term only depends on the path that we take in parameter space, this parameter dependent shift in the angle variable is called the \textit{Hannay change}.\newline
 
The \textit{Hannay angle}, is the Hannay change obtained when our initial parameter is the same as the final one $\lambda_i = \lambda_f$, meaning that we traversed a closed path in parameter space 
\begin{equation}
    \Delta \phi_H = \frac{\partial}{\partial I} \oint \mathcal{A}(I, \lambda) \cdot d \lambda 
\end{equation}
and it has the important property of being gauge invariant \cite{classical Hannay, Calkin}. \newline

Now that we have seen that Berry's phase has a classical analog, in the next section we will find one for the whole Quantum Geometric Tensor, including the Quantum Metric Tensor.

\section{Classical analog of the Quantum Geometric Tensor}

In this section we will formulate the classical analogs  of the quantum metric tensor and the Berry curvature  for classical integrable systems, since this implies that we are able to construct their action-angle variables. To do so we must consider \eqref{TGC general} for the ground state and varying only parameters
\begin{equation}
G_{i j}^{(0)}=\frac{-1}{\hslash^2}\int_{-\infty}^{t_0}dt_1\int_{t_0}^{\infty}dt_2[\expval{\mathcal{\hat{O}}_i(t_1)\mathcal{\hat{O}}_j(t_2)}_0-\expval{\mathcal{\hat{O}}_i(t_1)}_0\expval{\mathcal{\hat{O}}_j(t_2)}_0],
\end{equation}
from which we can obtain the Quantum Metric Tensor by taking its real part
\begin{equation}\label{QMT pathint}
g_{i j}^{(0)}(x)=-\frac{1}{\hslash^{2}} \int_{-\infty}^{0} \mathrm{~d} t_{1} \int_{0}^{\infty} \mathrm{d} t_{2}\left(\frac{1}{2}\left\langle\left[\mathcal{\hat{O}}_{i}\left(t_{1}\right), \mathcal{\hat{O}}_{j}\left(t_{2}\right)\right]_{+}\right\rangle_{0}-\left\langle\mathcal{\hat{O}}_{i}\left(t_{1}\right)\right\rangle_{0}\left\langle\mathcal{\hat{O}}_{j}\left(t_{2}\right)\right\rangle_{0}\right)
\end{equation}
and Berry's curvature with the imaginary one
\begin{equation}\label{BerryC pathint}
F_{i j}^{(0)}(x)=\frac{1}{i \hslash^{2}} \int_{-\infty}^{0} \mathrm{~d} t_{1} \int_{0}^{\infty} \mathrm{d} t_{2}\left\langle\left[\mathcal{\hat{O}}_{i}\left(t_{1}\right), \mathcal{\hat{O}}_{j}\left(t_{2}\right)\right]_{-}\right\rangle_{0}
\end{equation}
where $[\hat{a},\hat{b} ]_{+}$ and $[\hat{a},\hat{b}]_{-}$ are the anticommutator and the commutator, respectively, of the operators $\hat{a}$ and $\hat{b}$.\newline

In these equations the operators $\mathcal{\hat{O}}_{i}(t)$ are in terms of the phase space operators in Heisenberg's representation
\begin{equation}
\mathcal{\hat{O}}_{i}(t)=\mathcal{\hat{O}}_{i}(\hat{q}(t), \hat{p}(t) ; \lambda)=\left(\partial_{i} \hat{H}(\hat{q}(t), \hat{p}(t) ; \lambda)\right)_{\hat{q}(t), \hat{p}(t)}
\end{equation}
where $\hat{H}(\hat{q}(t), \hat{p}(t) ; \lambda)$ is the original Hamiltonian of the system before the perturbation that occurs at $t=0$. Now, these phase space operators can be written in terms of the ones from the Schrödinger's representation $\hat{q}(t=0)=\hat{q}_{0}$ and $\hat{p}(t=0)=\hat{p}_{0}$ so we can redefine our $\mathcal{\hat{O}}$ so that it only depends in these initial operators and in time,
\begin{equation}
\hat{\mathcal{O}}_{i}(t):=\mathtt{\hat{O}}_{i}\left(t, \hat{q}_{0}, \hat{p}_{0} ; \lambda\right)=\hat{\mathcal{O}}_{i}\left(\hat{q}\left(t, \hat{q}_{0}, \hat{p}_{0} ; \lambda\right), \hat{p}\left(t, \hat{q}_{0}, \hat{p}_{0} ; \lambda\right) ; \lambda\right)
\end{equation}
and with this in mind we write the desired expectation values, beginning by the individual one
\begin{align}
\left\langle\hat{\mathcal{O}}_{i}(t)\right\rangle_{0} &=\left\langle\psi_{0}(\lambda)\left|\hat{\mathcal{O}}_{i}(t)\right| \psi_{0}(\lambda)\right\rangle \\
&=\int \mathrm{d} q_{0} \psi_{0}^{*}\left(q_{0} ; \lambda\right) \mathcal{O}_{i}\left(t, q_{0},-i \hslash \frac{\partial}{\partial q_{0}} ; \lambda\right) \psi_{0}\left(q_{0} ; \lambda\right),
\end{align}
now for the one containing two operators we have
\begin{align}
&\nonumber \left\langle\left[\hat{\mathcal{O}}_{i}\left(t_{1}\right), \hat{\mathcal{O}}_{j}\left(t_{2}\right)\right]_{\pm}\right\rangle_{0} \\
&=\left\langle\psi_{0}(\lambda)\left|\left[\hat{\mathcal{O}}_{i}\left(t_{1}\right), \hat{\mathcal{O}}_{j}\left(t_{2}\right)\right]_{\pm}\right| \psi_{0}(\lambda)\right\rangle \\
&=\int \mathrm{d} q_{0} \psi_{0}^{*}\left(q_{0} ; \lambda\right)\left[\mathcal{O}\left(t_{1}, q_{0},-i \hslash \frac{\partial}{\partial q_{0}} ; \lambda\right), \mathcal{O}_{j}\left(t_{2}, q_{0},-i \hslash \frac{\partial}{\partial q_{0}} ; \lambda\right)\right]_{\pm} \psi_{0}\left(q_{0} ; \lambda\right),
\end{align}
where $\ket{\psi_0}$ is the ground state of the unperturbed system so $\psi_{0}\left(q_{0} ; \lambda\right) \equiv\left\langle q_{0} \mid \psi_{0}(x)\right\rangle$ is its normalized wave function dependent on the parameters denoted by $\lambda$, and  we are encapsulating all the integrals for every degree of freedom as $\int d q_{0} = \prod_{a=1}^{n} \int d q_{0}^{a}$.\newline

At this point we have everything we need to formulate the classical analog, the first step is to use the semiclassical approximation of the wave function $\psi_{0}\left(q_{0} ; \lambda\right)$:
\begin{equation}
\psi_{0}\left(q_{0} ; \lambda\right)=\sum_{\alpha} A_{(\alpha)}\left(q_{0}, I_{0} ; \lambda\right) \mathrm{e}^{\frac{i}{\hslash} S^{(\alpha)}\left(q_{0}, I_{0} ; \lambda\right)}
\end{equation}
which is in terms of $S^{(\alpha)}\left(q_{0}, I_{0} ; \lambda\right)$, being the generating function of the canonical transformation that takes us from $\left(q_{0}, p_{0}\right)$ to our action-angle variables the action-angle variables $\left(\phi_{0}, I_{0}\right)$ with $\phi_{0}=\left\{\phi_{0}^{ a}\right\}$ and $I_{0} \equiv I(t)=\left\{I_{a}\right\}$, one for each degree of freedom denoted by $a$, and also $b$ in the definition of the function $A$:
\begin{equation}
A_{(\alpha)}\left(q_{0}, I ; \lambda\right)= \sqrt{\frac{1}{(2 \pi)^{n}} \operatorname{det}\left(\frac{\partial \phi_{0}^{(\alpha) a}}{\partial q_{0}^{b}}\right)},
\end{equation}
and finally, $\alpha$ stands for the different branches of $S$. These branches correspond to different regions of solutions in phase space in which we can separate our theory of study, take the simple pendulum for example, as we said earlier it can have an oscillatory or rotatory behavior in phase space, and for each behavior there are different action-angle variables.\newline

Using this semicalssical approximation of the wave function within our expected values we obtain
\begin{equation}
\langle\hat{\mathcal{O}}(t)\rangle_{0}=\int \frac{\mathrm{d} q_{0}}{(2 \pi)^{n}} \sum_{\alpha} \operatorname{det}\left(\frac{\partial \phi_{0}^{(\alpha) a}}{\partial q_{0}^{b}}\right) \mathcal{O}_{i}\left(t, q_{0}, \frac{\partial S^{(\alpha)}}{\partial q_{0}} ; \lambda\right)+f(\hslash^2),
\end{equation}
and
\begin{equation}
\begin{aligned}
&\left\langle\left[\hat{\mathcal{O}}_{i}\left(t_{1}\right), \hat{\mathcal{O}}_{j}\left(t_{2}\right)\right]_{\pm}\right\rangle_{0} \\
&=\int \frac{\mathrm{d} q_{0}}{(2 \pi)^{n}} \sum_{\alpha} \operatorname{det}\left(\frac{\partial \phi_{0}^{(\alpha) a}}{\partial q_{0}^{b}}\right) \\
&\quad \times\left[\mathcal{O}_{i}\left(t_{1}, q_{0}, \frac{\partial S^{(\alpha)}}{\partial q_{0}} ; \lambda\right), \mathcal{O}_{j}\left(t_{2}, q_{0}, \frac{\partial S^{(\alpha)}}{\partial q_{0}} ; \lambda\right)\right]_{\pm}+f_{\pm}(\hslash^2)
\end{aligned}
\end{equation}
where $f(\hslash)$ and $f_{\pm}(\hslash)$ are functions at least proportional in second order to $\hslash$ which we are going to consider negligible with respect to our first terms.\newline

The next step is to substitute our commutators for Poisson brackets
\begin{equation}
[f(t_1),g(t_2)] \rightarrow i\hslash \left\{f\left(t_{1}\right), g\left(t_{2}\right)\right\}_{\left(q_{0}, p_{0}^{(\alpha)}\right)}=\sum_{a=1}^{n}\left(\frac{\partial f\left(t_{1}\right)}{\partial q_{0}^{a}} \frac{\partial g\left(t_{2}\right)}{\partial p_{a 0}^{(\alpha)}}-\frac{\partial f\left(t_{1}\right)}{\partial p_{a 0}^{(\alpha)}} \frac{\partial g\left(t_{2}\right)}{\partial q_{0}^{a}}\right),
\end{equation}
and the anticommutators for products of the corresponding functions\footnote{This  substitution is suited for bosonic operators, for fermionic operators  we must do the opposite, we replace the anticommutators by Poisson brackets and the commutators by the products of functions.}, we also follow the Hamilton-Jacobi rule stated at the beginning of the chapter which tells us to replace $p_{0}^{(\alpha)}$ by $\frac{\partial S^{(\alpha)}}{\partial q_{0}}$. Doing so we get
\begin{equation}
\left\langle\hat{\mathcal{O}}_{i}(t)\right\rangle_{0} \approx \int \frac{\mathrm{d} q_{0}}{(2 \pi)^{n}} \sum_{\alpha} \operatorname{det}\left(\frac{\partial \phi_{0}^{(\alpha) a}}{\partial q_{0}^{b}}\right) \mathcal{O}_{i}\left(t, q_{0}, p_{0}^{(\alpha)} ; \lambda\right)
\end{equation}
\begin{align}
\left\langle\left[\hat{\mathcal{O}}_{i}\left(t_{1}\right), \hat{\mathcal{O}}_{j}\left(t_{2}\right)\right]_{+}\right\rangle_{0} &\nonumber\approx 2 \int \frac{\mathrm{d} q_{0}}{(2 \pi)^{n}} \sum_{\alpha} \operatorname{det}\left(\frac{\partial \phi_{0}^{(\alpha) a}}{\partial q_{0}^{b}}\right) \\
& \times \mathcal{O}_{i}\left(t_{1}, q_{0}, p_{0}^{(\alpha)} ; \lambda\right) \mathcal{O}_{j}\left(t_{2}, q_{0}, p_{0}^{(\alpha)} ; \lambda\right)
\end{align}

\begin{align}
\left\langle\left[\hat{\mathcal{O}}_{i}\left(t_{1}\right), \hat{\mathcal{O}}_{j}\left(t_{2}\right)\right]_{-}\right\rangle_{0} \approx & i \hslash \int \frac{\mathrm{d} q_{0}}{(2 \pi)^{n}} \sum_{\alpha} \operatorname{det}\left(\frac{\partial \phi_{0}^{(\alpha) a}}{\partial q_{0}^{b}}\right) \\
&\quad \times\left\{\mathcal{O}_{i}\left(t_{1}, q_{0}, p_{0}^{(\alpha)} ; \lambda\right), \mathcal{O}_{j}\left(t_{2}, q_{0}, p_{0}^{(\alpha)} ; \lambda\right)\right\}_{\left(q_{0}, p_{0}^{(\alpha)}\right)}
\end{align}
Up to this point we have been keeping the choice of branch open with $\alpha$, but in order to make $\phi_{0}^{(\alpha)}$ single-valued we choose the one defined by $\phi_0 \in [0,2\pi]$, thus making the $\alpha$ label redundant, so we will omit it from now on. With this selection done we can conduct the change of variables $q_{0} \rightarrow \phi_{0}$ to leave everything in terms of averages with respect to the angle variable, analogously to what we did for Hannay's angle, obtaining 

\begin{equation}
    \left\langle\hat{\mathcal{O}}_{i}(t)\right\rangle_{0} \approx \frac{1}{(2 \pi)^{n}} \oint \mathrm{d} \phi_{0} \mathcal{O}_{i}(t)=\left\langle\mathcal{O}_{i}(t)\right\rangle 
\end{equation}

\begin{align}
\left\langle\left[\hat{\mathcal{O}}_{i}\left(t_{1}\right), \hat{\mathcal{O}}_{j}\left(t_{2}\right)\right]_{+}\right\rangle_{0} & \approx \frac{2}{(2 \pi)^{n}} \oint \mathrm{d} \phi_{0} \mathcal{O}_{i}\left(t_{1}\right) \mathcal{O}_{j}\left(t_{2}\right) \\
&=2\left\langle\mathcal{O}_{i}\left(t_{1}\right) \mathcal{O}_{j}\left(t_{2}\right)\right\rangle 
\end{align}
\begin{align}
    \left\langle\left[\hat{\mathcal{O}}_{i}\left(t_{1}\right), \hat{\mathcal{O}}_{j}\left(t_{2}\right)\right]_{-}\right\rangle_{0} & \approx \frac{i \hslash}{(2 \pi)^{n}} \oint \mathrm{d} \phi_{0}\left\{\mathcal{O}_{i}\left(t_{1}\right), \mathcal{O}_{j}\left(t_{2}\right)\right\}_{\left(q_{0}, p_{0}\right)} \\
&=i \hslash\left\langle\left\{\mathcal{O}_{i}\left(t_{1}\right), \mathcal{O}_{j}\left(t_{2}\right)\right\}_{\left(q_{0}, p_{0}\right)}\right\rangle
\end{align}
where alongside the reduced notation $\mathcal{O}_{i}(t)=\mathcal{O}_{i}\left(t, q_{0}, p_{0} ; \lambda\right)$ we also used the following definition for the average of a function
\begin{equation}\label{prom clasico}
\langle f\rangle=\frac{1}{(2 \pi)^{n}} \oint \mathrm{d} \phi_{0} f = \frac{1}{(2 \pi)^{n}} \prod_{a=1}^{n} \int_{0}^{2 \pi} d \phi_{0}^{a} f.
\end{equation}

Now that the right hand side of the expected values has been simplified this far we can confidently say that classical functions $\mathcal{O}_{i}\left(t, q_{0}, p_{0} ; \lambda\right)$ are given by
\begin{equation}
\mathcal{O}_{i}(t)=\mathcal{O}_{i}(q(t), p(t) ; \lambda)=\left(\partial_{i} H(q(t), p(t) ; \lambda)\right)_{q(t), p(t)}
\end{equation}
with $H(q(t), p(t) ; \lambda)$ being the classical counterpart of the Hamiltonian operator $\hat{H}(\hat{q}(t), \hat{p}(t) ; \lambda)$, where the classical variables $q(t)$ and $p(t)$ are expressed in terms of the initial conditions $q_{0}=q(t=0), p_{0}=p(t=0)$ by solving the Hamilton equations of motion. Substituting this results  in \eqref{QMT pathint} we get our \textit{classical analog for the Quantum Metric Tensor }
\begin{equation}
g_{i j}^{(0)}(\lambda) \approx \frac{1}{\hslash^{2}} g_{i j}(I ; \lambda)
\end{equation}
where
\begin{equation}
g_{i j}(I ; \lambda)=-\int_{-\infty}^{0} \mathrm{~d} t_{1} \int_{0}^{\infty} \mathrm{d} t_{2}\left(\left\langle\mathcal{O}_{i}\left(t_{1}\right) \mathcal{O}_{j}\left(t_{2}\right)\right\rangle-\left\langle\mathcal{O}_{i}\left(t_{1}\right)\right\rangle\left\langle\mathcal{O}_{j}\left(t_{2}\right)\right\rangle\right).
\end{equation}
This classical analog allows us to measure the distance on parameter space between two points in phase space corresponding to infinitesimally different parameters.

Finally, plugging the operators in \eqref{BerryC pathint} we obtain the \textit{classical analog of the Berry curvature} 
\begin{equation}
F_{i j}^{(0)}(\lambda) \approx \frac{1}{\hslash} F_{i j}(I ; \lambda)
\end{equation}
where
\begin{equation}
F_{i j}(I ; \lambda)=\int_{-\infty}^{0} \mathrm{~d} t_{1} \int_{0}^{\infty} \mathrm{d} t_{2}\left\langle\left\{\mathcal{O}_{i}\left(t_{1}\right), \mathcal{O}_{j}\left(t_{2}\right)\right\}_{\left(q_{0}, p_{0}\right)}\right\rangle.
\end{equation}
which in reality should is the curvature of Hannay's connection, since with it we can calculate Hannay's angle. \newline

It should be noted that the analog of the QMT involves  a factor of $1 / \hslash^{2}$  while the one for Berry's curvature only has $1 / \hslash$. These different factors can be traced back to the replacement of the commutators by the Poisson brackets, which introduces $\hslash$, while in the replacement of the anticommutators does not \cite{classical QMT, classical QGT}.\newline

With this examples we have seen that quantities created with an emphasis on quantum mechanics also have an applicability in classical mechanics, in the next chapter we will follow these ideas to construct the classical analog of the quantum covariance matrix an its derived quantities.

\chapter{Classical analogs of the Quantum Covariance Matrix, purity and entropy}

In this final chapter we will generate and study the classical analogs of the quantum covariance matrix, which if the state is Gaussian, carries the complete information about the purity, linear entropy and von Neumann entropy of our system, so we will generate classical analogs for these quantities as well. 

\section{Classical analog of the Quantum Covariance Matrix}

To generate a classical analog of the Quantum Covariance Matrix we will follow the same assumptions that in the previous chapter, namely that our classical system has to be integrable, for the action-angle variables $I=\left\{I_{a}\right\}$ and $\varphi=\left\{\varphi_{a}\right\}$ to exist, and that we have chosen our branch of the action $S$.\newline 

We will make use of the Wigner formalism so that the expectation value of an operator $\hat{\mathbf{O}}(\hat{\mathbf{q}}, \hat{\mathbf{p}})$ can be written as
\begin{equation}
\langle\hat{\mathbf{O}}\rangle_{m}=\int_{-\infty}^{\infty} \mathrm{d}^{N} q \mathrm{~d}^{N} p W_{m} \mathcal{O},
\end{equation}
where $W_{m}$ is the Wigner function in its original formulation \eqref{Wigner Original}, and $O_W$ is the \textit{Weyl transform} \cite{PaperWigner} of $\hat{\mathbf{O}}$, which are respectively given by
\begin{equation}
W_{m}(q, p)=\frac{1}{(2 \pi \hslash)^{N}} \int_{-\infty}^{\infty} \mathrm{d}^{N} z \mathrm{e}^{-\frac{\mathrm{i} p \cdot z}{\hslash}} \psi_{m}\left(q+\frac{z}{2}\right) \psi_{m}^{*}\left(q-\frac{z}{2}\right),
\end{equation}
\begin{equation}
O_W(q, p)=\int_{-\infty}^{\infty} \mathrm{d}^{N} z \mathrm{e}^{-\frac{\mathrm{i} p \cdot z}{\hslash}}\left\langle q+\frac{z}{2}|\hat{\mathbf{O}}(\hat{\mathbf{q}}, \hat{\mathbf{p}})| q-\frac{z}{2}\right\rangle
\end{equation}
notice that in the Wigner function the factor of $1/2$ in the exponential is not included,  thus we must only use $(2\pi \hslash)$ in the denominator, and also that here for utility we will put in the subindex the state of the wavefunction instead of the density matrix.
Also, we have use the simplified notation $p \cdot z=\sum_{a=1}^{N} p_{a} z_{a}$.\newline

When we apply the classical approximation, denoted by $\simeq$ and consisting of making $\hslash \rightarrow 0$ and  $m \rightarrow\infty$, the product $\hslash m$ becomes a constant with units of action that we will denote by $I_m$, then the Wigner function $W_{m}(q, p)$ takes the form of a delta function as \cite{classical Berry}
\begin{equation}
W_{m}(q, p) \simeq \frac{1}{(2 \pi)^{N}} \delta\left(I(q, p)-I_{m}\right).
\end{equation}
With these mathematical tools we can now get the classical approximation of the quantum covariance matrix \eqref{qcov matrix} by appliyng them to the expected values, let us begin with the one containing just a single operator

\begin{align}
\left\langle\hat{\mathbf{q}}_{a}\right\rangle_{m} & \simeq \int_{-\infty}^{\infty} \mathrm{d}^{N} q \mathrm{~d}^{N} p \frac{1}{(2 \pi)^{N}} \delta\left(I(q, p)-I_{m}\right) q_{a} \\
&=\frac{1}{(2 \pi)^{N}} \int_{0}^{\infty} \mathrm{d}^{N} I \int_{0}^{2 \pi} \mathrm{d}^{N} \varphi \delta\left(I-I_{m}\right) q_{a}(I, \varphi) \\
&=\frac{1}{(2 \pi)^{N}} \int_{0}^{2 \pi} \mathrm{d}^{N} \varphi q_{a}\left(I_{m}, \varphi\right) \\
&=\left\langle q_{a}\right\rangle_{\mathrm{cl}},
\end{align}
from which we learned that our expected values can be approximated by the average in terms of the action angle variables, so all the rest of expected values needed can be written as
\begin{align}
\left\langle\hat{\mathbf{q}}_{a} \hat{\mathbf{q}}_{b}\right\rangle_{m} & \simeq\left\langle q_{a} q_{b}\right\rangle_{\mathrm{cl}} \\
\left\langle\hat{\mathbf{p}}_{a}\right\rangle_{m} & \simeq\left\langle p_{a}\right\rangle_{\mathrm{cl}} \\
\left\langle\hat{\mathbf{p}}_{a} \hat{\mathbf{p}}_{b}\right\rangle_{m} & \simeq\left\langle p_{a} p_{b}\right\rangle_{\mathrm{cl}} \\
\frac{1}{2}\left\langle\hat{\mathbf{q}}_{a} \hat{\mathbf{p}}_{b}+\hat{\mathbf{p}}_{b} \hat{\mathbf{q}}_{a}\right\rangle_{m} & \simeq\left\langle q_{a} p_{b}\right\rangle_{\mathrm{cl}}.
\end{align}

So the classical analog of our quantum covariance matrix is simply 
\begin{equation}
\sigma \simeq \sigma^{\mathrm{cl}} 
\end{equation}
with matrix elements
\begin{equation}
\sigma_{\alpha \beta}^{\mathrm{cl}}:=\left\langle r_{\alpha} r_{\beta}\right\rangle_{\mathrm{cl}}-\left\langle r_{\alpha}\right\rangle_{\mathrm{cl}}\left\langle r_{\beta}\right\rangle_{\mathrm{cl}},
\end{equation}
taking the expectation values as stated above.\newline

It should be noted that we have not needed for our state to be Gaussian, this will be the case only until the next section.\newline

\section{Classical analog of the Purity and Entropy}

Using the classical analog of the classical covariance matrix we can construct classical analogs of the purity, linear entropy and von Neumann entropy for Gaussian states, since as we stated in Chapter $3$, all the information needed is contained within it.\newline

Let us begin with the purity, using $\sigma_{(n)} \simeq$ $\sigma_{(n)}^{\mathrm{cl}}$, \eqref{purity gaussian} and the Bohr-Sommerfeld quantization rule for the action variables $\hslash / 2 \rightarrow I_{k}$, we define the following classical analog
\begin{equation}\label{purity classical}
\mu^{\mathrm{cl}}\left(a_{1}, a_{2}, \ldots, a_{n}\right):=\frac{1}{\sqrt{\operatorname{det} \sigma_{(n)}^{\mathrm{cl}}}} \prod_{k=1}^{n} I_{a_{k}},
\end{equation}
where the action variable $I_{a_{k}}$ is associated with the $k$-th normal mode. Now for the linear entropy, since its simply $S_L = 1 - \mu$, we naturally define its classical analog as 
\begin{equation}\label{lin ent c}
S_{L}^{\mathrm{cl}}\left(a_{1}, a_{2}, \ldots, a_{n}\right):=1-\mu^{\mathrm{cl}}\left(a_{1}, a_{2}, \ldots, a_{n}\right).
\end{equation}
We must remark that because $\mu^{\mathrm{cl}}$ and $S_{L}^{\mathrm{cl}}$ are classical functions, we do not need have solved the quantum system to calculate them.\newline

However, we might be inclined to avoid the Bohr-Sommerfeld quantization rule to have a completely classical definition, or even to have more closely related equations between the quantum and classical versions by eliminating the product of action variables in \eqref{purity classical} that does not appear on \eqref{purity gaussian}. To do so we can make every action variable in \eqref{purity classical} and \eqref{lin ent c} equal to a real positive constant $\alpha$ getting
\begin{align}
\tilde{\mu}^{\mathrm{cl}}\left(a_{1}, a_{2}, \ldots, a_{n}\right) &:=\lim _{I_{k} \rightarrow \alpha} \mu^{\mathrm{cl}}\left(a_{1}, a_{2}, \ldots, a_{n}\right) \\
&=\alpha^{n} \lim _{I_{k} \rightarrow \alpha} \frac{1}{\sqrt{\operatorname{det} \sigma_{(n)}^{\mathrm{cl}}}} \\
\tilde{S}_{L}^{\mathrm{cl}}\left(a_{1}, a_{2}, \ldots, a_{n}\right) &:=1-\tilde{\mu}^{\mathrm{cl}}\left(a_{1}, a_{2}, \ldots, a_{n}\right)
\end{align}
which we can consider classical analogs of the purity and linear quantum entropy respectively. \newline

Using once again our classical analog of the covariance matrix and and the Bohr-Sommerfeld quantization rule for the action variables, we define the classical function related to the von Neumann entropy
\begin{equation}
S^{\mathrm{cl}}\left(a_{1}, a_{2}, \ldots, a_{n}\right):=\sum_{k=1}^{n} \mathcal{S}^{\mathrm{cl}}\left(\nu_{k}\right)
\end{equation} 
where
\begin{equation}\label{vn c}
\begin{aligned}
\mathcal{S}^{\mathrm{cl}}\left(\nu_{k}\right):=&\left(\nu_{k}+\frac{1}{2}\right) \ln \left(\nu_{k}+\frac{1}{2}\right) -\left(\nu_{k}-\frac{1}{2}\right) \ln \left(\nu_{k}-\frac{1}{2}\right)
\end{aligned}
\end{equation}
and although might look similar to \eqref{von Neumann gaussian} the difference resides in the symplectic eigenvalues, since now the are obtained with the classical analog of the quantum covariance matrix
\begin{equation}
    \nu_{k}:=\nu_{k}^{\mathrm{cl}} / 2 I_{a_{k}},
\end{equation}
where $\sigma_{k}^{\mathrm{cl}}$ are these symplectic eigenvalues of $\sigma_{(n)}^{\mathrm{cl}}$.
For example, one particle with one degree of freedom will have
\begin{equation}
\nu_{1}=\frac{1}{2 I_{a_{1}}} \sqrt{\sigma_{p_{a_{1}} p_{a_{1}}}^{\mathrm{cl}} \sigma_{q_{a_{1}} q_{a_{1}}}^{\mathrm{cl}}-\left(\sigma_{q_{a_{1}} p_{a_{1}}}^{\mathrm{cl}}\right)^{2}} .
\end{equation}
that turns out rather similar when comparing with \eqref{vn 1dof nu}, except for action variable in the denominator. If we once again make all the action variables equal to $\alpha$ we get the function
\begin{equation}
\begin{aligned}
\tilde{S}^{\mathrm{cl}}\left(a_{1}, a_{2}, \ldots, a_{n}\right) &:=\sum_{k=1}^{n} \mathcal{S}^{\mathrm{cl}}\left(\tilde{\nu}_{k}\right) \\
\tilde{\nu}_{k} &:=\lim _{I_{k} \rightarrow \beta} \nu_{k}
\end{aligned}
\end{equation}
where $\mathcal{S}^{\mathrm{cl}}$ is \eqref{vn c} using $\tilde{nu}_k$ insted of $\nu_k$ and $\beta$ is a real positive constant that will not matter in the end as we shall see in the exampleswhich, as we will see, disappears during the calculation (as in the classical analog of the purity).

As final remark, the classical analog of the purity can be written in terms of the symplectic eigenvalues $\tilde{\sigma}_{k}$ as
\begin{equation}
\tilde{\mu}^{\mathrm{cl}}\left(a_{1}, a_{2}, \ldots, a_{n}\right)=\left(\frac{1}{2^{n}}\right) \prod_{k=1}^{n} \tilde{\sigma}_{k}^{-1}.
\end{equation}

In the next section we will use these definitions to calculate all the classical counterparts of the quantum quantities that we studied in Chapter $3$, and we will explore the meaning of the classical analog of the von Neumann entropy within this context since our results will turn out to be  the same.

\section{The coupled oscilators revisited}

To see how this classical analogs apply to a particular system we will once again study our two coupled harmonic oscillators whose Hamiltonian is
$$
\hat{H}=\frac{1}{2}\left[\hat{p}_{1}^{2}+\hat{p}_{2}^{2}+k\left(\hat{q}_{1}^{2}+\hat{q}_{2}^{2}\right)+k^{\prime}\left(\hat{q}_{1}^{2}-\hat{q}_{2}^{2}\right)\right]
$$
For this system we will have
\begin{align}
U &=\frac{1}{\sqrt{2}}\left(\begin{array}{cc}
1 & 1 \\
1 & -1
\end{array}\right) \\
W &=\left(\begin{array}{cc}
\omega_{1} & 0 \\
0 & \omega_{2}
\end{array}\right) \\
I &=\left(\begin{array}{cc}
I_{1} & 0 \\
0 & I_{2}
\end{array}\right)
\end{align}

so our classical average are
\begin{equation}
\sigma_{p_{a} p_{b}}^{(\text {cl})} =U^{\top} W I U 
\end{equation}
\begin{equation}
\sigma_{q_{a} q_{b}}^{(\text {cl})} =U^{\top} W^{-1} I U,
\end{equation}
with
\begin{equation}
    I =\operatorname{diag}\left\{I_{1}, I_{2}\right\}
\end{equation}

\begin{equation}
\sigma_{p_{a} p_{b}}^{(\text {cl})} =U^{\top} \Omega I U=\frac{1}{2}\left(\begin{array}{ll}
I_{1} \omega_{1}+I_{2} \omega_{2} & I_{1} \omega_{1}-I_{2} \omega_{2} \\
I_{1} \omega_{1}-I_{2} \omega_{2} & I_{1} \omega_{1}+I_{2} \omega_{2}
\end{array}\right)
\end{equation}
\begin{equation}
\sigma_{q_{a} q_{b}}^{(\text {cl})} =U^{\top} \Omega^{-1} I U=\frac{1}{2}\left(\begin{array}{cc}
\frac{I_{1}}{\omega_{1}}+\frac{I_{2}}{\omega_{2}} & \frac{I_{1}}{\omega_{1}}-\frac{I_{2}}{\omega_{2}} \\
\frac{I_{1}}{\omega_{1}}-\frac{I_{2}}{\omega_{2}} & \frac{I_{1}}{\omega_{1}}+\frac{I_{2}}{\omega_{2}}
\end{array}\right)
\end{equation}
if we focus on the first particle
\begin{equation}
\begin{aligned}
\sigma_{p_{1} p_{1}}^{(\text {cl})} &=\frac{1}{2}\left(I_{1} \omega_{1}+I_{2} \omega_{2}\right) \\
\sigma_{q_{1} q_{1}}^{(\text {cl})} &=\frac{1}{2}\left(\frac{I_{1}}{\omega_{1}}+\frac{I_{2}}{\omega_{2}}\right),
\end{aligned}
\end{equation}
therefore the classical symplectic eigenvalue for any of the particles is
\begin{equation}
\tilde{\nu}_{1}^{\text {class }}=\frac{1}{4 I_{1}} \sqrt{\frac{\left(I_{1} \omega_{1}+I_{2} \omega_{2}\right)\left(I_{2} \omega_{1}+I_{1} \omega_{2}\right)}{\omega_{1} \omega_{2}}}
\end{equation}
and applying the Bohr-Sommerfeld $I_1 = I_2 = \hslash/2$ quantization rule we get
\begin{equation}
\tilde{\nu}_{1}^{\text {cl}} \approx \frac{\omega_{1}+\omega_{2}}{4 \sqrt{\omega_{1} \omega_{2}}}
\end{equation}
from which we get that the purity is
\begin{equation}
\mu(1)=\frac{2 \sqrt{\omega_{1} \omega_{2}}}{\omega_{1}+\omega_{2}}
\end{equation}
which is exactly what we got from the quantum procedure. For the entropy 
\begin{equation}
S_{1}=\left(\frac{\omega_{1}+\omega_{2}}{4 \sqrt{\omega_{1} \omega_{2}}}+\frac{1}{2}\right) \ln \left(\frac{\omega_{1}+\omega_{2}}{4 \sqrt{\omega_{1} \omega_{2}}}+\frac{1}{2}\right)-\left(\frac{\omega_{1}+\omega_{2}}{4 \sqrt{\omega_{1} \omega_{2}}}-\frac{1}{2}\right) \ln \left(\frac{\omega_{1}+\omega_{2}}{4 \sqrt{\omega_{1} \omega_{2}}}-\frac{1}{2}\right)
\end{equation}
and thus both our results are exactly the same.\newline

To better understand why we obtain the same mathematical results of the purity and entropy from a classical point of view, we can observe that we have both local and global information when describing our system in terms of action-angle variables and since all the variables are correlated in this case the classical analogy of the von Neumann entropy provides us a measure of "non-separability" of the individual subsystems in phase space.

\chapter*{Conclusions} \chaptermark{Conclusions} 
	\addcontentsline{toc}{chapter}{Conclusions}

    In this thesis, we have thoroughly studied two apparently different perspectives of quantum information geometry, parameter space's point of view and the quantum covariance matrix. However, with \eqref{qcov matrix}, \eqref{qcov qgt 1}, \eqref{qcov qgt 2} and \eqref{qcov qgt 3} we showed that the two are closely related and thus should be studied together to understand the dynamics of quantum systems. \newline
    
    In chapter 3, we presented the standard method to calculate the purity, linear entropy, and von Neumann entropy of a quantum system that uses the density matrix. Subsequently, we introduced the quantum covariance matrix formalism and showed that, at least for Gaussian states, it reproduces the same results while being much simpler than the former.\newline
    
    This leaves open one intriguing question that should be explored in the following work. What exactly is the information not contained within the quantum covariance matrix that is needed to calculate the purity and entropy of non-Gaussian states? There is little literature exploring non-Gaussian states' entanglement, and this seems
    like an appropriate starting point.\newline
    
    In the final chapters, we developed classical analogs for all these quantum quantities and showed that if our state is Gaussian, it does not matter if we use the classical or quantum approach; we get the same results. This, in turn, implies that the genuinely quantum part of purity and entropy is contained in this non-Gaussian information.\newline
    
    Using our classical analog of the von Neumann entropy, we can get a measure of how inseparable is our classical system. Nonetheless, in quantum mechanics, there are already explicit separability criteria that must be met in order to do so \cite{separability}, then it should be possible to investigate how these conditions apply to our classical analogs and if it is possible to follow the same procedure of introducing the action-angle variables.\newline

\appendix
\chapter{Review of a few relevant probability and statistics concepts}

To understand the concept of the quantum covariance matrix is useful to briefly review some of the most important concepts in probability and statistics.

\section{Variance and standard deviation}

Although the mean or expectation value of a distribution is an useful summary of the information of our sample, it does not tells us very much about the distribution nor its range. For example, a random variable $X$ with possible values $\{-20,40,12,0,-12,-8\}$ has a mean of $2$, which is the same mean as the one from the constant random variable $Y=2$. To distinguish how different the distribution of $X$ is from the distribution of $Y$, we would require some quantity that measures how spread out the distributions are. The variance is one tool to do so.\newline

Let $X$ be a random variable with finite mean $\mu=E(X)$ ($E$ stands for expected value). Then the \textit{variance} of $X$, will be denoted by $\operatorname{Var}(X)$, and is defined as follows:
\begin{equation}
\operatorname{Var}(X)=E\left[(X-\mu)^{2}\right] .
\end{equation}
Which can be expressed in a simpler manner:
\begin{equation}
\begin{aligned}
\operatorname{Var}(X) &=E\left[(X-\mu)^{2}\right] \\
&=E\left(X^{2}\right)-2 \mu E(X)+\mu^{2} \\
&=E\left(X^{2}\right)-\mu^{2}
\end{aligned}
\end{equation}

As we can see, the variance has units of $ [X]^2$, therefore we would like some other quantity that relates more easily to $X$, for this purpose we define the \textit{ standard deviation} of $X$ as the nonnegative square root of $\operatorname{Var}(X)$.\newline

Regularly when dealing with only one random variable, the standard deviation is denoted by the symbol $\sigma$, and the variance is denoted by $\sigma^{2}$. If instead we are dealing with more than one random variable, to avoid confusion we include the name of the respective random variable in the subscript, e.g., $\sigma_{X}$ would be the standard deviation of $X$ while $\sigma_{Y}^{2}$ would be the variance of $Y$.

\section{Covariance and correlation}

When working with two random variables we could calculate all the quantities from the previous section but they would not provide any information about how the two variables are related or more specifically, about their tendency to vary together rather than independently.\newline

To understand how much the two random variables depend on each other we can utilize the covariance and correlation as fist endeavors to measure that dependence. However, it should be noted that these concepts can only interpret a particular type of dependence between the variables, which is linear dependence.\newline

Let $X$ and $Y$ be random variables having the finite expectation values $E(X)=\mu_{X}$ and $E(Y)=\mu_{Y}$ respectively, then the \textit{covariance} of $X$ and $Y$, which is denoted by $\operatorname{Cov}(X, Y)$, is defined as

\begin{align}
\operatorname{Cov}(X, Y)&=E\left[\left(X-\mu_{X}\right)\left(Y-\mu_{Y}\right)\right]
\\&=E(X Y)-\mu_{X} \mu_{Y} .
\end{align}

The covariance between $X$ and $Y$ intends to measure how one tends to increase while the other increases or decreases. If both grow or decline alongside each other the covariance will be positive, on the other hand if one increases while the other one decreases the covariance will be negative. Finally, if there is no connection between the growths of both variables then the covariance will be zero.\newline

Even if $\operatorname{Cov}(X, Y)$ provides us a number that somewhat measures how $X$ and $Y$ vary together, its magnitude does not carry that much significance since it is influenced by the overall magnitudes of $X$ and $Y$ individually. To generate a measure which gives us a "sense of how big" is the association between $X$ and $Y$ we will apply to the covariance a similar procedure of that in which we normalize wave functions in physics to give it its probabilistic interpretation.\newline

Considering our two random variables $X$ and $Y$ with corresponding finite and not null standard deviations $\sigma_{X}$ and $\sigma_{Y}$, we define the \textit{correlation} between them, regularly denoted by $\rho(X, Y)$ (we will not do so since we are reserving the symbol for the density matrix), as follows:
\begin{equation}
Cor(X, Y)=\frac{\operatorname{Cov}(X, Y)}{\sigma_{X} \sigma_{Y}}
\end{equation}
which has the possible values:
\begin{equation}
-1 \leq \rho(X, Y) \leq 1.
\end{equation}

As a side note, there are two important inequalities involving the expected values and variances, the Schwarz Inequality which is
\begin{equation}
[E(X Y)]^{2} \leq E\left(X^{2}\right) E\left(Y^{2}\right) .
\end{equation}
and the Cauchy-Schwarz Inequality 
\begin{equation}
[\operatorname{Cov}(X, Y)]^{2} \leq \sigma_{X}^{2} \sigma_{Y}^{2},
\end{equation}
that tells us that the covariance is delimited by the individual variances \cite{Probability}.

\clearpage
\thispagestyle{empty}
\begin{flushright}
    	\begin{figure}[b]
		\includegraphics[width=0.7\linewidth]{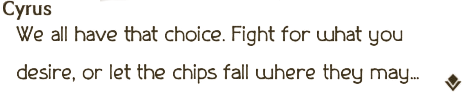}
		\caption*{}
	\end{figure}
	
\end{flushright}

\end{document}